\documentclass{emulateapj}

\usepackage{natbib}

\shorttitle{Anatomy of helical relativistic jets: The case of S5~0836+710}
\shortauthors{Perucho et al.}

\begin{document}

\title{Anatomy of helical extragalactic jets: The case of S5~0836+710}

\author{M. Perucho\altaffilmark{1,3}}
\email{mailto: manel.perucho@uv.es}
\affil{$^1$Departament d'Astronomia i Astrof\'{\i}sica, Universitat de Val\`encia, 46100, Burjassot (Val\`encia), 
Spain}

 \author{Y.Y. Kovalev\altaffilmark{1,2,3}}
\affil{$^2$Astro Space Center of Lebedev Physical Institute, 117997 Moscow, Russia}

\author{A.P. Lobanov\altaffilmark{3}}
\affil{$^3$Max-Planck-Institut f\"ur Radioastronomie, Auf dem H\"ugel 69, 53121 Bonn, Germany}

 \author{P.E. Hardee\altaffilmark{4}}
 \affil{$^4$Department of Physics \& Astronomy, The University of Alabama, Tuscaloosa, AL 35487, USA} 
 \author{I. Agudo\altaffilmark{5,6}}
\affil{$^5$Instituto de Astrof\'{\i}sica de Andaluc\'{\i}a, Apartado 3004, 18080, Granada, Spain}
 \affil{$^6$Institute for Astrophysical Research, Boston University, 725 Commonwealth Avenue, Boston, MA 02215, 
USA}

\begin{abstract}
Helical structures are common in extragalactic jets. They are usually attributed
in the literature to periodical phenomena in the source (e.g., precession). In
this work, we use VLBI data of the radio-jet in the quasar \object{S5~0836+710} 
and hypothesize that the ridge-line of helical jets like this corresponds to a
pressure maximum in the jet and assume that the helically twisted pressure
maximum is the result of a helical wave pattern. For our study, we use
observations of the jet in \object{S5~0836+710} at different
frequencies and epochs. The results show that the structures observed are
physical and not generated artificially by the observing arrays. Our hypothesis
that the observed intensity ridge-line can correspond to a helically twisted
pressure maximum is confirmed by our observational tests. This interpretation
allows us to explain jet misalignment between parsec and kiloparsec scales when
the viewing angle is small, and also brings us to the conclusion that
high-frequency observations may show only a small region of the jet flow
concentrated around the maximum pressure ridge-line observed at low frequencies.
Our work provides a potential explanation for the apparent transversal
superluminal speeds observed in several extragalactic jets by means of 
transversal shift of an apparent core position with time.
\end{abstract}

\keywords{galaxies: jets - hydrodynamics - instabilities - quasars: individual:
\object{S5~0836+710}}

\section{Introduction}\label{sec:intro}

Jets in Active Galactic Nuclei (AGN) involve some of the most energetic
processes in the Universe. The way they 
propagate and interact with the ambient medium and, in particular, their
stability properties have been thoroughly 
studied \citep[see][for reviews]{ha06,ha11,pe11}. These jets are subject to the
growth of different instabilities and 
many of the structures observed like knots, bendings, helices, have been
interpreted as a result of this physical process. Although there could be other
ways to produce these structures, such as interactions with clumps of dense gas or
precession of the central engine, these in turn can also give rise to the
growth of the instabilities via coupling to any of the unstable modes
\citep[see, e.g.][for numerical studies of coupling to Kelvin-Helmholtz and current-driven instabilities 
in relativistic flows]
{pe+05,mi07,mb09,mi09,mi10,pe+10,mi11}. Our interest in understanding how this process works in
jets lies in the possibility of obtaining the physical parameters of the flow
(velocity, gas density and sound speed), via modeling the observed structures
\citep[e.g.,][]{lz01,hr+05,he+11}. The Very Long Baseline Interferometry (VLBI)
technique has provided high-resolution observations that produce highly detailed
images of the structure of AGN jets, and that even resolves them transversally
\citep{lz01}. 

Radio images of many extragalactic jets from AGN present helical patterns on
different spatial scales. There are a number of possible processes that can
cause these structures, such as periodicity in the direction of ejection of the
flow due to precession. Precession can be  produced by the gravitational effect
of the accretion disk on the compact object in the center of the AGN or in a
binary-black-hole object \citep[see, e.g.,][]{li03,st03,lr05,ba06,sa06}. If this periodical motion or any other initial
perturbation of the relativistic jet couples to an unstable mode of a magneto-hydrodynamical
instability like Kelvin-Helmholtz \citep[KH,][]{pe+05,mi07,pe+10} or current driven 
\citep[CD][]{mi09,mb09,mi10,mi11} instabilities, the
perturbation may grow in amplitude as it is advected downstream with the jet
flow. Thus, the growing helices can give us indirect information about their
triggering process or eventually, about the physical parameters of the flow. 

\citet{har03} compared the intrinsic and observed nature of helical structures
via pseudo-synchrotron images of artificially built helical jets from linear KH
theory, following \citet{har00}. The results of this work were applied to the jet in 3C~120 and allowed
estimates to be made of some of the physical parameters of the jet. This work
was extended in \citet{hr+05}, where the authors constrained the jet parameters
more tightly and reproduced the basic  observed structure of 3C~120 by modeling
the helical structures as KH modes. This type of analysis has also been applied
to the jet in M~87 by \citet{he+11} who were able to constrain physical
parameters in the jet and surrounding cocoon. Recently, \citet{cma09} have
proposed a method (cross-entropy optimizer) to fit helical patterns produced by
precession in jets. The parameters obtained by this method are the jet velocity,
the jet viewing angle, and the jet opening angle. The authors generated
artificial data-sets, which were successfully fitted by their method. 

Another problem related to observations of VLBI-jets is the measurement of
superluminal transversal motion of components. \citet{bip08} has proposed
non-ballistic motion of physical components and misidentification as the origin of those
measurements, although the latter is not a valid explanation for the case of densely
time sampled jets \citep[e.g.,the case of NRAO 150 and OJ287,][]{ag07,ag10,ag11}.
In this context, it is interesting to note that components fitted in radio-jets do not necessarily represent 
physical entities, but just a way to model the continuous structure of those jets. Only in some cases it is 
possible to associate the components that have been fitted to physical structures, such as perturbations 
crossing the jet or standing features, both usually related to shocks in the literature.

In this paper, we present a deep analysis of multi-wavelength and multi-epoch
observations of the helical jet in \object{S5~0836+710} (0836+710 hereafter). 
Our analysis is based on the assumption that the maximum intensity of the radio emission 
coincides with
the location of the pressure maximum in the jet. This is a reasonable assumption
if the velocity of the emitting region of the flow is fairly symmetric and there
are no strong beaming asymmetries across the jet. The pseudo-synchrotron images
produced in \citet{har03}, \citet{hr+05} and \citet{he+11} also indicate that
the maximum intensity in helically perturbed jets comes from the location where
the pressure is at its maximum. Here, we present different tests that confirm
this point from the observational data. Our analysis does not put any constraint
on the origin of the helices, but we hypothesize that the initial helical motion
couples to either a KH or CD instability.

The luminous quasar \object{S5 0836+710} at a redshift $z=2.16$ hosts a powerful
radio jet extending up to kiloparsec scales \citep{hu92}. At this redshift,
$1\,\rm{mas} \simeq 8.4\,\rm{pc}$ \citep[see MOJAVE database and][]{wr06}\footnote{https://www.physics.purdue.edu/astro/mojave/, which 
uses $\Lambda CDM$ Cosmology from WMAP 5 year results, and Ned Wrights Cosmology Calculator
http://www.astro.ucla.edu/\\$\sim$wright/CosmoCalc.html}. VLBI monitoring of the
source \citep{ot98} has yielded estimates of the bulk Lorentz factor
$\gamma_\mathrm{j}=12$ and the viewing angle $\alpha_\mathrm{j}=3^\circ$ of the
flow. The presence of an instability developing in the jet is suggested by the kink structures observed
on milliarcsecond scales with ground VLBI \citep{kr90}. \citet{lo98} observed
the source at 5~GHz with VSOP\footnote{VLBI Space Observatory Program, a Japanese-led 
space VLBI mission operated in 1997-2007 by the Institute of Space and Astronautical Science, Sagamihara, 
Japan http://www.vsop.isas.jaxa.jp/top.html} and also reported oscillations of the ridge-line.
Identifying these structures with KH modes, they were able to
derive an estimate of the basic parameters of the flow. High dynamic range VSOP
and VLBA (Very Long Baseline Array of National Radio Astronomy Observatory, USA)
observations of 0836+710 at $1.6\,\rm{GHz}$ indicated the presence of an
oscillation at a wavelength as long as $\sim100\,\rm{mas}$ \citep{lo06}, which
cannot be readily reconciled 
with the jet parameters given in \citet{lo98}. \citet{pl07} have shown that the
presence of a shear layer allows fitting all the observed oscillation
wavelengths within a single set of parameters, assuming that they are produced
by KH instability growing in a cylindrical outflow. In this picture, the
longest mode corresponds to a surface mode growing in the outer layers, whereas
the shorter wavelengths are identified with body modes developing in the inner
radii of the jet. This result is reviewed here, in the light of additional
information. 

The paper is structured as follows. In Section~2 we present the results from
observations that will be used in later sections for our study. In Section~3 we
describe the implications deduced from this set of multi-frequency and 
multi-epoch observations. In Section~4 we present a model to fit the ridge-line
structures in relativistic jets. 
In Section~5 we discuss some properties of the jet inferred from the observations. 
Section~6 is devoted to the nature of the observed superluminal transversal
motions in parsec-scale AGN jets in the context of our work. Finally, we present
our summary in Section~7.

\section{Multi-epoch and multi-frequency observations of the jet in
0836+710}\label{obs}

\begin{figure*}[!t]
 \centering
    \includegraphics[clip,angle=0,width=0.4\textwidth]{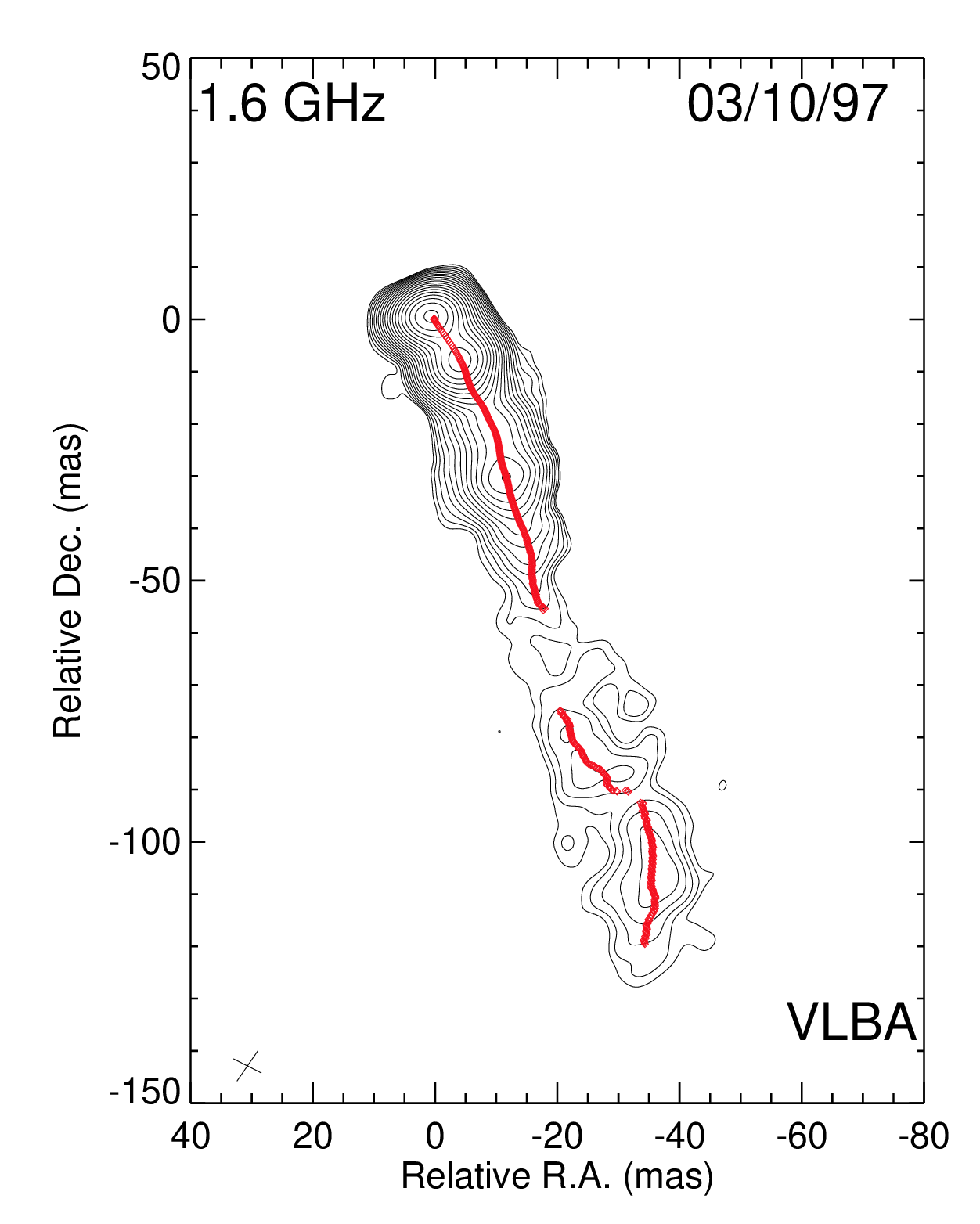}
    \includegraphics[clip,angle=0,width=0.4\textwidth]{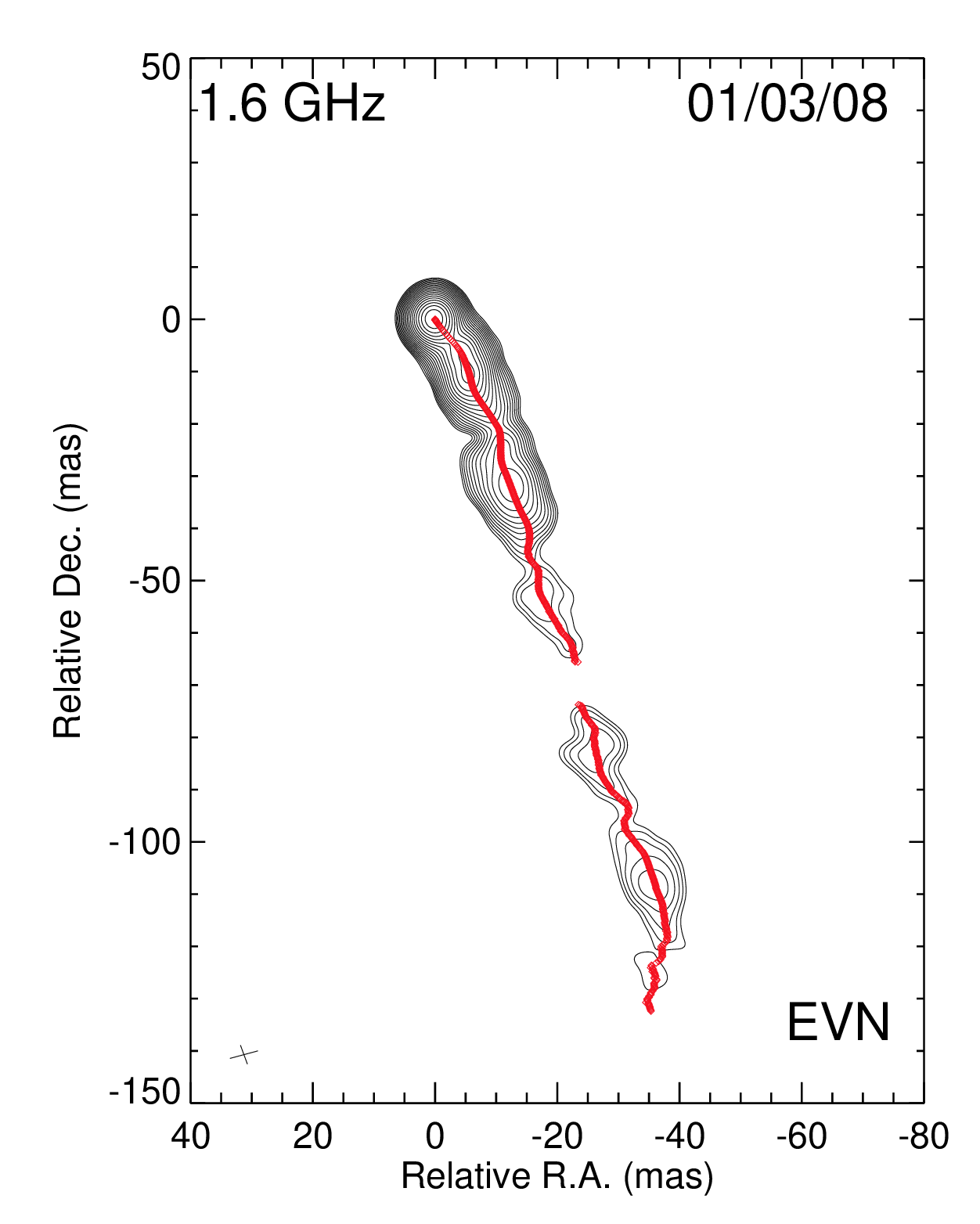}\\
    \includegraphics[clip,angle=0,width=0.4\textwidth]{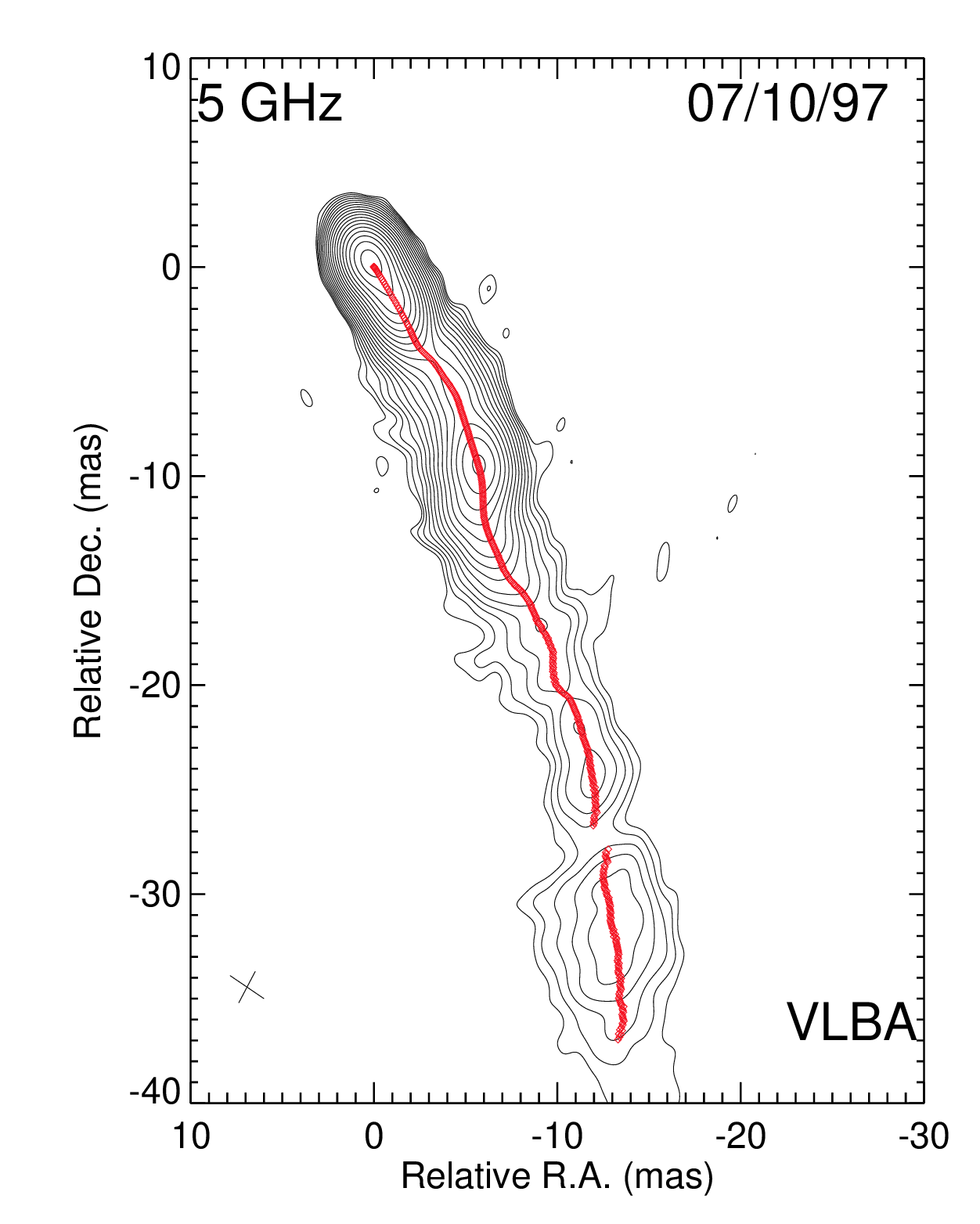}
    \includegraphics[clip,angle=0,width=0.4\textwidth]{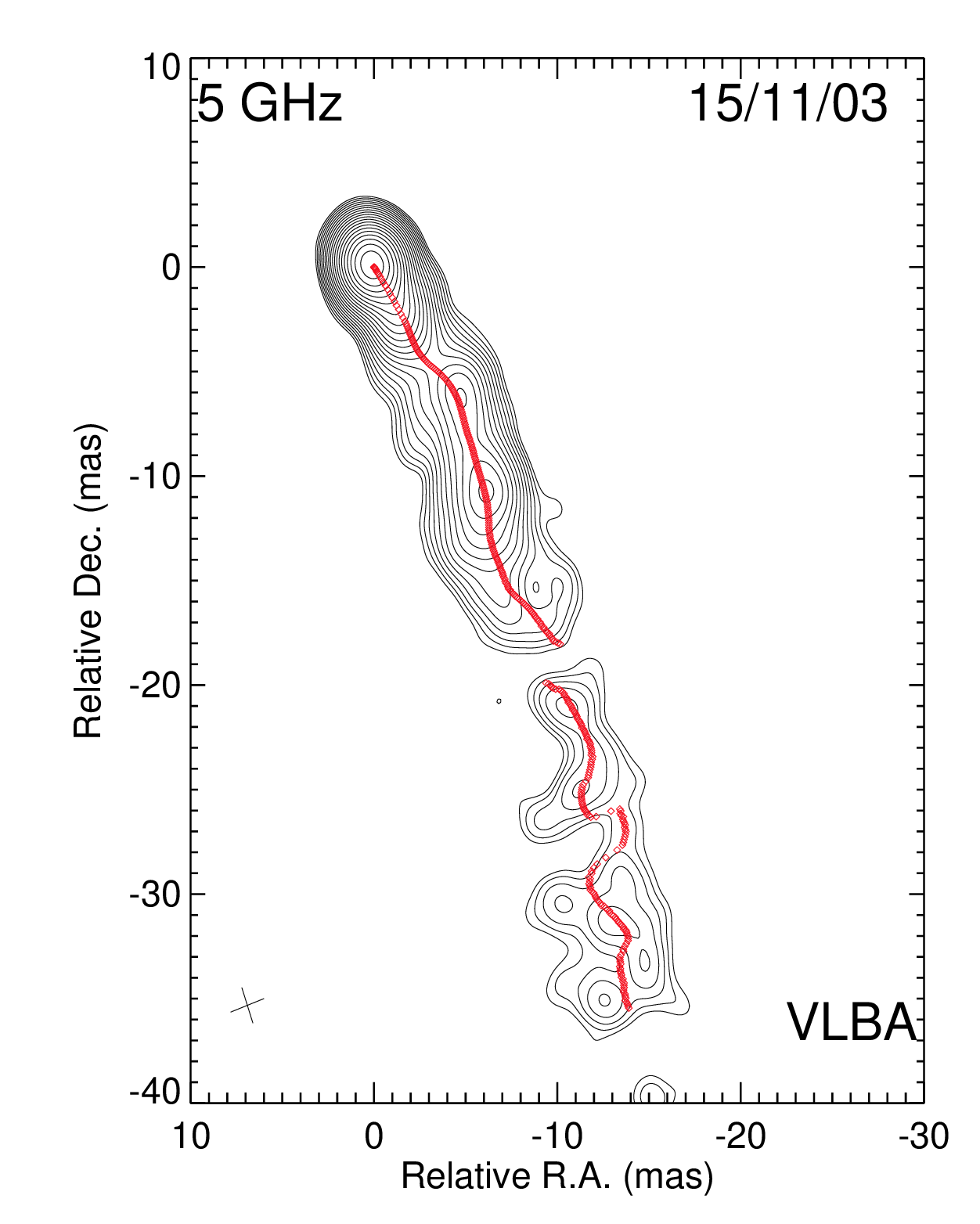}
\caption{Radio map of the jet in 0836+710 at 1.6~GHz (top) in 1997 and 2008 and
5~GHz (bottom) in 1997 and 
2003. Isocontours range from 2 mJy to 1.45 Jy in the 1.6~GHz images and from 1.5
mJy to 1.08 Jy in the 
5~GHz ones. The red diamonds indicate the position of the ridge-line.}
\label{fig:lmaps}
\end{figure*}

\begin{figure*}[!t]
    \centering
    \includegraphics[clip,angle=0,width=0.3\textwidth]{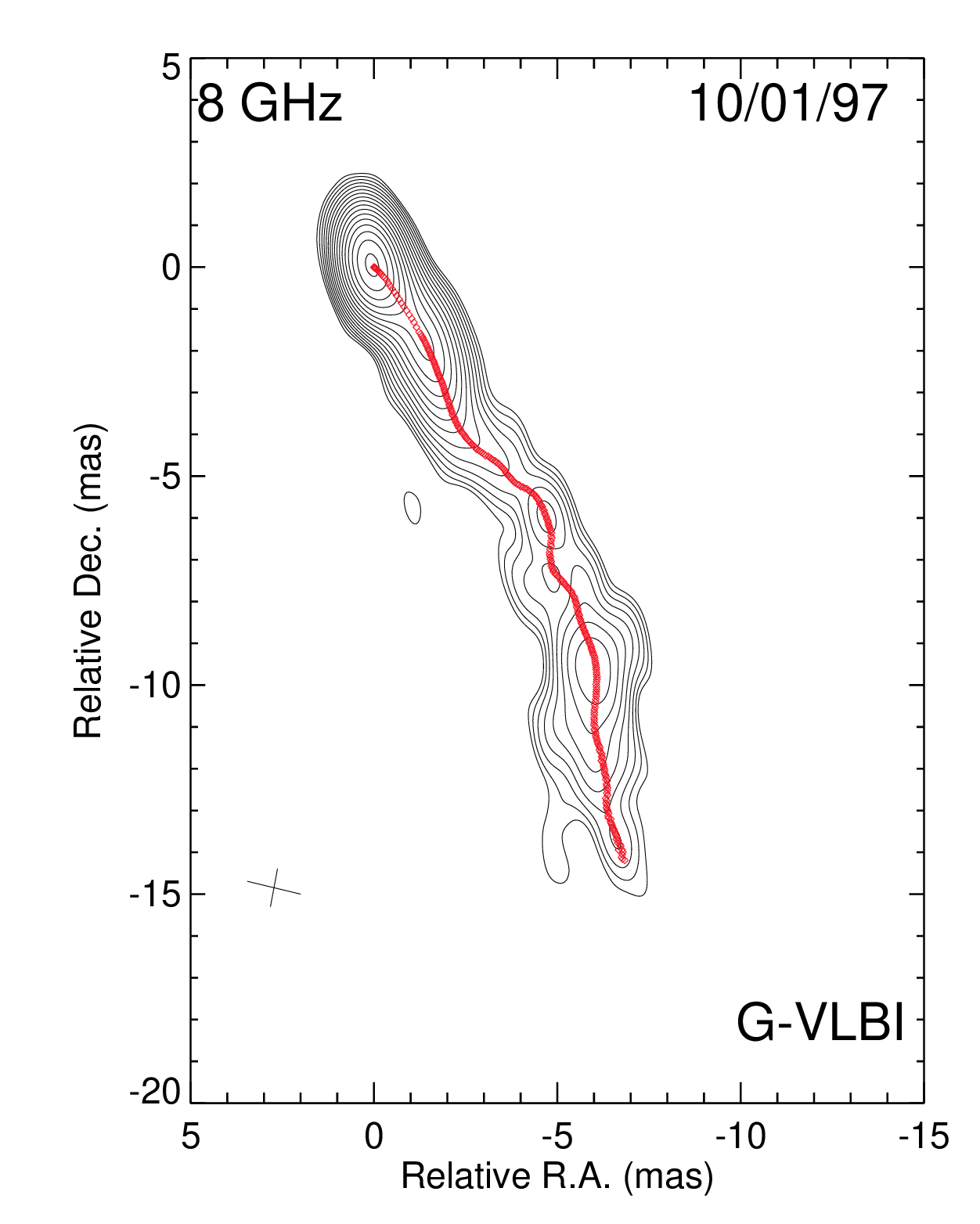}
    \includegraphics[clip,angle=0,width=0.3\textwidth]{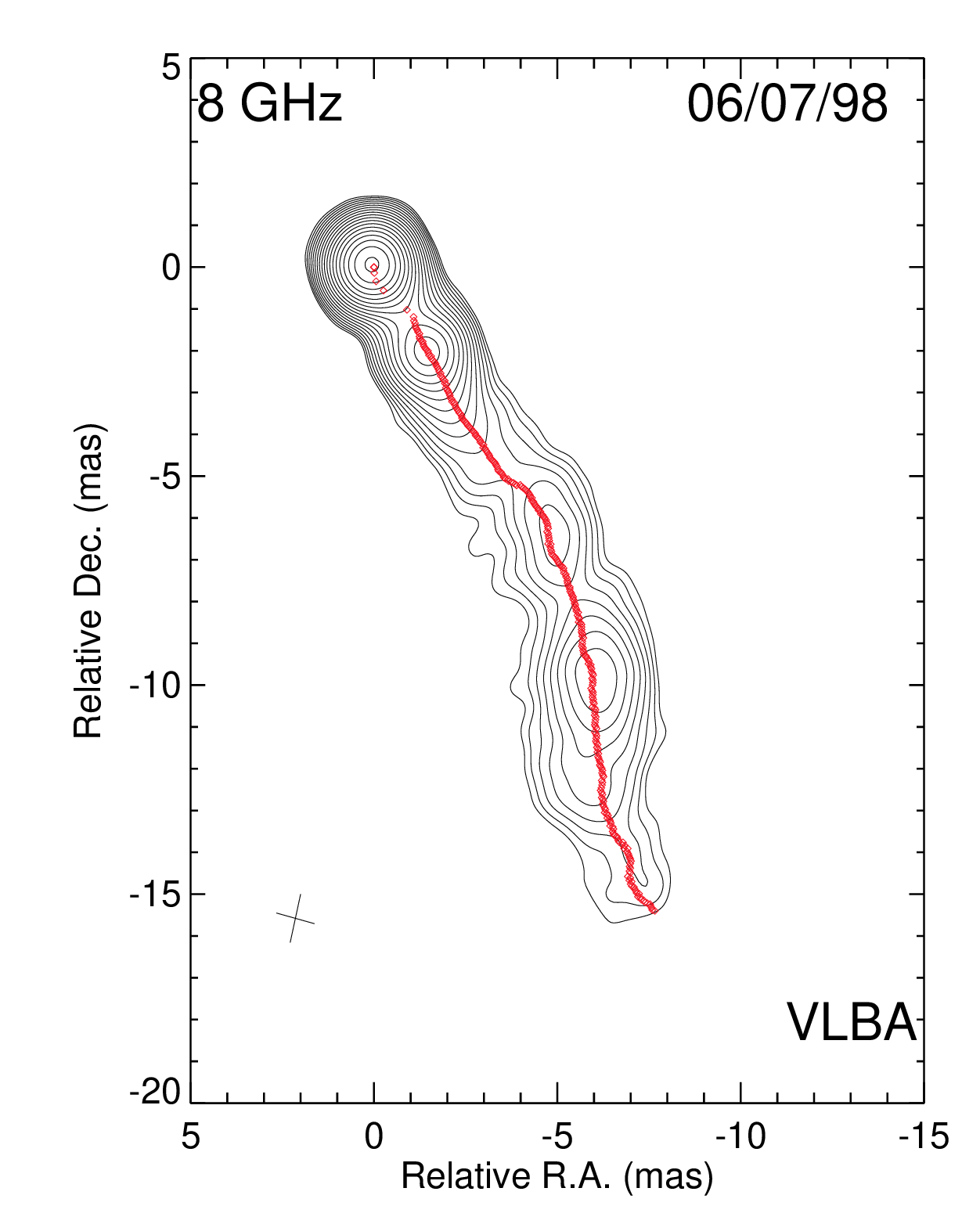}
    \includegraphics[clip,angle=0,width=0.3\textwidth]{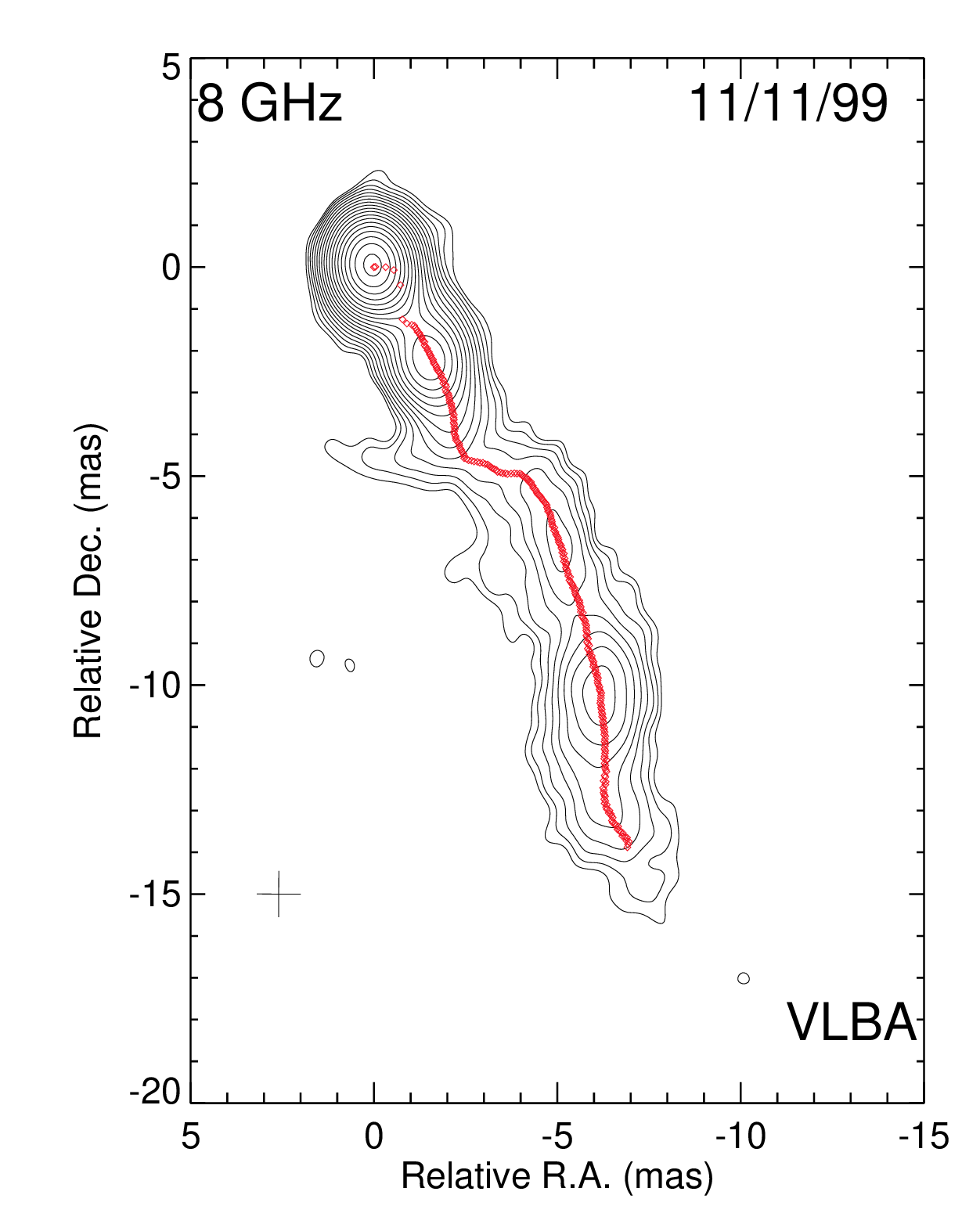}\\
    \includegraphics[clip,angle=0,width=0.3\textwidth]{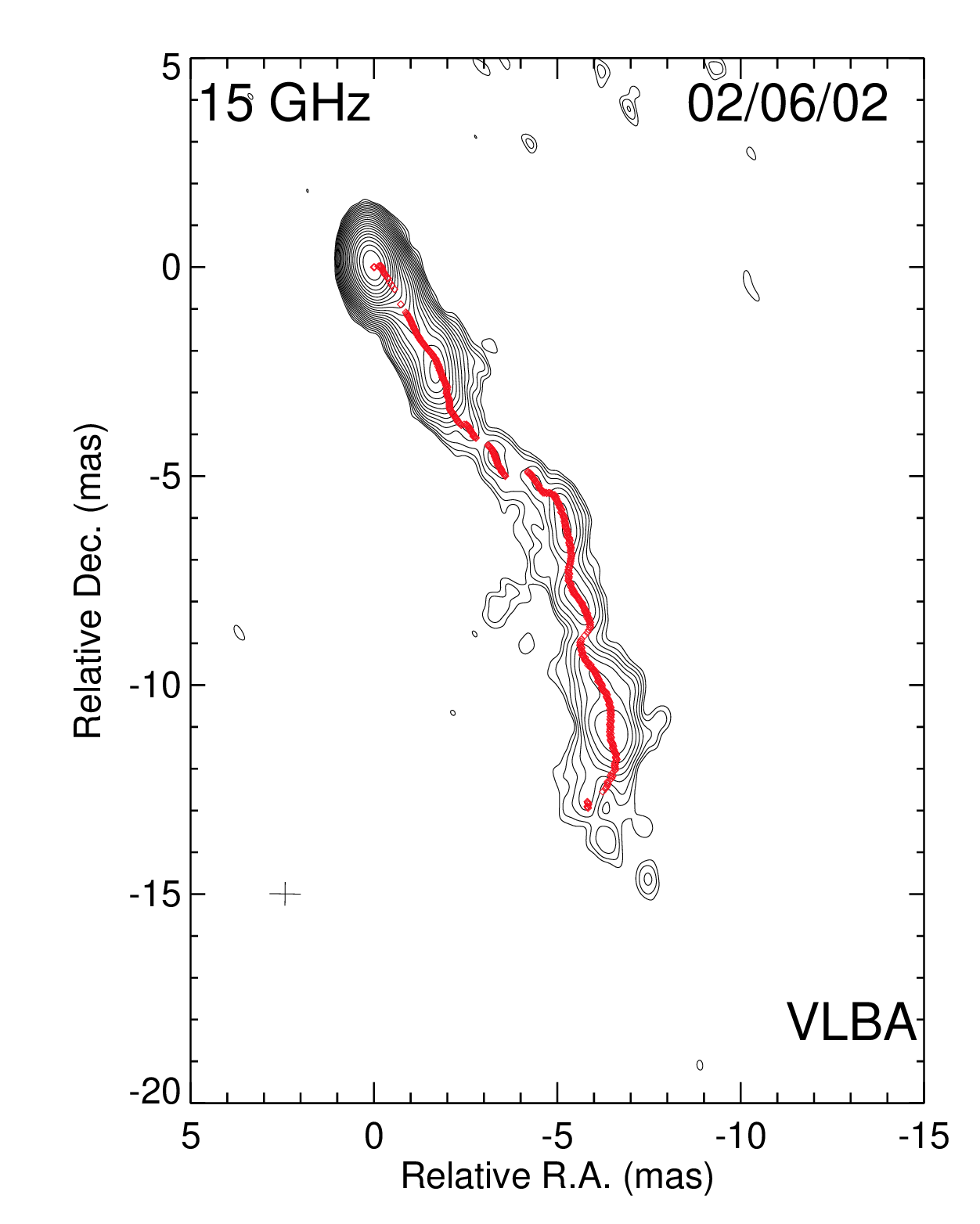}
    \includegraphics[clip,angle=0,width=0.3\textwidth]{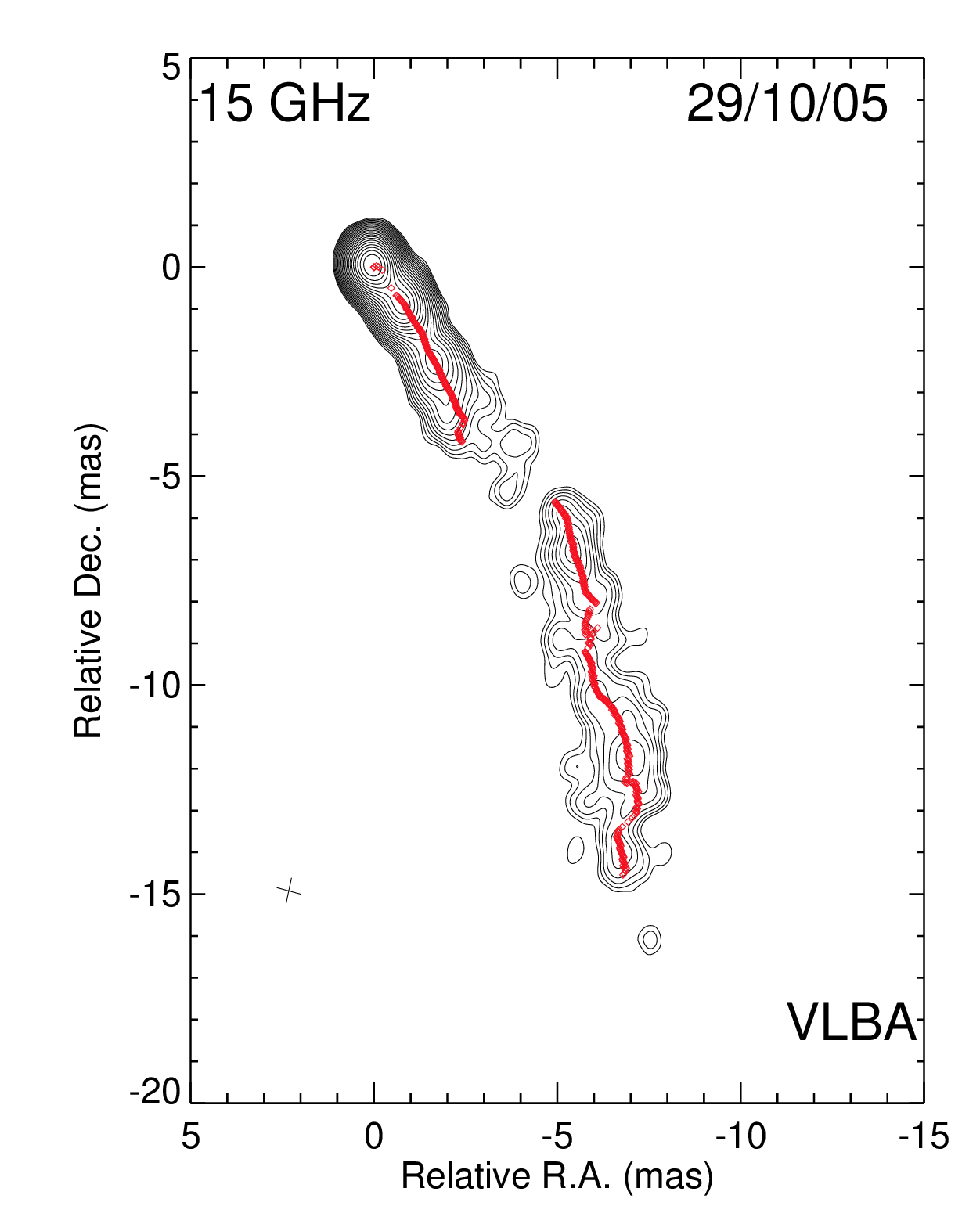}
    \includegraphics[clip,angle=0,width=0.3\textwidth]{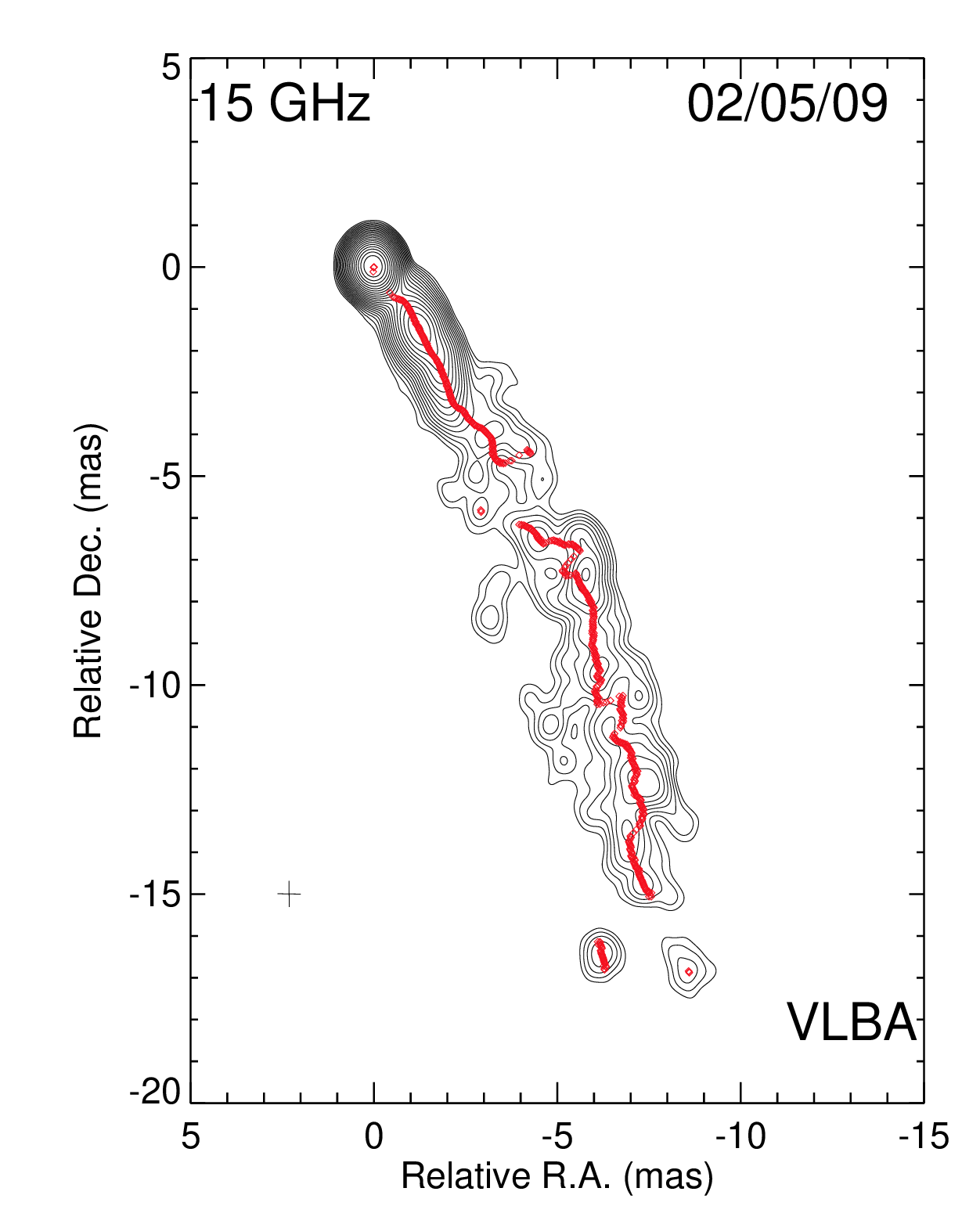}
\caption{Same as \ref{fig:lmaps} for 8~GHz (top) in 1997, 1998 and 1999 and
15~GHz (bottom) in 2002, 2005 and 2009.  Isocontours range from 2.5 mJy to 1.8
Jy in the 8~GHz images and from 0.5 mJy to 0.7 Jy in the 15~GHz ones.}
\label{fig:xmaps}
 \end{figure*}

\begin{figure*}[!t]
   \centering
    \includegraphics[clip,angle=0,width=0.4\textwidth]{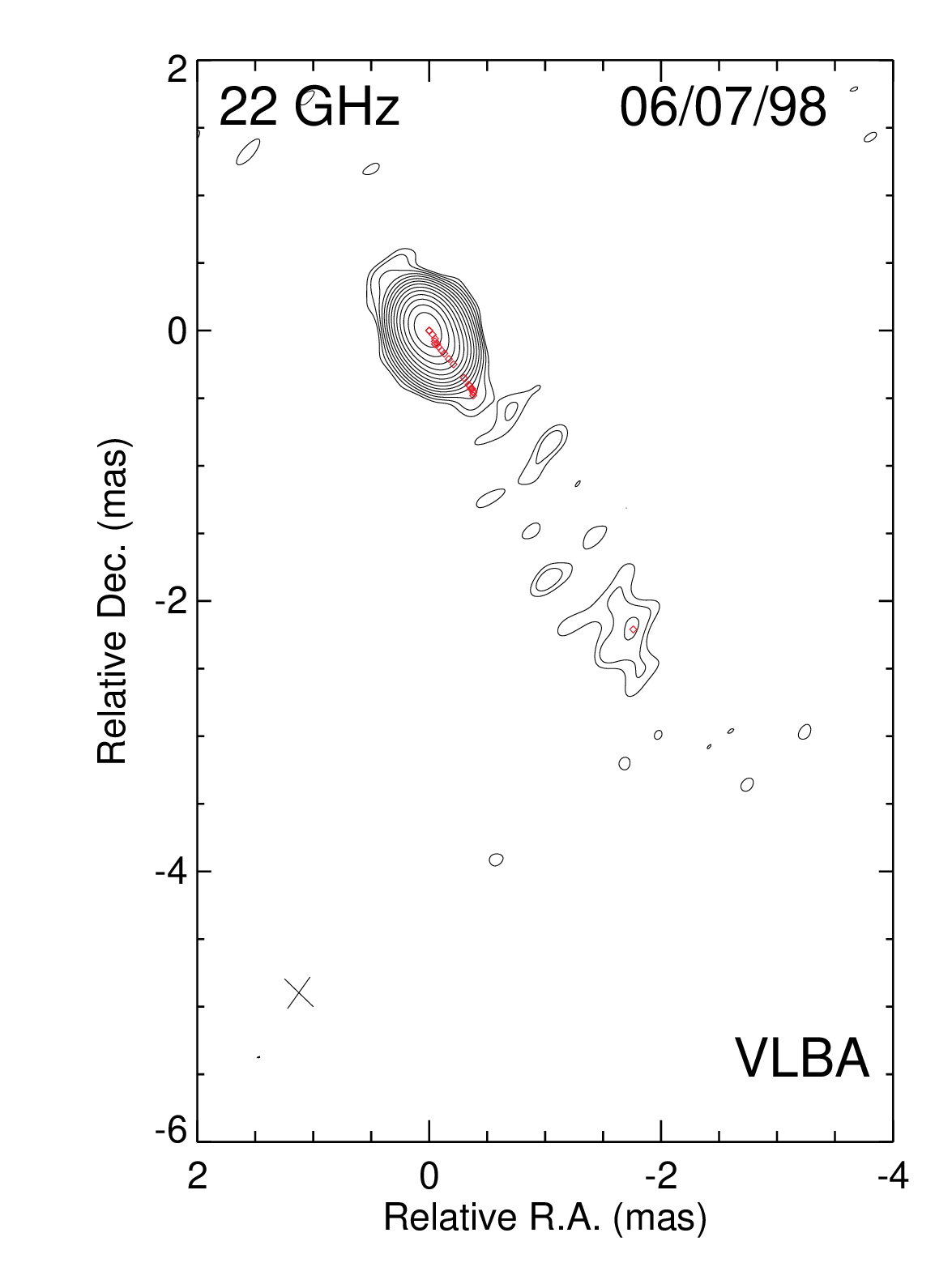}
    \includegraphics[clip,angle=0,width=0.4\textwidth]{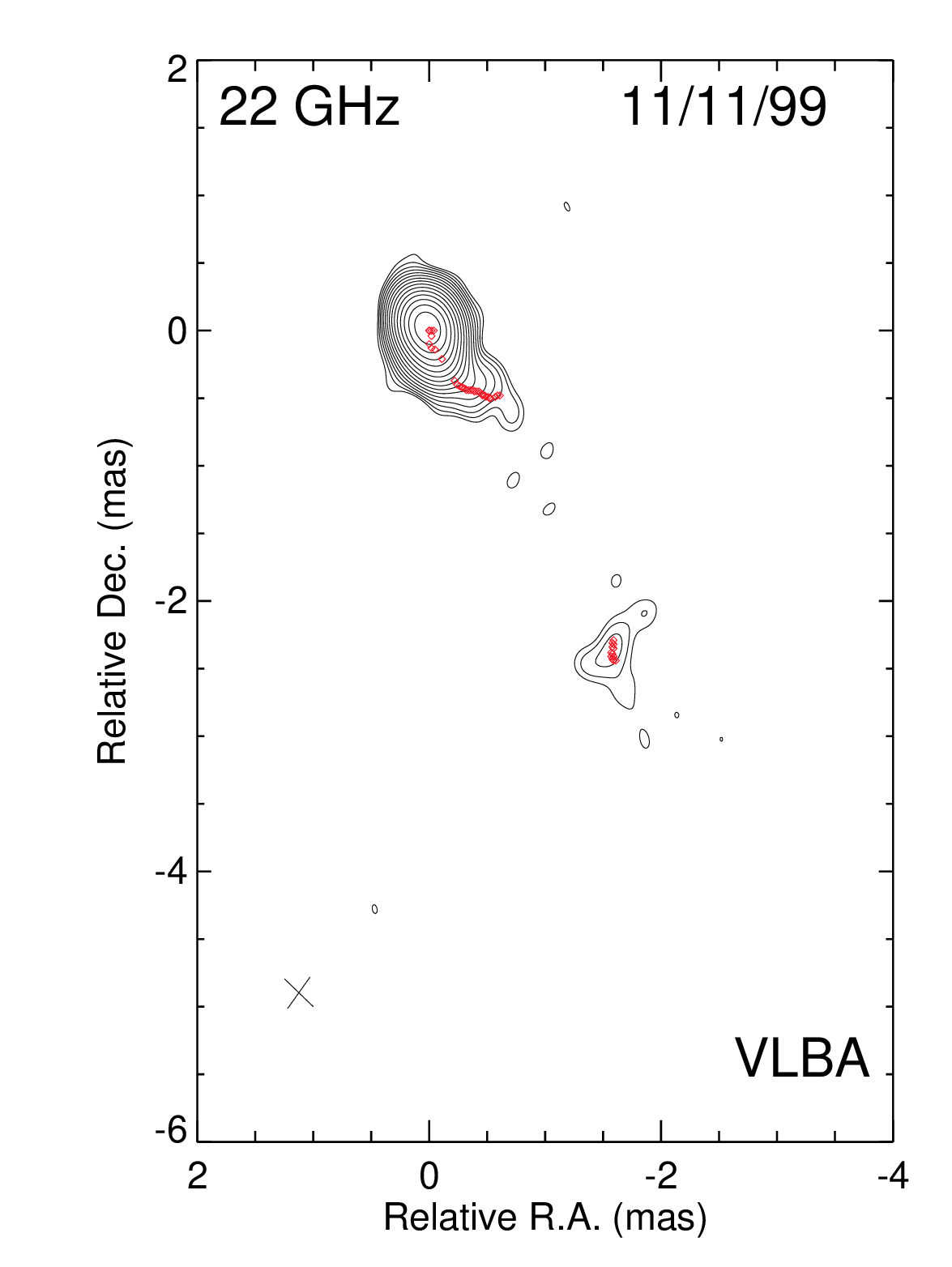}\\
    \includegraphics[clip,angle=0,width=0.4\textwidth]{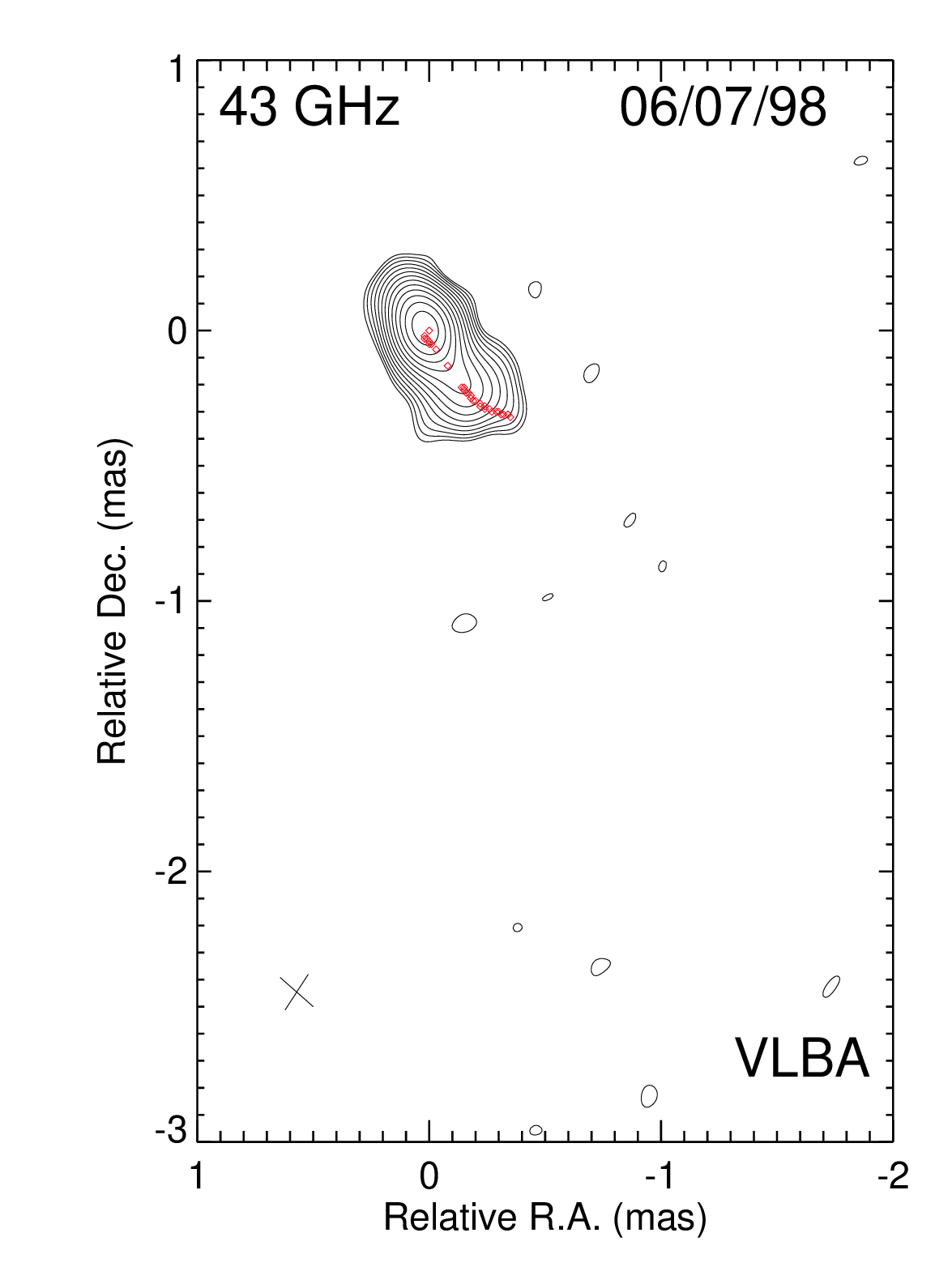}
    \includegraphics[clip,angle=0,width=0.4\textwidth]{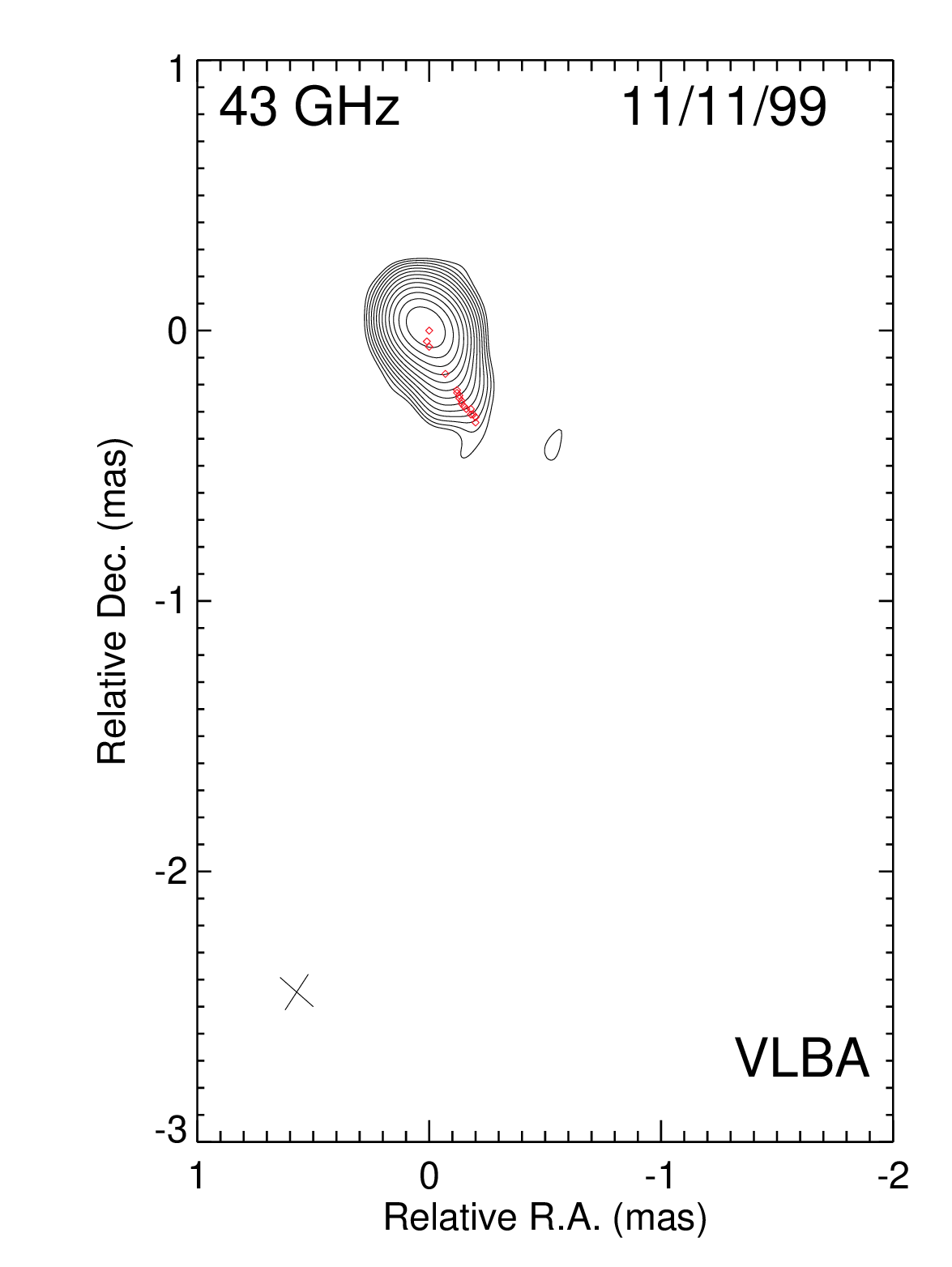}
\caption{Same as \ref{fig:lmaps} for 22~GHz (top) and 43~GHz (bottom) in 1998
and 1999.
Isocontours range from 7 mJy to 5 Jy in the 22~GHz images and from 8 mJy to 5.8
Jy in the 43~GHz ones.}
\label{fig:kmaps}
\end{figure*}

Observations of the jet at different frequencies and epochs are critical to our
study. The different frequencies 
provide information on the structures arising at different spatial scales within
the jet \citep{pl07}. 
The different epochs provide a means to determine structure motions. 
Table~1 shows the frequencies and epochs that have been used in this work: VLBA
and VSOP at 1.6 and 5~GHz \citep{lo06} at three and two different epochs,
respectively, two epochs at 1.6~GHz from EVN\footnote{The European VLBI Network is a joint facility of 
European, Chinese, South African and other radio astronomy institutes funded by their national research councils.}, one epoch of simultaneous global
VLBI (including VLBA) observations at 2 and 8~GHz (01/1997, Pushkarev \&
Kovalev, submitted), two epochs from VLBA at 8~GHz, three epochs from
VLBA at 22 and 43~GHz, and 13 epochs, between 1998 and 2009, from the
2\,cm~VLBA/MOJAVE database at 15~GHz. Figures~\ref{fig:lmaps} to \ref{fig:kmaps}
show several  of these maps. 
\begin{deluxetable*}{cccccccc} \label{tab1}
\tablecolumns{8}
\tablewidth{0pc}
\tablecaption{Observation Epochs and Frequencies}
\tablehead{
\colhead{YEAR} & \colhead{1.6 GHz} & \colhead{2 GHz} &  \colhead{5 GHz} & 
\colhead{8 GHz} & \colhead{15 GHz} & \colhead{22 GHz} & \colhead{43 GHz}}
\startdata
1997& VLBA & G-VLBI & VLBA & G-VLBI &\nodata &\nodata &\nodata \\
1998& VLBA & \nodata & \nodata & VLBA & VLBA & VLBA &VLBA \\
1999& \nodata &\nodata &\nodata& VLBA & VLBA & VLBA &VLBA\\
2000& \nodata &\nodata &\nodata&\nodata&VLBA &\nodata&\nodata\\
2001& \nodata &\nodata &\nodata&\nodata&VLBA &\nodata&\nodata\\
2002& \nodata &\nodata &\nodata&\nodata&VLBA &\nodata&\nodata\\
2003& VLBA& \nodata & VLBA &\nodata&VLBA &VLBA &VLBA\\
2004& \nodata &\nodata &\nodata&\nodata&VLBA &\nodata&\nodata\\
2005& \nodata &\nodata &\nodata&\nodata&VLBA &\nodata&\nodata\\
2006& \nodata &\nodata &\nodata&\nodata&VLBA &\nodata&\nodata\\
2007& EVN &\nodata &\nodata&\nodata&VLBA &\nodata&\nodata\\
2008& EVN &\nodata &\nodata&\nodata& VLBA &\nodata&\nodata\\
2009& \nodata &\nodata &\nodata&\nodata&VLBA &\nodata&\nodata\\
\enddata
\end{deluxetable*}

The top two panels of Figure~\ref{fig:lmaps} show observations at 1.6~GHz
separated by about 10.17~yr  but using two different instruments. There is clear
indication of a long wavelength oscillation and it can be seen that there are no
significant changes at these large scales over this period. The bottom two
panels show observations at 5~GHz separated by about 6.83~yr, both with the
VLBA. There is some evidence for a long wavelength oscillation in both images
but there are significant changes to the structure  with a suggestion of a
shorter wavelength oscillation in the 2003 epoch at larger core distance (right
bottom panel).

The top three panels of Figure~\ref{fig:xmaps} show observations at 8~GHz, 19
and 16 months apart, spanning 
about 3~yr. The first image was obtained using Global-VLBI but the subsequent
two images were obtained with the VLBA. These panels show only negligible
difference, so there is little change in the jet structure over this 3~yr
period. The bottom three panels show observations at 15~GHz, separated by 40 and
43 months and spanning about 7~yr. All three show evidence for some long
wavelength oscillation, longer than the observed jet, that does not change
significantly over this period. The only significant difference is  that the
first image clearly
shows a shorter wavelength oscillation far from the core, whereas this shorter
wavelength oscillation is weak at later times.

The top two panels of Figure~\ref{fig:kmaps} show observations at 22~GHz with
1.5~yr separation. Both show a straight jet along the first miliarcsecond, with
some curvature appearing in the second image at larger distances from the core.
The bottom two panels show observations at 43~GHz with the same 1.5~yr
separation and show a curved jet in the first epoch and a straighter structure
in the second epoch. Note that the curved structure in the first epoch at 43~GHz
could propagate out to larger scale with time and appear as the curved jet at
larger scale in the 22~GHz second epoch. This would require a motion of about
0.2 mas in 1.5~yr ($0.13$~mas/yr~$\simeq3.5\,c$ apparent velocity). If the
change in curvature were related to one quarter of a wave period, it would imply
an observed periodicity of $\sim 6$~yr.
 
The ridge-line of the jet was obtained by determining the center of the jet
emission (fitted by a Gaussian) at a given radial distance from the core. This
calculation was performed at the different epochs and frequencies.
It was done radially outwards to obtain a complete picture of the ridge-line of
the jet. 
The computed ridge-lines are indicated as red diamonds on
Figures~\ref{fig:lmaps}-\ref{fig:kmaps}. 
   
\begin{figure*}[!t]
 \centering
    \includegraphics[clip,angle=0,width=0.45\textwidth]{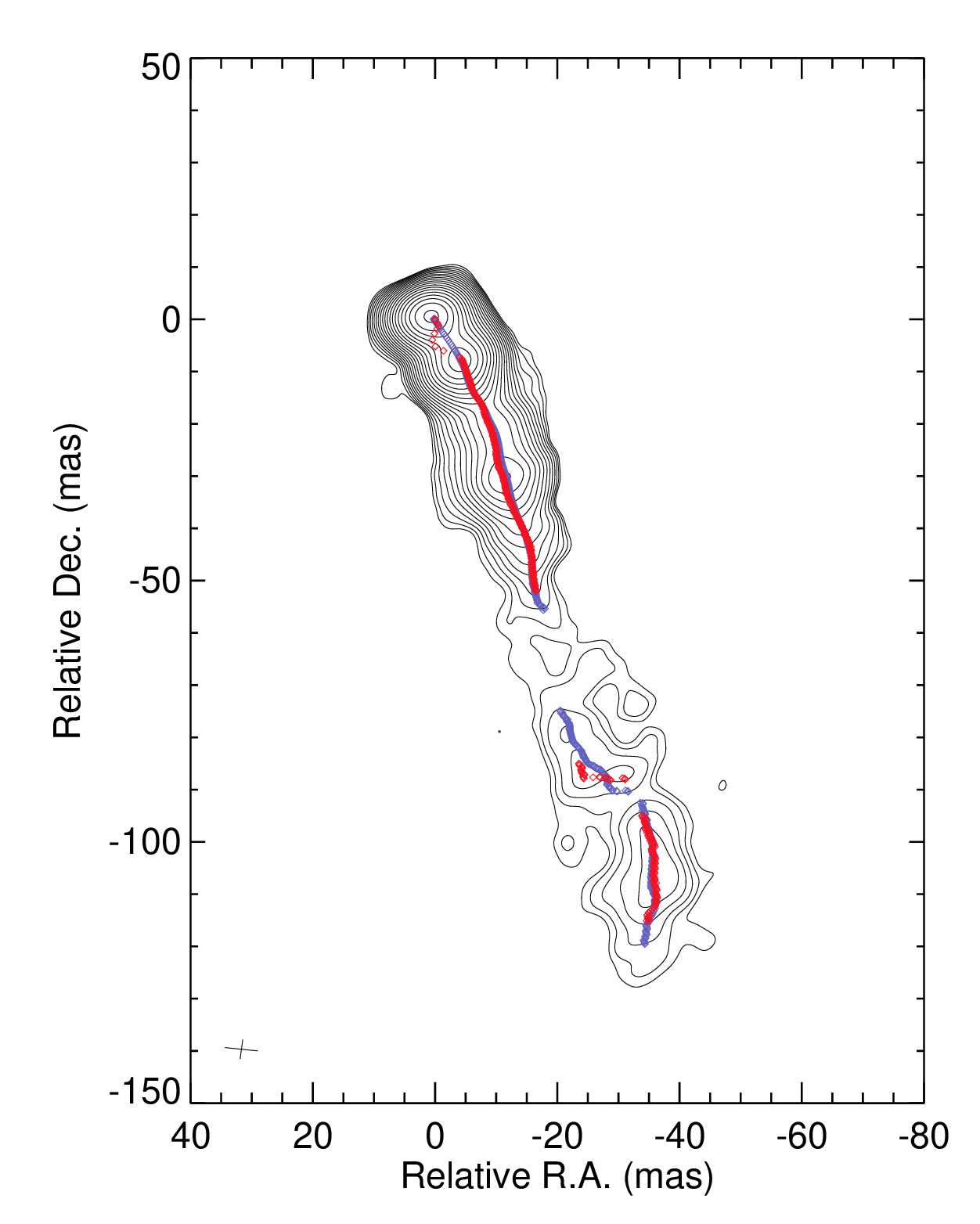}
    \includegraphics[clip,angle=0,width=0.45\textwidth]{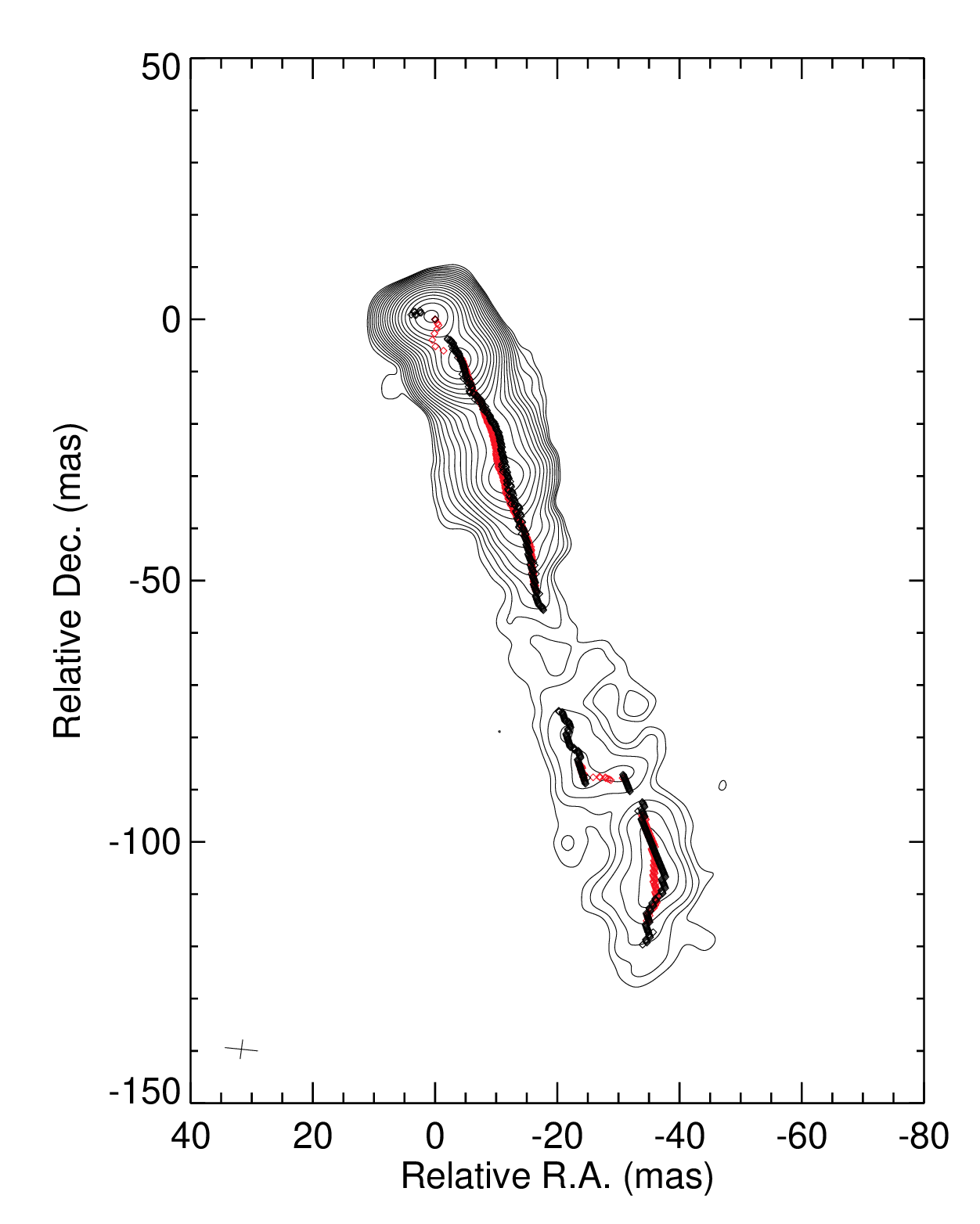}
\caption{Radio map of the jet in 0836+710 at 1.6~GHz in 1997. The left panel
shows the comparison between the ridge-lines computed from a Gaussian fit (red
diamonds) and the geometrical center of the jet emission profile (blue diamonds). The right panel
shows the comparison with the position of the maximum emission (black
diamonds).}
\label{fig:ridges}
\end{figure*}

We have tested possible deviations in the computed ridge-line by obtaining it
using two other approaches: from the location of the maximum emission and from the 
geometrical center of the emission profile in a transversal slice above a certain image
rms cutoff level. As an example, 
Figure~\ref{fig:ridges} shows the ridge-lines
found by the three different methods for the 1.6~GHz image of the jet in 1997
(also shown in Fig.~\ref{fig:lmaps}). The lines are overlaid for comparison and
we find that there are no significant differences between the geometrical
center of the emission profile, position of maximum emission and the center of jet emission as found by
a Gaussian fit (as that shown in Fig.~\ref{fig:lmaps}). This coincidence will be reviewed below
in the light of higher resolution observations at 15~GHz.

In these comparisons, errors in the displacements are assumed to be
one fifth of the FWHM of the resolving beam in the slice direction
(this assumption is also used throughout the paper). We note that, in
our specific case, this is a conservative assumption, as the
positional errors are expected to be $\sim 2
\mathrm{FWHM}/\mathrm{SNR}$ \citep[where SNR described the peak-to-rms
signal to noise in the slice; {\em cf.},][]{co97}.
$\mathrm{SNR} \ge 10$ is obtained in all slices covering the entire
relevant section of the jet, a conservative error estimate is
justified to be used for the purpose of comparing the ridge lines
obtained by the three different methods. With this approach, we find
that all three methods are consistent with each other, hence implying
that the resulting individual ridge lines are different by less than
20\% of the FWHM. A more detailed measure of positional errors of the
ridge-line, taking into account the effect of the SNR and the
transverse width of the jet, will be employed in future
work (Perucho et al., in preparation) in order to provide accurate comparisons 
with predictions from theoretical models.

\section{Observational evidence}\label{fobs}

 In this section we discuss the ridge-line positions at different frequencies and epochs. 
In general, the apparent position of the opaque core depends on observing
frequency and can shift with frequency on the sky \citep{lob98}.
However, we align images by apparent position of the core and do not take
this effect into account while comparing the ridge-lines. This is
justified by the small expected shifts ($\leq 1\,\rm{mas}$) as compared to the
length of the structures that are compared (typically tens of mas). See
for typical amount of the shift the statistical studies by \citet{ko08} and \citet{so11}.

\subsection{On the nature of ridge-lines}

\begin{figure*}[!t]
       \centering
    \includegraphics[clip,angle=0,width=0.45\textwidth]{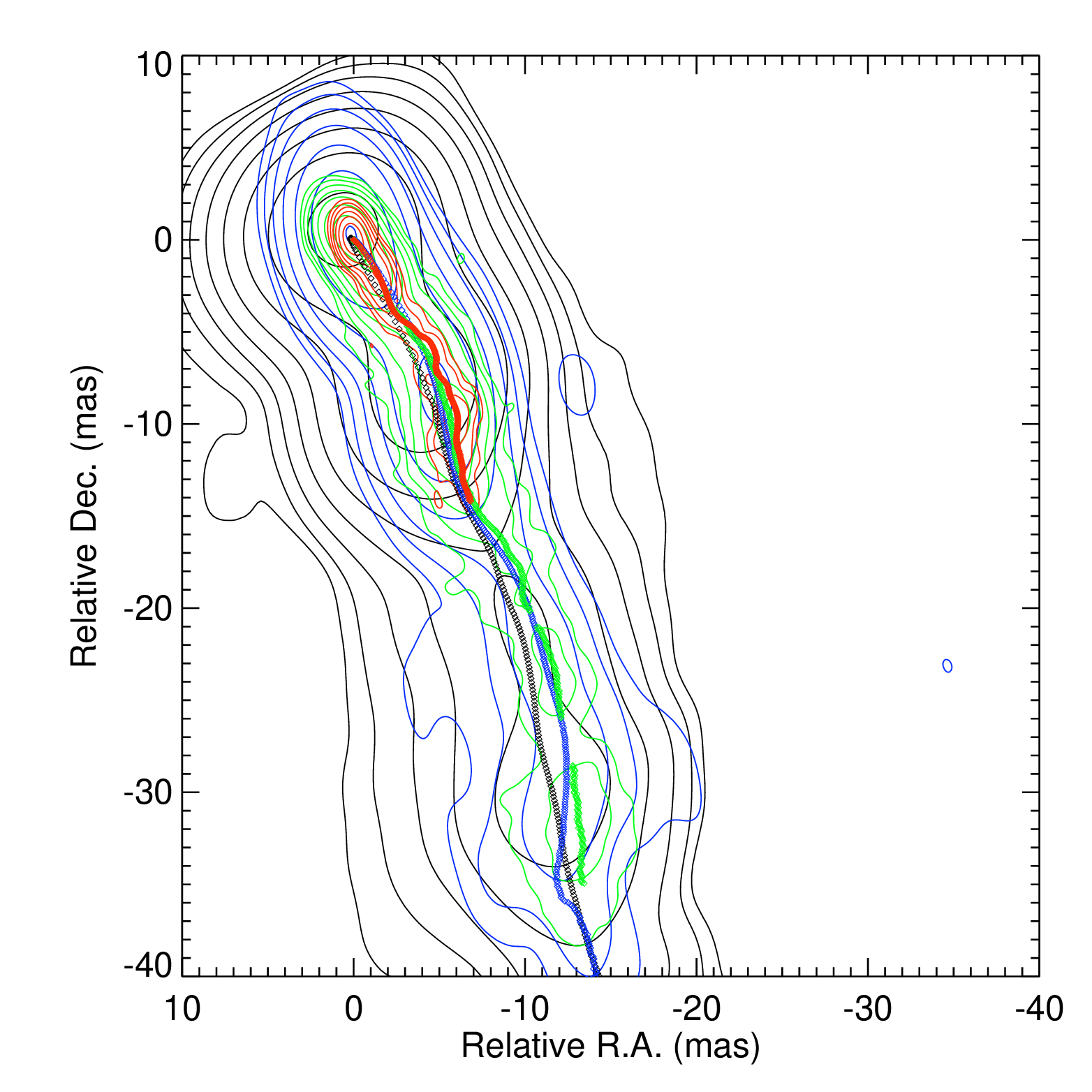}
    \includegraphics[clip,angle=0,width=0.45\textwidth]{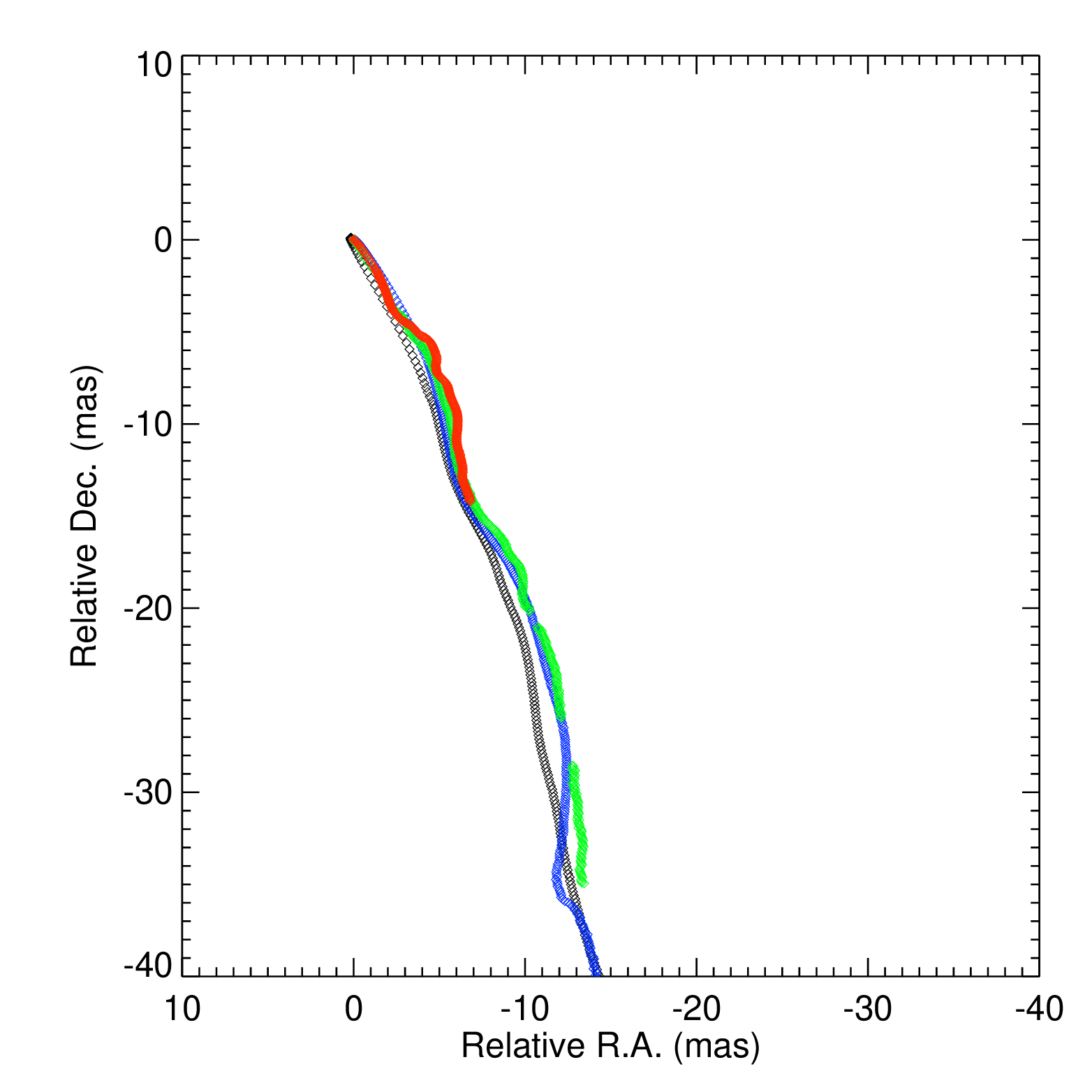}
\caption{Radio maps at the 1.6 (black contours), 2 (blue contours), 5 (green
contours) and 8~GHz (red contours) of the jet in 
0836+710 in 1997. The ridge-lines at the different frequencies are indicated by
diamonds in the same colour as the contours and are shown 
without contours on the right.}
    \label{fig:r97}
    \end{figure*}

\begin{figure*}[!t]
       \centering
    \includegraphics[clip,angle=0,width=0.45\textwidth]{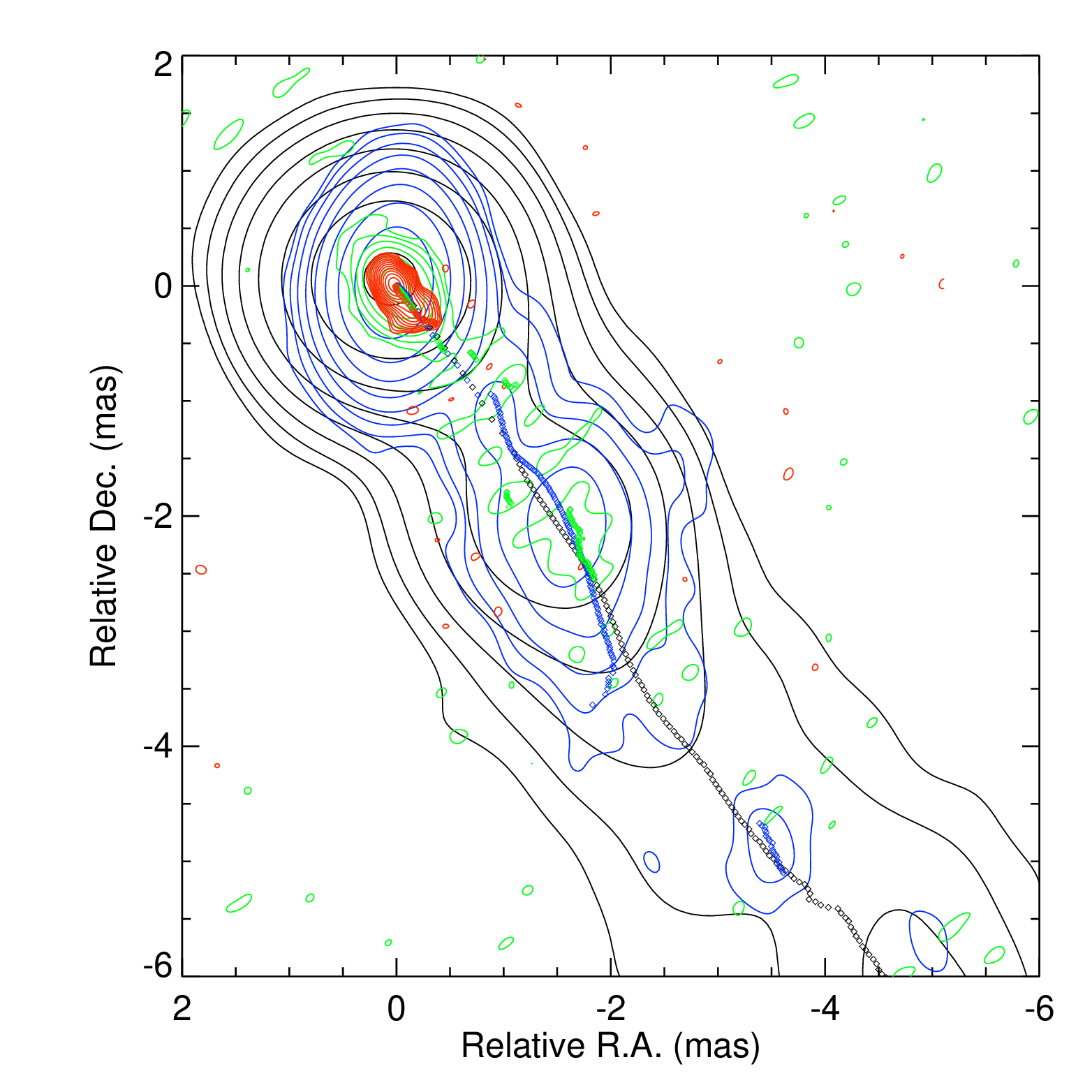}
    \includegraphics[clip,angle=0,width=0.45\textwidth]{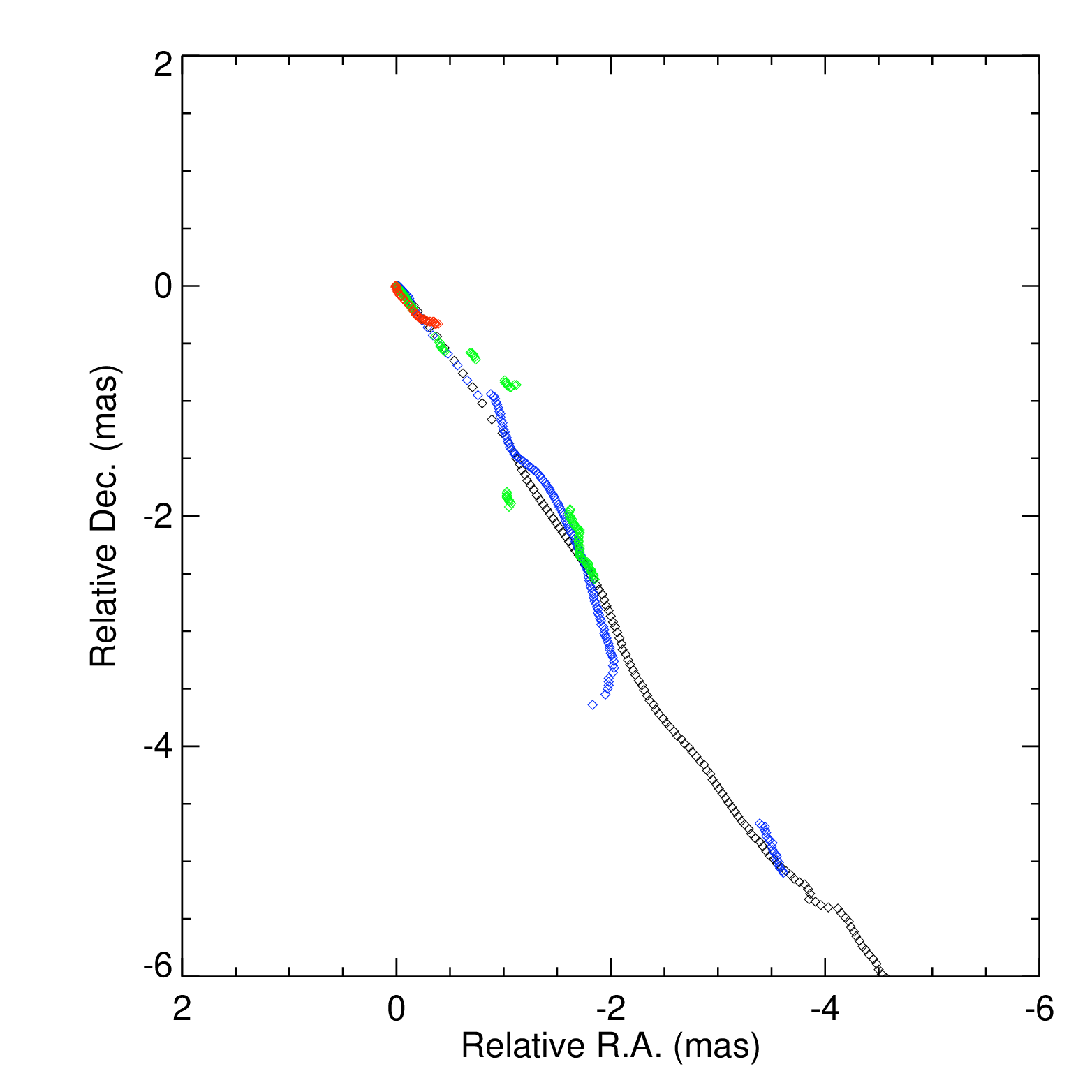}
\caption{Radio maps at the 8 (black contours), 15 (blue contours), 22 (green
contours) and 43~GHz (red contours) of the jet in 0836+710 
in 1998; the ridge-lines at the different frequencies are indicated by diamonds
in the same colour as the contours and are shown 
without contours on the right.}
    \label{fig:r98}
    \end{figure*}

\begin{figure*}[!t]
       \centering
    \includegraphics[clip,angle=0,width=0.45\textwidth]{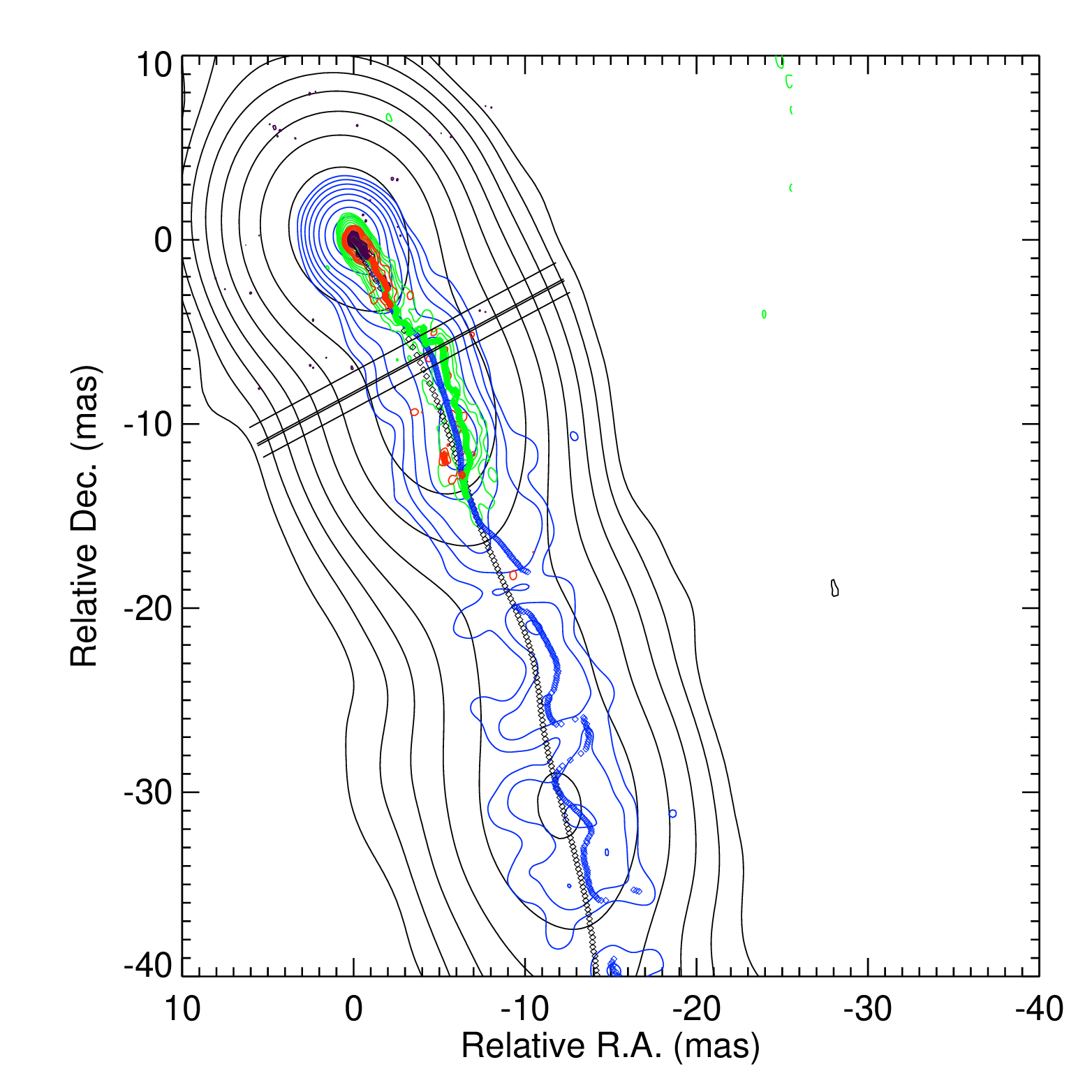}
    \includegraphics[clip,angle=0,width=0.45\textwidth]{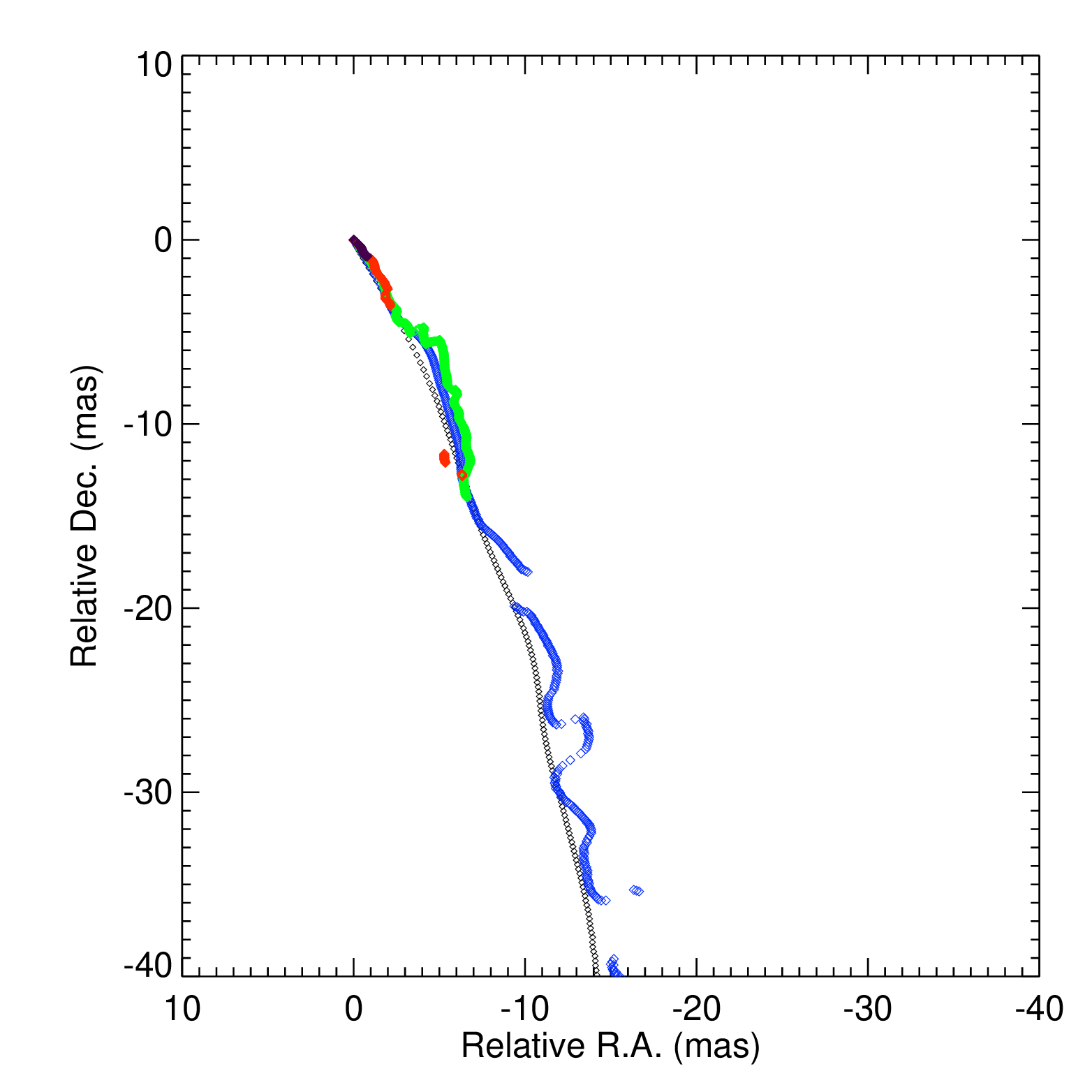}
    \includegraphics[clip,angle=0,width=0.45\textwidth]{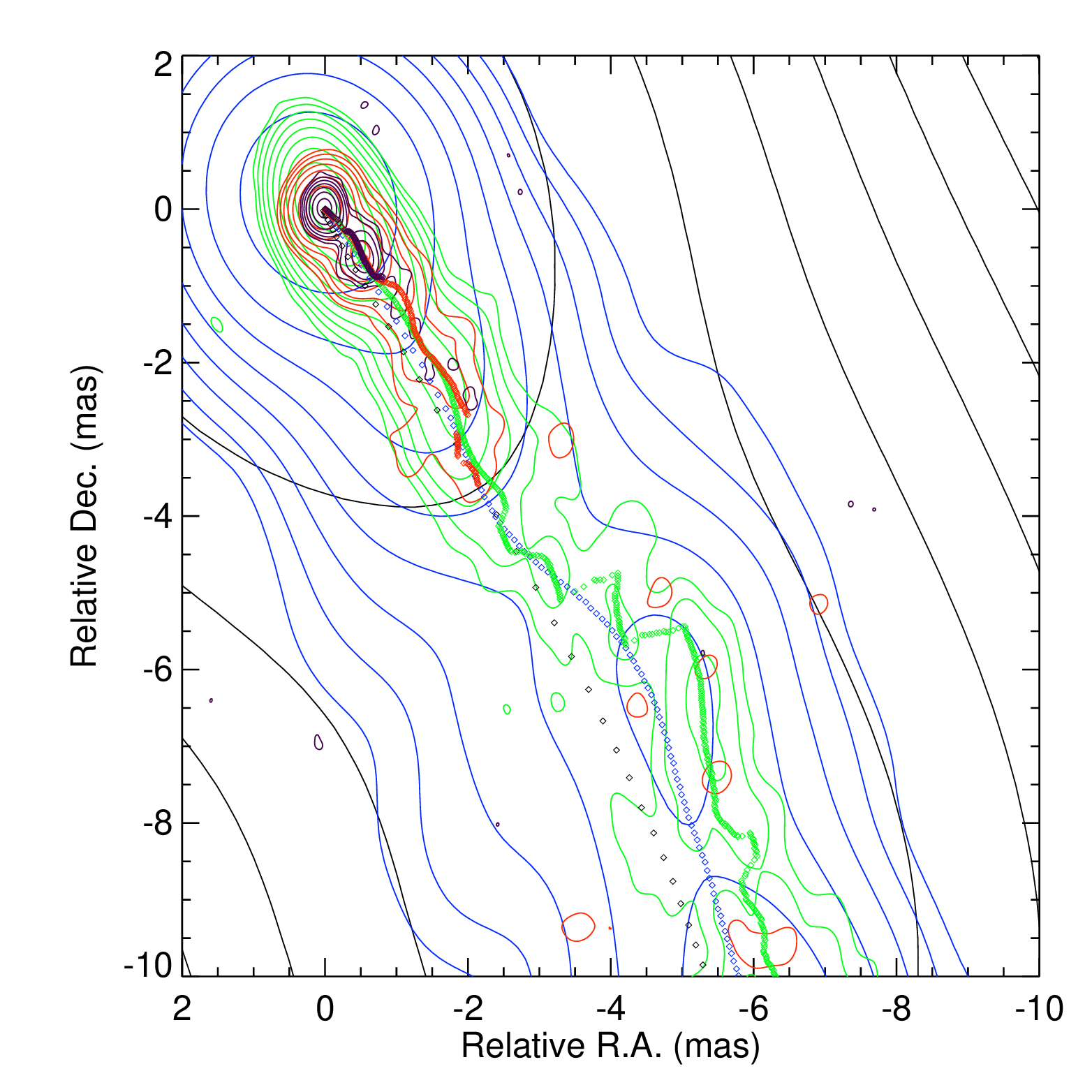}
    \includegraphics[clip,angle=0,width=0.45\textwidth]{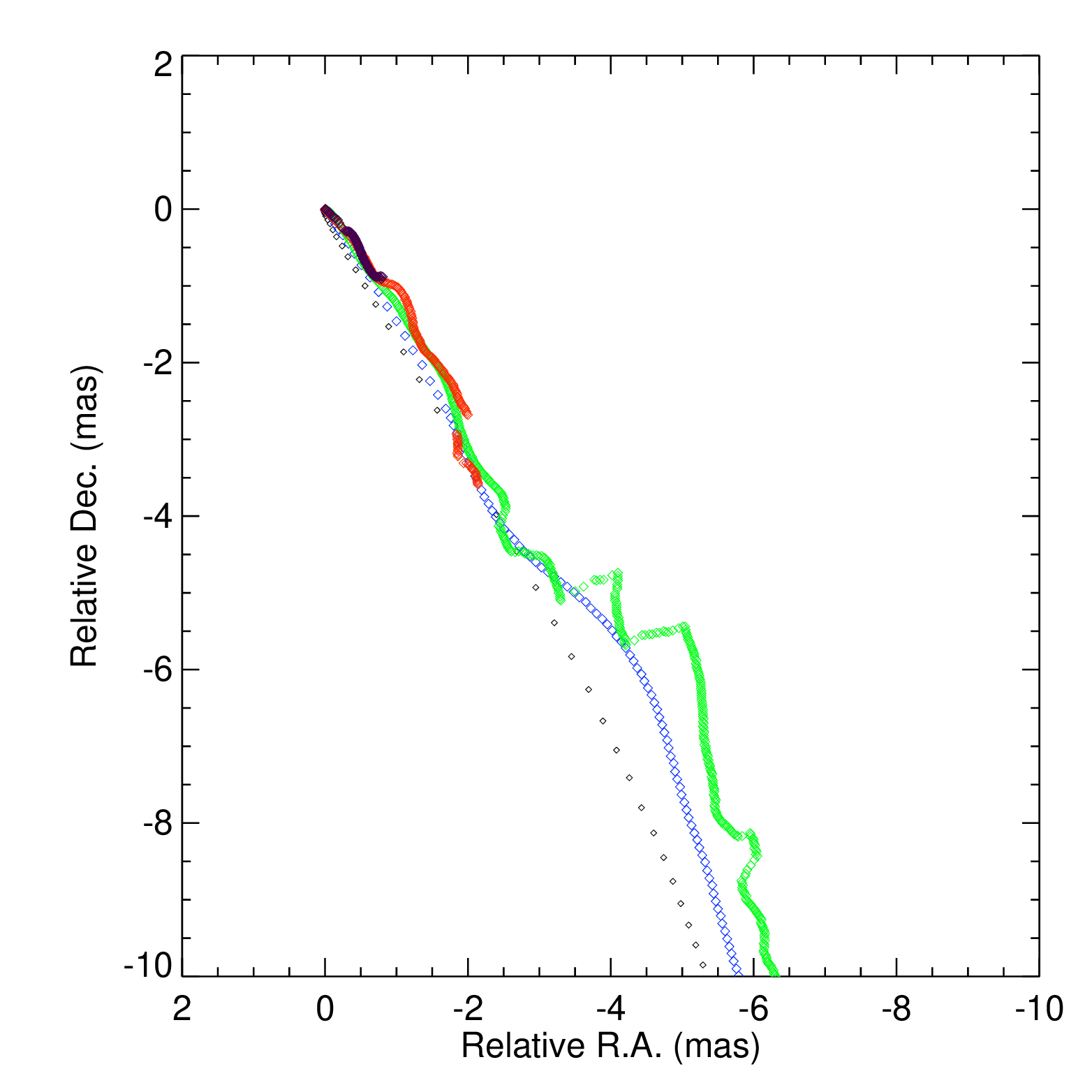}
\caption{Radio maps at the 1.6 (black contours), 5 (blue contours), 15 (green
contours), 22 (red contours) and 
43~GHz (violet contours) of the jet in 0836+710 in 2003. The ridge-lines at the
different frequencies are indicated by diamonds in the same colour as the
contours and are shown without contours on the right. The black lines in the top-left panel indicate
the locations of the cuts shown in Fig.~\ref{fig:cuts}. The bottom panels show an
enlarged image of the inner region.}
    \label{fig:r03}
    \end{figure*}

Figures~\ref{fig:r97}-\ref{fig:r03} display a combination of the radio emission
contours for different frequencies 
along with their ridge-lines for observations performed in 1997, 1998 and 2003.
Figure~\ref{fig:r97} shows that the ridge-lines indicated at successively higher
frequencies appear as oscillations on top of the ridge-lines obtained at
the lower frequencies: the green 5~GHz and red 8~GHz ridge-lines wrap around the
blue 2~GHz ridge-line. Differences between the 1.6 and 2~GHz ridge-lines are
within the errors. 

Figure~\ref{fig:r98} shows the same kind of behavior at higher frequencies in 1998,
with the longest ridge-line belonging to the 8~GHz image and the ridge-lines at
15, 22 and 43~GHz appearing as oscillations on top of the 8~GHz ridge-line.

Finally, Figure~\ref{fig:r03} shows the broadest range of frequencies, from 1.6
to 43~GHz, with the bottom
panels showing an enlarged image of the inner region. Again, the ridge-lines of
the high-frequency images tend to lie on top of the lower frequency ridge-lines,
except for the 15~GHz image ridge-line that deviates significantly 
from the 1.6~GHz ridge-line far from the core. However, at these distances, the
signal-to-noise ratio is low in the 15~GHz image, which should translate into
larger errors, and the difference is always within the error of the 1.6~GHz
ridge-line ($\simeq 1$~mas).

In summary we find that the ridge-lines show remarkable coincidence independent
of the epoch and the array. 
Significant differences would be expected if oscillation of the ridge-line were
array or epoch-dependent. 
This result confirms the real nature of the structures observed in extragalactic
jets by means of VLBI observations.

If the jet is transversally resolved, the maximum in the emission, or the
position of the ridge-line can be due to two main factors: 1) differential
Doppler boosting caused by velocity gradients in the jet, or 2) an increase in
the local pressure and number of emitting particles. If the position of this maximum is
due to differential Doppler boosting, 
changes in the flow direction would have an influence on the local emission from
the different regions. This should result, depending on the internal velocity
gradient, in irregular patterns, contrary to the results obtained. In addition,
relatively large changes in speed and/or flow direction are required in order to
get significant differential Doppler boosting within the jet. Thus, the most
reasonable hypothesis is that the ridge-line corresponds to a pressure maximum.
In the following sections we present evidence that supports this interpretation.

\subsection{Ridge-lines and transversal structure}

A straight axis can be defined for the jets, by a line starting at the core and
defined by the symmetry axis of the the longest-observed wavelength
oscillation (assumed to result from helical motion). Note that this line, which will be referred to 
as \emph{the jet axis} or simply \emph{the axis}, is not the same as the center or the maximum of the jet 
emission at each radial distance, used above to determine the position of the ridge-line. Within our interpretation, 
the axis describes a mean or reference direction of the jet -- the direction of flow propagation in absence of instabilities.
This axis needs to be defined in order to be able to estimate the amplitudes and wavelength of individual instability modes, as done in \citet{lz01}.
The mean overall jet direction $\chi$, as defined by the jet axis, is modified by a helical instability \citep[see][for a discussion of the effect of different KH modes on jet structure and next sections]{har00}. Variations of the ridge lines observed on smaller scales are likely to be caused by shorter unstable wavelengths. In order to estimate accurately wavelengths and amplitudes of these oscillations, the effect of the longest wavelength has to be taken
into account, either by modeling it with an oscillatory term or by assuming
a mean local jet direction at each relevant spatial scale. As the small-scale
ridge line variations occur on scales substantially shorter than the
longest observed wavelength, we assume local mean jet
directions $\chi_\nu$ for observations at different frequencies (relying on
the fact that the spatial scales investigated depend largely on the observing
frequency). If the jet axis is computed locally, for the different frequencies, the comparison shows that the absolute value of the jet position angle ($\chi_\nu$) increases with the frequency, where this angle is measured from the Declination axis, with $0\arcdeg$ corresponding to the positive values.
 
\begin{figure}[!t]
\includegraphics[clip,angle=-90,width=0.85\columnwidth]{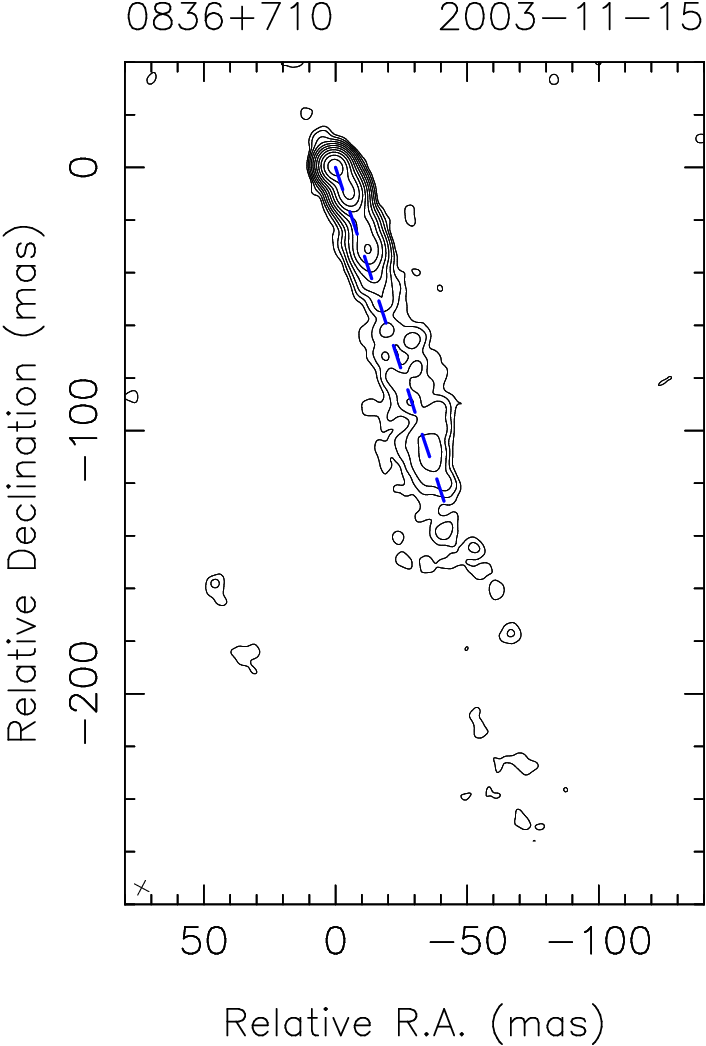}
\caption{1.6 GHz map of the jet in 2003. The dashed-blue line indicates the
position angle of the jet axis 
with respect to the positive Relative Declination. (198$\arcdeg$).}
 \label{fig:axl}
\end{figure}

\begin{figure}[!t]
\includegraphics[clip,angle=-90,width=0.85\columnwidth]{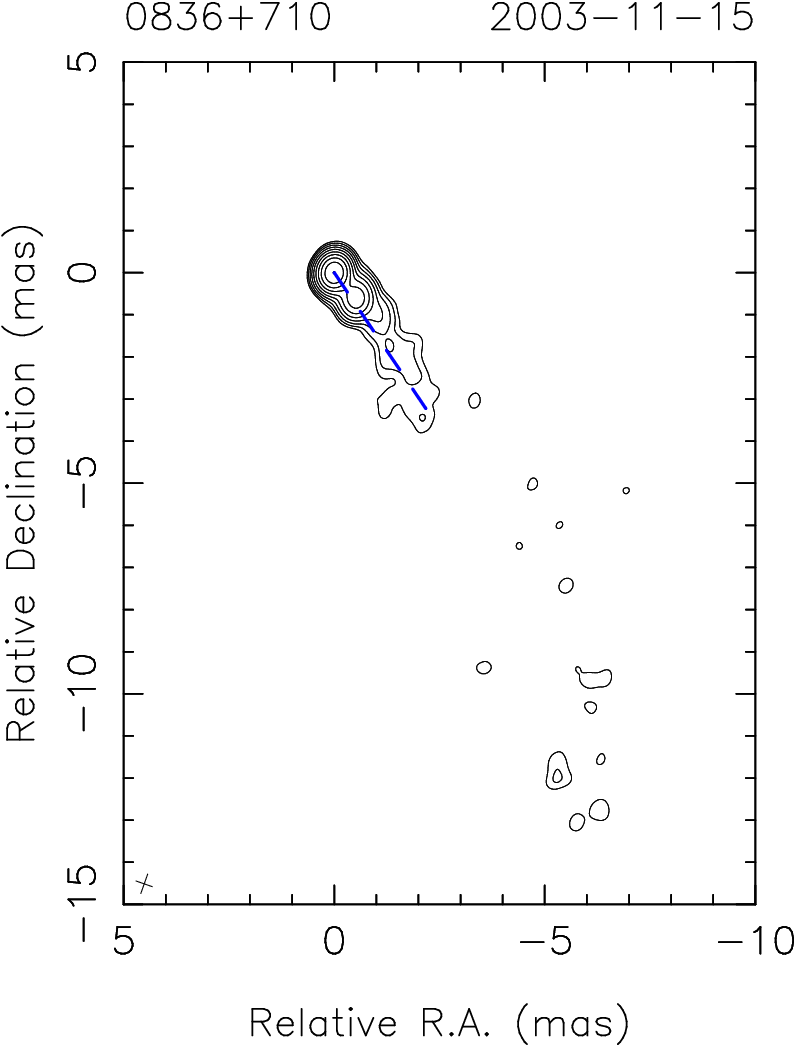}
\caption{22 GHz map of the jet in 2003. The dashed-blue line indicates the
position angle of the jet axis 
with respect to the positive Relative Declination. (214$\arcdeg$).}
    \label{fig:axk}
    \end{figure}

The angle $\chi_\nu$ does not change with time at each frequency. Figures~\ref{fig:axl} and \ref{fig:axk} show this
effect, with the position angle of the jet ranging from 198$\arcdeg$ at 1.6~GHz
to 214$\arcdeg$ at 22~GHz\footnote{It is impossible to determine the jet
position angle at 43~GHz due to its short length.}, indicated by  the dashed
blue lines. The values of the different position angles at the different
frequencies are listed in Table~2.

Recall that there is a clear superposition of high frequency ridge-lines on top
of the low frequency ridge-lines (Figs.~\ref{fig:r97}-\ref{fig:r03}). This can
be interpreted as the radio emitting regions at high-frequencies being
concentrated in a relatively small region around the lower frequency ridge-line
with the small scale oscillations seen at higher frequency developing on top of
the long scale ones seen at lower frequency. This interpretation provides a
natural explanation of parsec-to kiloparsec scale jet misalignment for
helically twisted jets pointing 
close to the line of sight. However, this raises a question as to whether the
lowest-frequency emission observed is embedded in an even broader region, with
lower or no emission. If this was the case, we would not be observing the whole
cross-section of the jet at high-frequencies, but only a small region around the
highest-emitting locations. Thus, the apparent helix would be tracing a
high-emission region inside the jet, but not the whole jet.

To test the possibility of a frequency dependent transversal structure in the
emitting region, we computed the jet 
observed opening angles and widths for all the available frequencies at
different positions along the jet axis following \citet{pu+09}. Note that, as 
stated above, the direction of the observed jet changes with the frequency and
this should be taken into account to obtain these parameters at each frequency. Thus,
the appropriate position angle of the frequency-dependent axis, $\chi_\nu$, was used for
each frequency. The results of this analysis are summarized in Table~2, where
average values for each frequency are given. Not all epochs were used, as some
were too noisy for this study. The errors come from the dispersion of the values
for all epochs. We find that the opening angles are very similar at all
frequencies, with an average value of $13.8\arcdeg \pm 1.9$. At the highest
frequencies, 22 and 43~GHz, we have only considered one epoch (2003) due to
problems with the other two epochs (1998, 1999). The resulting values are
subject to larger errors as the opening angles are obtained with fewer cuts. If
we disregard these high frequency results, the average opening angle is
$12.1\arcdeg \pm 0.8$, which at a $3\arcdeg$ viewing angle implies an intrinsic
opening angle of $0.63\arcdeg \pm0.04$. 
This value coincides with that obtained at 15~GHz, for which we have the largest
sample and best quality data. 
The small or negligible differences between the opening angles at different
frequencies tells us that either all 
transversal regions in the jet expand at the same rate or that, with the present
resolution it is still difficult to 
tell whether there is a transversal frequency profile for this jet. 
    
\begin{deluxetable*}{lcccccc} \label{tab2}
\tablecolumns{7}
\tablewidth{0pc}
\tablecaption{Observed Jet properties.}
\tablehead{
&\colhead{1.6~GHz} &  \colhead{5~GHz} & 
\colhead{8~GHz} & \colhead{15~GHz}  & \colhead{22~GHz} & \colhead{43~GHz}}
\startdata
Average opening angle&$(12.3\pm 1.2)^\circ$ & $(13.5\pm 1.3)^\circ$ & $(12.1\pm
0.8)^\circ$ &  $(10.5\pm 0.9)^\circ$ & 
$18.1^\circ$& $16.6^\circ$ \\
Position angle ($\chi$)&$198\arcdeg$&$202\arcdeg$&$206\arcdeg$& $210\arcdeg$&
$214\arcdeg$& - \\
Oscillation wavelengths (0-10 mas) & 10 - 80 & 10 & 10 & 10 & - & - \\
Oscillation wavelengths (10-35 mas) & 20 - 80 & 20 & 7-20 & 20 & - & - \\
Oscillation wavelengths ($>$35 mas) & 40 - 80 & - & - & - & - & - \\
\enddata
\end{deluxetable*}

\subsection{Possible secondary peak and implications} 
\begin{figure*}[!t]
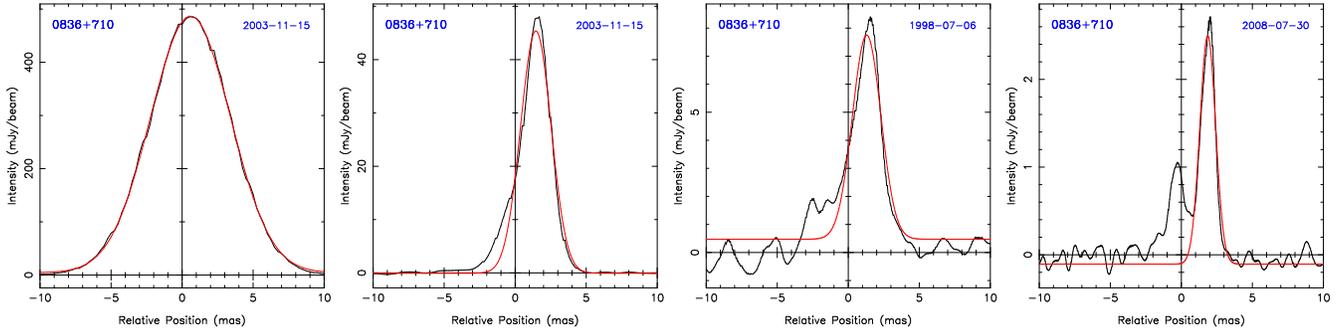

    \includegraphics[trim=0cm 0cm 0cm 1.6cm,clip,angle=0,width=0.24\textwidth]{fig10a.ps} 
    \includegraphics[trim=0cm 0cm 0cm 1.6cm,clip,angle=0,width=0.24\textwidth]{fig10b.ps}
    \includegraphics[trim=0cm 0cm 0cm 1.6cm,clip,angle=0,width=0.24\textwidth]{fig10c.ps}
    \includegraphics[trim=0cm 0cm 0cm 1.6cm,clip,angle=0,width=0.24\textwidth]{fig10d.ps}
\caption{Profiles and Gaussian fits for 1.6 and 5~GHz (2003 epochs), 8~GHz (1998) and 15~GHz (2006), from left to right.
The location of the cuts is shown in the top-left panel of Fig.~\ref{fig:r03} (the first two panels here are taken 
from the epoch in that Figure, whereas the other two have been plotted together for the sake of clarity).
In the left panel (1.6~GHz), the cut is done 7.50~mas from the core along the axis (PA=$-153.4\arcdeg$), 
the peak intensity is 482.5~mJy/bm, the peak relative position is 0.591~mas, and the FWHM of the Gaussian is 6.630~mas. 
In the second panel from the left (5~GHz), the cut is done 7.40~mas from the core along the axis (PA=$-153.4\arcdeg$), 
the peak intensity is 45.5~mJy/bm, the peak relative position is 1.454~mas, and the FWHM of the Gaussian is 2.533~mas.
In the third panel (8~GHz), the cut is done 6.40~mas from the core along the axis (PA=$-153.4\arcdeg$), 
the peak intensity is 7.3~mJy/bm, the peak relative position is 1.298~mas, and the FWHM of the Gaussian is 2.362~mas. 
In the right panel (15~GHz), the cut is done 8.20~mas from the core along the axis (PA=$-153.4\arcdeg$), 
the peak intensity is 2.6~mJy/bm, the primary peak relative position is 1.859~mas, and the FWHM of the Gaussian is 
1.31~mas.}
    \label{fig:cuts}
    \end{figure*}

Figure~\ref{fig:cuts} shows profiles of the jet brightness at distances from the core between
6.4, 7.5 and 8.2~mas for different frequencies and epochs. These profiles do not vary significantly with the 
position angle used, so we have taken $\chi_8$, i.e., the position angle of the jet at 8~GHz as a 
characteristic value. The positions of these profiles are indicated in Fig.~\ref{fig:r03}. In all the plots, the observed 
emission maximum is displaced (by $1\,-\,2$~mas) with respect to the reference axis,
beyond the errors in the determination of the peak. We will see in Section~\ref{wv} that the measured values of the 
relative shift between epochs deviate from zero due to possible evolution of the ridge-lines. In most cases, single Gaussians provide good fits to the profiles and the center of jet emission coincides with the computed ridge-line. However,
Figure~\ref{fig:cuts}  shows that the transversal profiles at 8 and 15~GHz contain 
a secondary peak at the relative offset of $-2\,-\,-1$~mas, which is not seen at the
other frequencies at the same position. In order to verify that the second peak is
real, a stacked image\footnote{This image is a composition from all the
epochs, which reinforces real features and smoothes noisy ones.} from the MOJAVE
database was used, and the same double peak structure in the profile was
observed at this distance. In addition, if the image at 15 GHz is restored with the resolution of the 5 GHz data,
the secondary peak disappears and the resulting profiles are very
similar to the 5~GHz ones (Fig.~\ref{fig:cuts}). Finally, the 5~GHz image (shown
in Fig.~\ref{fig:lmaps}) is accurately reproduced by the map of the jet at
15~GHz when convolved with the 5~GHz beam.

At 15~GHz, where the two peaks are observed, the stronger peak is slightly displaced
from the center of the emitting region (-3 to +4 mas, approximately), which is located around 2~mas to the right
from the coordinate axis (jet axis, see Fig.~\ref{fig:cuts}). The secondary peak is less important than the one to the right. 
This structure could be explained by superposition of a smaller amplitude elliptical
mode and a dominant helical mode, or by limb-brightening of the jet.

A growing instability in the linear regime implies a straight, expanding jet with the unstable pattern developing in its interior or on the surface/layer separating the jet from its surrounding medium \citep[e.g.,][]{har00,pe+06}. Thus, within this frame, the center of the jet emission and the ridge-line associated to the maximum emission do not necessarily coincide, if the latter is produced by the pressure maxima. The local direction of the flow can change when the jet is affected by a helical mode and the maximum in pressure is displaced from the center of the flow, as shown in previous works \citep{har00,pe+06}. The fact that the peak in emission at 15~GHz (Figure~\ref{fig:cuts}) is shifted from the center of the emitting jet supports the idea that the peak in emission does not coincide with the center of the jet flow and is more related to pressure differences in its cross-section. However, only if the jet is resolved, this feature is fully captured. This is the reason why, at 1.6~GHz, where the jet is not well resolved, the different methods to determine the ridge-lines give similar results (see Section~\ref{obs} and Figure~\ref{fig:ridges}).

\subsection{Observed oscillation wavelengths}

\begin{figure}[!t]
\includegraphics[clip,angle=0,width=0.9\columnwidth]{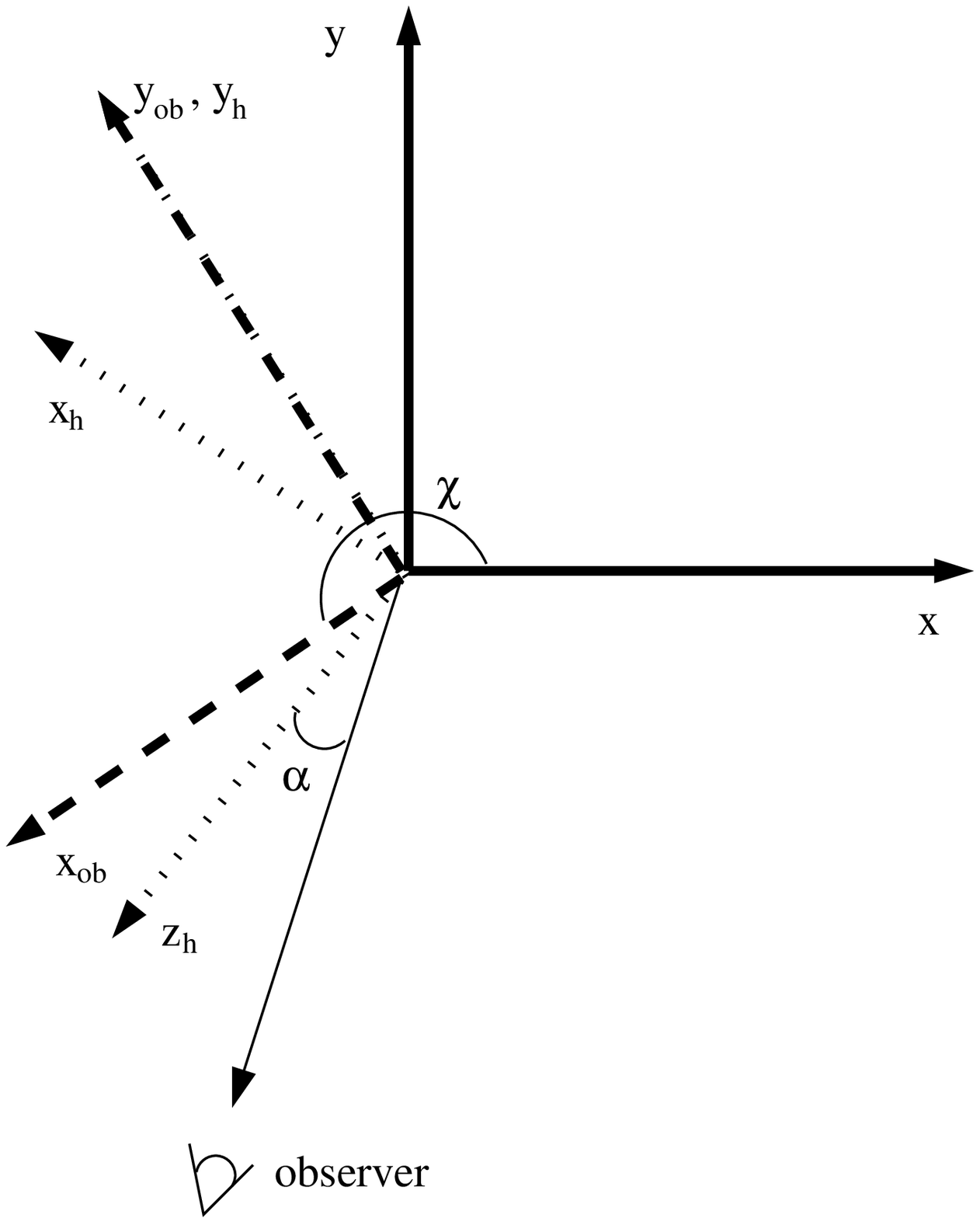}
\caption{Coordinate systems used in the text. The intrinsic coordinates of the
helix are $x_h$, $y_h$ and $z_h$. The observed coordinates are $x_{ob}$ and
$y_{ob}$. $x$, $y$, $x_{ob}$, and the axis including 
$y_{ob}$ and $y_h$ fall on the plane of the sky. Note that this image is
mirrored with respect to the radio maps in 
the paper.}
    \label{fig:coords}
    \end{figure}

Applying a rotation of the jet position angle to the coordinates of the
ridge-line on the plane of the sky, 
the structure is brought to a horizontal axis (x-axis in cartesian coordinates)
along the positive Declination axis, making the detection of patterns much easier: 
\begin{eqnarray}\label{rot}
  x = x_{ob}\,cos(2\pi-\chi) - y_{ob}\,sin(2\pi-\chi) \nonumber \\
  y = x_{ob}\,sin(2\pi-\chi) + y_{ob}\,cos(2\pi-\chi),
\end{eqnarray}
where $x_{ob}$ and $y_{ob}$ are the measured coordinates, $\chi$ is the position
angle of the jet in the plane 
of the sky defined in Section~\ref{fobs} and $x$ and $y$ are the rotated
coordinates. Note that once rotated, the axis $y_{ob}$ falls on the negative $y$
axis. Figure~\ref{fig:coords} shows the different coordinate systems used in
this work. The coordinate system $x,\, y$ is only used for wavelength detection
purposes. In the next sections we will only refer to $x_{ob},\, y_{ob}$
(observed positions) and $x_h,\, y_h,\, z_h$ (intrinsic coordinate system). 

\begin{figure*}[!t]
      \centering
   \includegraphics[clip,angle=0,width=\textwidth]{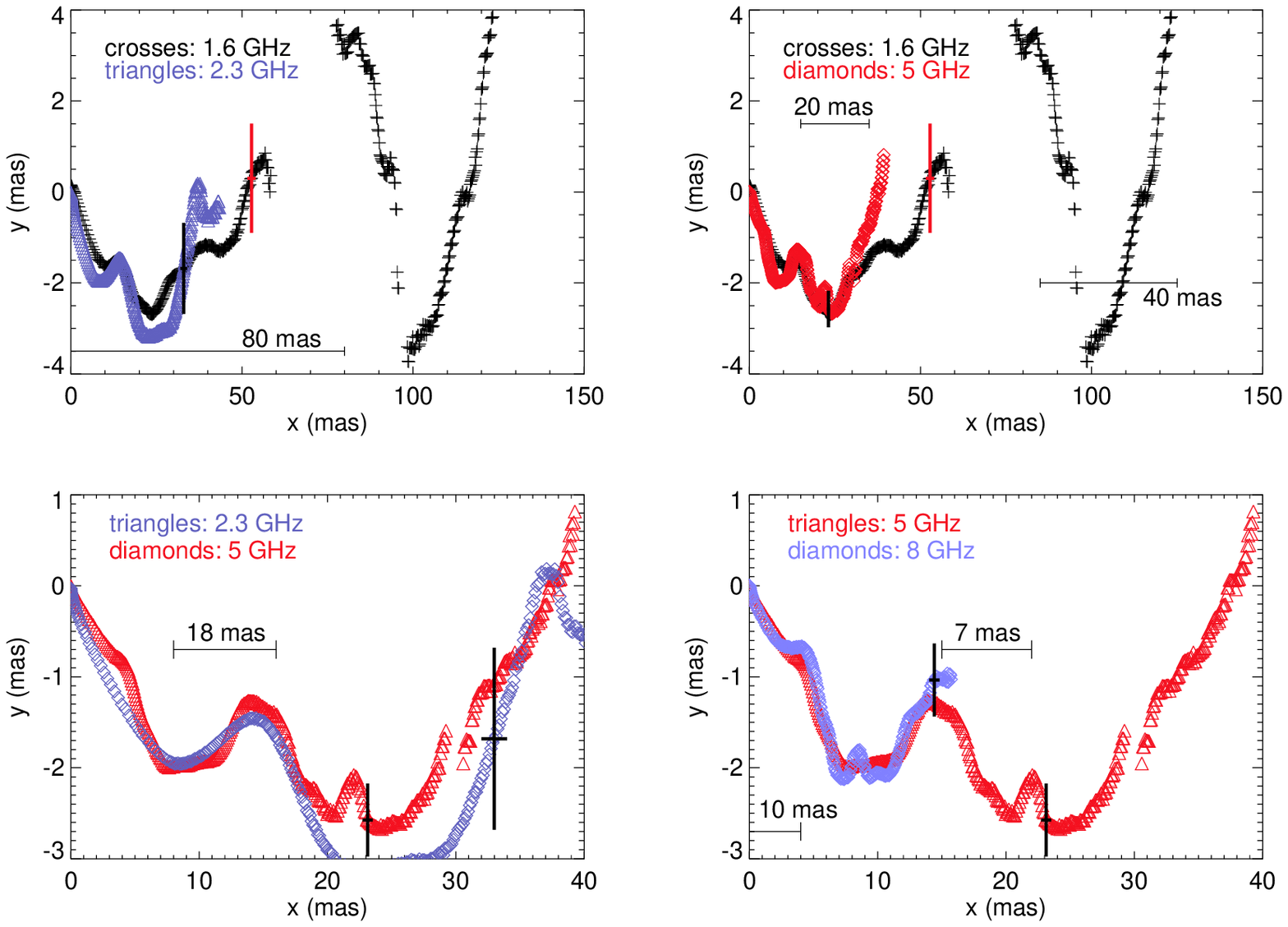}
   \caption{Transverse displacement of the jet ridge-line around the jet axis at several frequencies
in 1997. 1.6, 5 and 8~GHz images 
were obtained with VLBA data, whereas the 2~GHz one was obtained with Global
VLBI. We indicate the
wavelengths by the observed sizes between maxima or minima (half a wavelength in
the case of the 
10 and 18 mas oscillations). The black or red crosses show the error bars (one
fifth of the beam size  in the direction of the displacements), at randomly
selected points, to be applied to each data 
point in the ridge-line at a given frequency. The errors are not shown at all
points for the sake of clarity.} 
   \label{fig:ridge97}
   \end{figure*}

\begin{figure*}[!t]
      \centering
   \includegraphics[clip,angle=0,width=\textwidth]{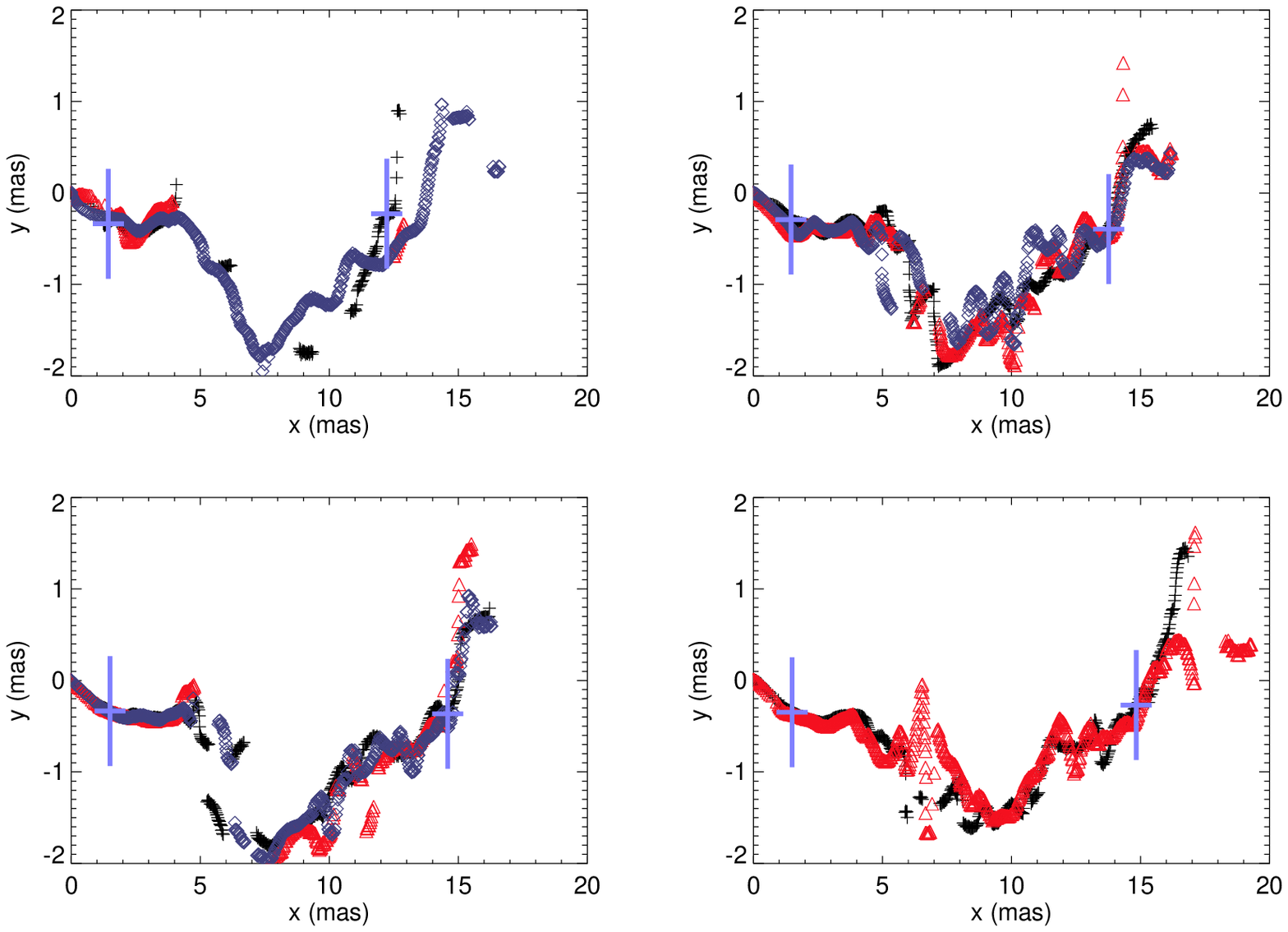}
   \caption{Transverse displacement of the jet ridge-line at 15~GHz and different
epochs around the position angle of the jet at this frequency, ranging from 1998 to 
2009. In the upper left panel, black crosses represent the ridge-line from March
1998, red triangles from June 2000 and blue diamonds from June 2002. In the upper right panel, black
crosses represent the ridge-line from March 2003, red triangles from September
2004 and blue diamonds from May 2005. In the lower left panel, black crosses
represent the ridge-line from October 2005, red triangles from September 2006
and blue diamonds from August 2007. In the lower right panel, black crosses
represent the ridge-line from July 2008 epoch and red triangles from May 2009.
As in Fig.~\ref{fig:ridge97}, the blue crosses show the error bars (one fifth of
the beam size  in the direction of the displacements), at randomly selected
points, to be applied to each data point in the ridge-lines. The errors are not
shown at all points for the sake of clarity.}
   \label{fig:ridgeu}
   \end{figure*}

Figure~\ref{fig:ridge97} shows the transverse location of
the ridge-lines relative to the jet axis ($\chi=198\arcdeg$) at a given epoch. The short ridge-lines fall on top of the long ones and do not show symmetry around the axis as they display, at most, half of the longest-observed wavelength, which determines the jet axis. In contrast, Figure~\ref{fig:ridgeu}  
shows the location of the ridge-line for different epochs at a given frequency (15~GHz), using the position angle of the jet
at this frequency ($\chi_{15}=210\arcdeg$) for the sake of clarity and to facilitate the detection of wavelengths
in this plot.

\begin{figure}[!t]
      \centering
   \includegraphics[clip,angle=0,width=0.37\textwidth]{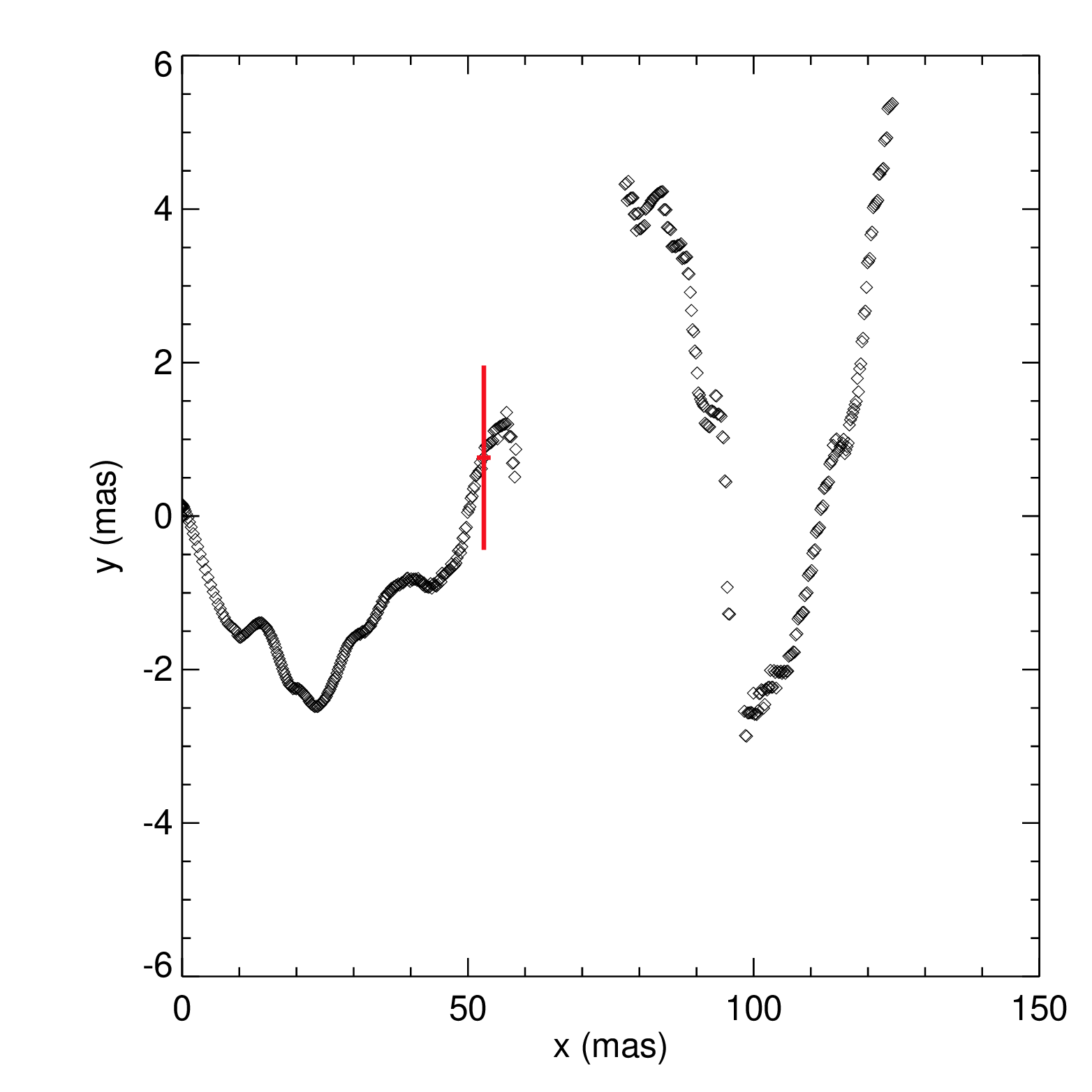}
   \includegraphics[clip,angle=0,width=0.37\textwidth]{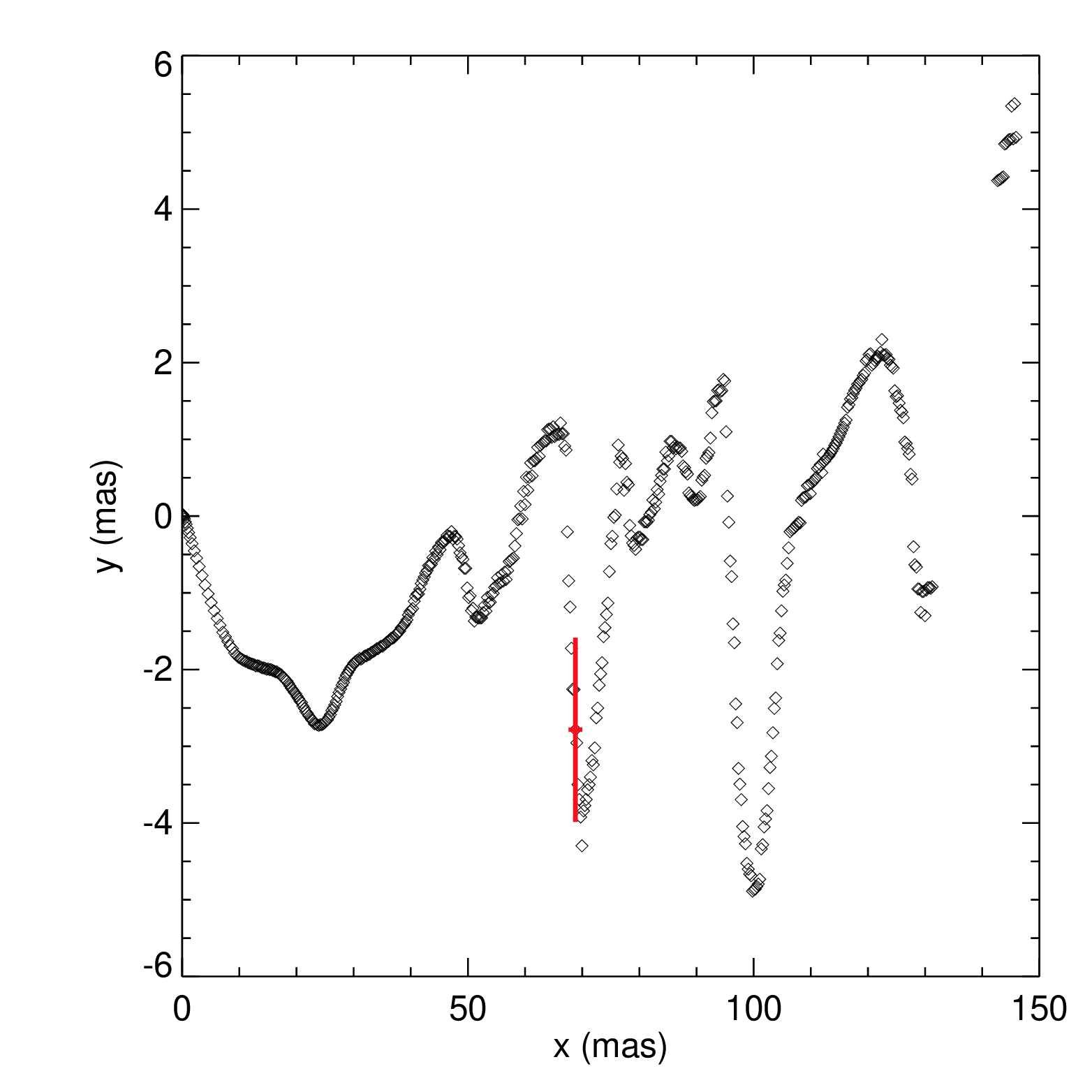}
   \includegraphics[clip,angle=0,width=0.37\textwidth]{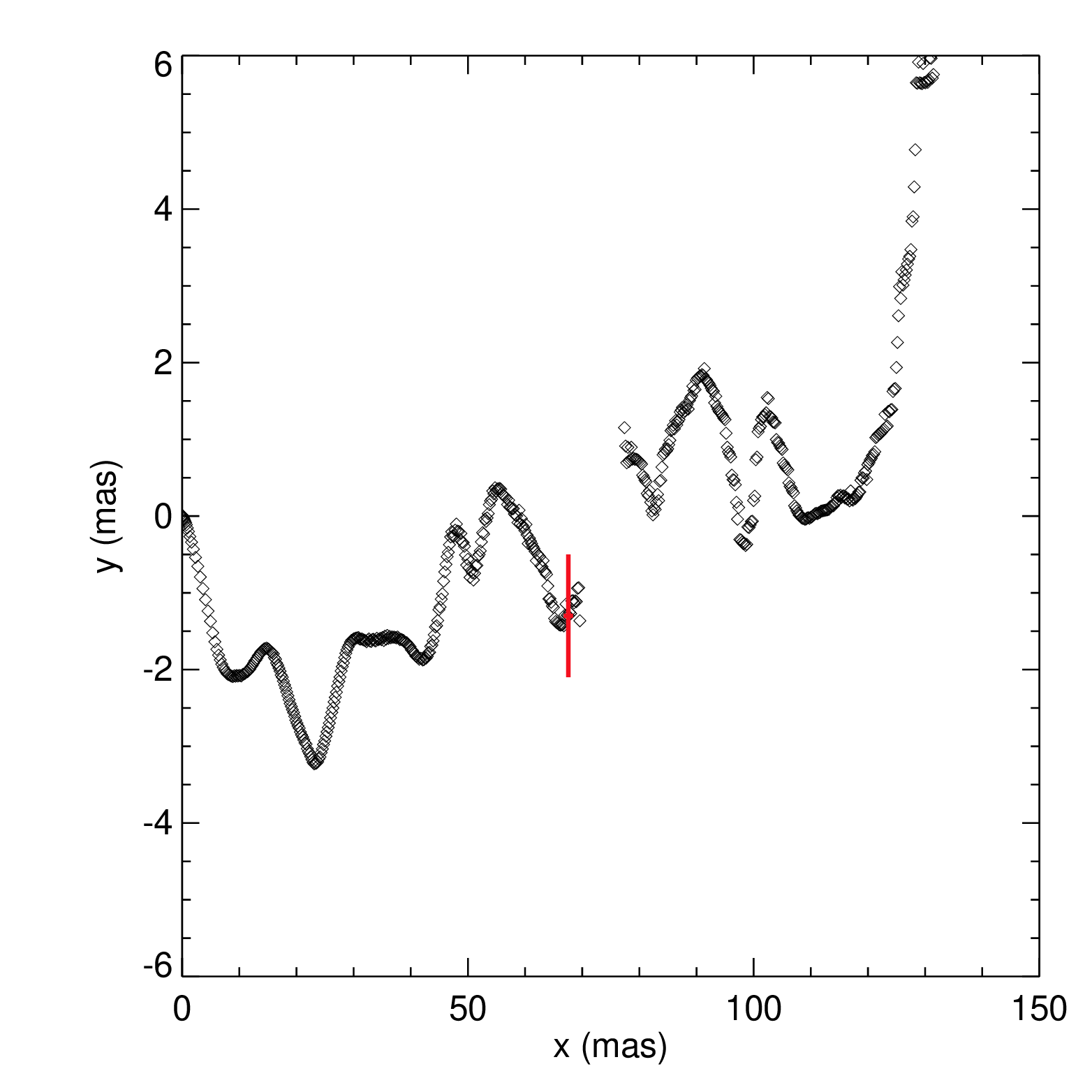}
   \caption{Rotated ridge-lines obtained for 1.6~GHz in 1997 (top), 2003
(center) and 2008 (bottom). 
As in Fig.~\ref{fig:ridge97}, the red crosses show the error bars (one fifth of
the beam size in the direction
of the displacements), at randomly selected points, to be applied to each data
point in the ridge-lines. 
The errors are not represented for all points for the sake of clarity.} 
   \label{fig:ridgel}
   \end{figure}

At 1.6~GHz, the ridge-line of the jet is dominated by a long wavelength
oscillation of $\simeq$80~mas (Fig.~\ref{fig:ridge97}). A small scale modulation
of 10~mas is also observed in the inner region ($<6$~mas to the core, as indicated in 
the bottom right panel of Fig.~\ref{fig:ridge97}). A $\simeq20$~mas structure is observed in the
inner 40~mas at 2 and 5~GHz (e.g., left bottom panel in Fig.~\ref{fig:ridge97}).
 Farther downstream, a high-amplitude ($\simeq$4 mas) structure with wavelength
$\simeq$40~mas is the most remarkable feature (e.g., upper right panel in
Fig.~\ref{fig:ridge97}).  There is an even shorter, small-amplitude wave, with
wavelength 5~mas, observed at 8 and 15~GHz in the inner 15~mas, followed by
another with wavelength 7-8~mas between 20 and 30~mas  
(indicated in the bottom right panel of Fig.~\ref{fig:ridge97}). Beyond 30~mas
the data for the higher 
frequencies have low signal-to-noise. Figure~\ref{fig:ridgel} shows the
ridge-lines at 1.6~GHz at different epochs (1997, 2003 and 2008). Although the
main structures are very similar, the long wavelength is not so clearly observed
in the last epochs and short-wavelengths seem to dominate the structure at
larger distances from the core. 

The ridge-line at 1.6~GHz shows two peaks starting at 80~mas from the core in
the first epoch 
(Fig.~\ref{fig:ridge97}, upper-left panel). The first peak does not appear in
the next epochs 
(see Fig.~\ref{fig:ridgel}). In addition, there is a low signal-to-noise in this
region. This leads to 
doubts as to the real nature of the first peak. The second increase in the
amplitude (120-140~mas) 
is present in all  three epochs.

From the radio maps, we see that the jet emission drops when the ridge-line
perturbation rises in the x-y 
plot at 1.6~GHz. This can be interpreted as the jet veering slightly away from
the line of sight. Moreover, the jet at higher frequencies is seen only up to
30-40 mas, i.e., exactly where this effect should be expected if the ridge-line
veers away from the observer. This deviation is, nevertheless, very small
because the amplitude of the perturbation is small compared to the axial
wave-length ($A \sim 4$~mas and $\lambda_{\rm{obs}} \simeq 80$~mas, see
Fig.~\ref{fig:ridge97}).

\subsection{Wave speeds} \label{wv}

\begin{figure*}[!t]
      \centering
   \includegraphics[clip,angle=0,width=0.45\textwidth]{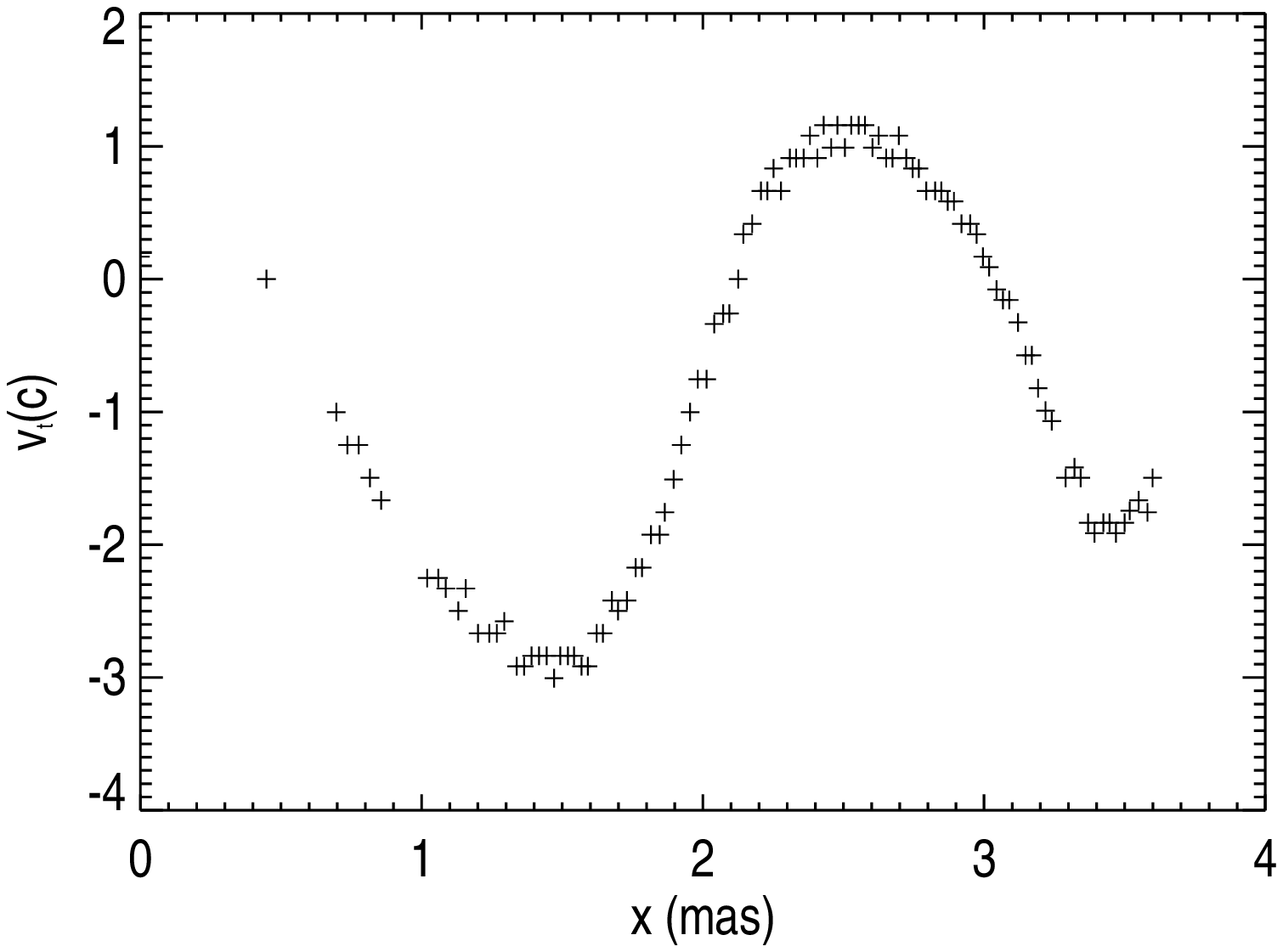}
   \includegraphics[clip,angle=0,width=0.45\textwidth]{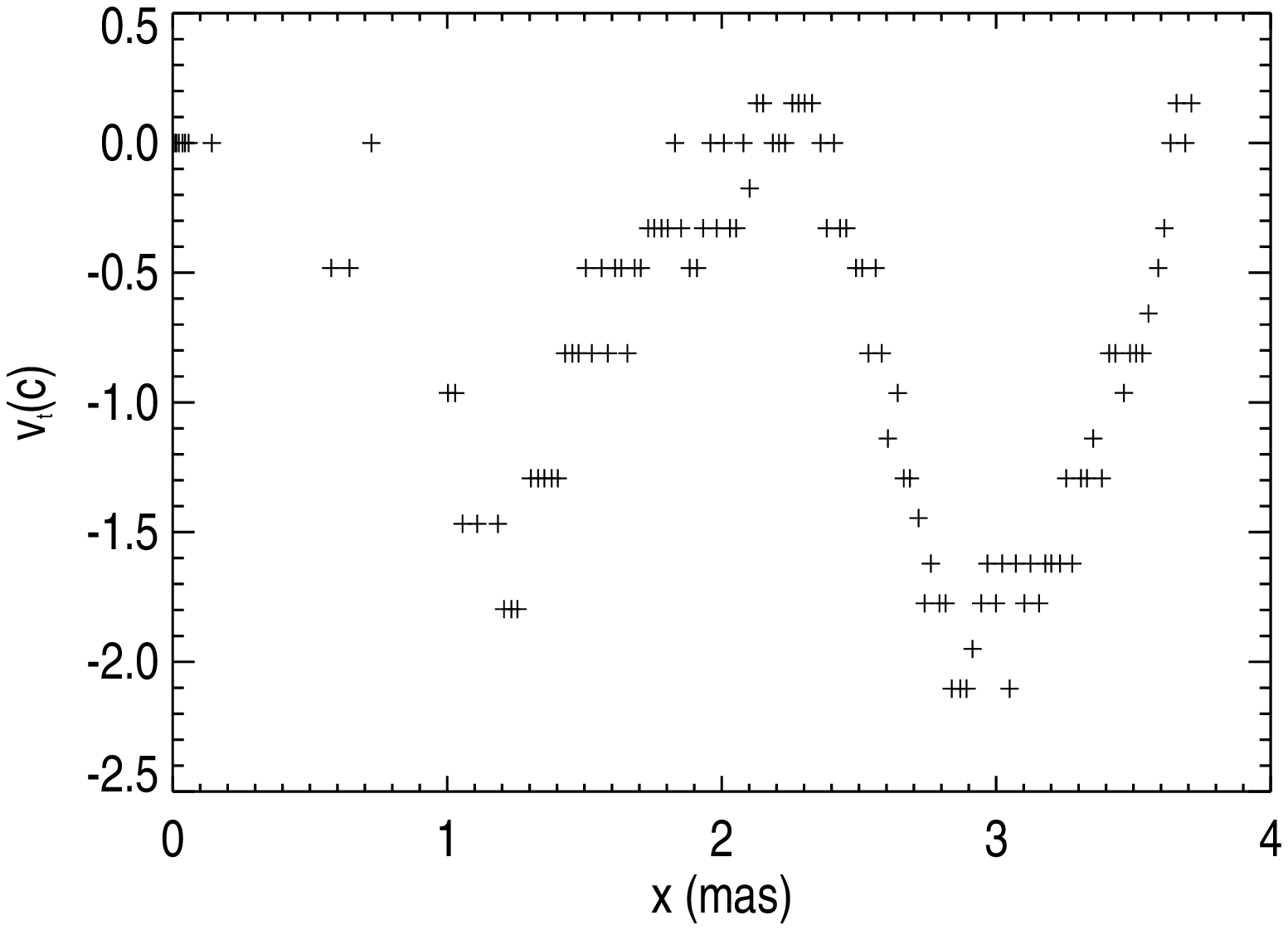}
   \caption{Transversal velocities measured from 15~GHz VLBA data between epochs
2002-2003 and 2008-2009, respectively.} 
   \label{fig:velu}
   \end{figure*}

We have tried to measure the speed of the waves, both transversally and axially.
 To obtain a transversal motion we take fixed points in the $x_{ob}$ coordinate,
defined by Equation~\ref{rot} and Figure~\ref{fig:coords}, and derive the
displacements in the $y_{ob}$ coordinate between two epochs. By doing this, we
would expect an oscillation in the transversal velocities if the apparent
sinusoidal wave is moving spatially along the $x_{ob}$-axis or if a standing
sinusoidal wave pattern is growing temporally in amplitude. The first and second
images in Figure~\ref{fig:velu} give the transverse velocities that are found
using the difference in ridge-line position between 15~GHz VLBA observations
made in 2002 and 2003, and made in 2008 and 2009 (see Table~1). 
Figure~\ref{fig:velu}  shows transversal velocities from the radio core out to
about 4~mas. The images for this part of the jet in 2002 and 2009 can be found in
the bottom leftmost and rightmost panels in Figure \ref{fig:xmaps}. In
Figure~\ref{fig:velu} we see that the transversal velocity is zero at the radio
core since here it is assumed that the radio core is at a fixed position. In
this regard note that the oscillation in transversal velocities is not symmetric
around zero and we will return to a more detailed discussion of this and other
transversal velocity issues in Section \ref{revisit}. The oscillation pattern
could be produced by a growing standing wave or a moving wave. However, the
velocities obtained are larger than $c$ and for a standing wave would imply
amplitude growth at greater than light-speed and thus not possible. On the other
hand, it is possible to find a transverse speed greater than $c$ for a moving
wave and we will show under what conditions in the next sections. Additionally,
we observe an identical region of the jet that the oscillation pattern
in transversal velocity has shifted between the epoch pairs along the
$x_{ob}$-axis by $\sim 0.4$~mas and includes only negative values for the
transversal velocity. This shift indicates motion of the pattern along the
$x_{ob}$-axis that would not occur for a temporally growing standing wave. We
will return to the reasons for the lack of similar maxima in the positive and
negative velocities in Section \ref{revisit}. However, the errors in the
ridge-line position are usually larger than the equivalent distance travelled by
light in the interval between epochs. This results in values larger than $c$
even if the true values are less than c. Actually, we have found that the maxima
in transversal velocity are larger for shorter time-intervals between epochs,
where less real motion is expected and differences in positions are dominated by
errors. Thus, we cannot conclusively rule out a temporally growing standing wave
based on the transverse velocities in Figure~\ref{fig:velu}, but examination of
the differences between the 2002 and 2009 15~GHz panels in Figure
\ref{fig:xmaps} allows us to conclude that a moving wave pattern is the most
likely cause of the transversal velocities.

It is also possible to compute the axial speed, using fixed $y_{ob}$ position
and computing the difference in the 
projected axial coordinate between epochs. We would expect a constant value for
this speed if the wave is moving. In fact we do not obtain a constant velocity.  
Although the evolution of the ridge-lines at 15~GHz (see Fig.~\ref{fig:ridgeu}) 
shows that the small peak at 4~mas in the left top panel seems to move to the 
right with time, no regular velocity pattern has been found in relation with 
this motion. The reason for this might be errors in determining the axial locations
due to blending of the axial and the second transversal spatial coordinates on the plane 
of the sky. In addition, whereas in the case of
only one dominant wavelength, the measure could be easy, the superposition of the various 
different wavelengths with different potential speeds would lead to variability in
displacements from epoch to epoch. In the next sections, we provide the equations that 
couple the transverse and axial speeds with the amplitude and wavelengths. However, a detailed
derivation of these equations will be presented elsewhere (Perucho et al., in preparation).
 
\section{From observations to theory}\label{sec4}

\subsection{Observed and intrinsic structures} \label{obshel}

Past work has shown that oscillations like that observed in 0836+710 can be
produced by three-dimensional helices developing in a relativistic outflow. In
this case the position of the ridge-line  will trace the accompanying helically
twisted pressure maximum along the jet, e.g., the evidence presented in
Section~\ref{fobs}. In order to 
model these structures we have to take into account projection and relativistic
effects due to the possible 
propagation of the waves with a high velocity. 

From Figure~\ref{fig:coords}, we can see that the coordinate $y_{ob}$ is not
affected by projection or light travel 
time-delay and coincides with the corresponding coordinate of the intrinsic
helix $y_h$. However, coordinate $x_{ob}$ is affected by both projection and
light travel-time effects and depends on both the transversal 
coordinate $x_h$ and on the position $z_h$ along the axis of the helix and the
jet itself.

The observed helical wavelength is related to the intrinsic one\footnote{We
will call \textit{intrinsic} a 
perpendicular position (viewing angle of $90^\circ$) of an observer at rest with
respect to the jet flow.} via the viewing angle, $\alpha$, and the intrinsic
wave speed, $\beta_w$, following\footnote{cf. \cite{pl07}, where cosmological
effects were mistakenly considered: the redshift effect in the frequency also
affects the measured velocities, but not the measured wavelengths.}
 \begin{equation}\label{wave}
   \lambda_{obs} = \lambda_{int}\,\frac{\sin \alpha}{(1-\beta_w \cos \alpha)}\,,
 \end{equation}

The intrinsic equations describing a helix moving along the $z_h$ axis with
speed $v_w$ are:
 \begin{eqnarray}\label{hel}
  x_h =
A_0\,e^{z_h/\lambda_i}\cos\left(\frac{2\,\pi(z_h-v_w\,t)}{\lambda}
+\varphi_0\right) \nonumber \\
  y_h =
A_0\,e^{z_h/\lambda_i}\sin\left(\frac{2\,\pi(z_h-v_w\,t)}{\lambda}
+\varphi_0\right)
\end{eqnarray}
where $A_0$ is the initial amplitude of the helix, $\lambda_i$ is the growth
length of the helix if it couples to an 
instability, $\lambda$ is the wavelength, and $\varphi_0$ is the phase. These
four variables plus the wave velocity are the unknowns that determine the
structure of the wave. Finally, $t$ is the time, and $z_h$ is the axial
coordinate. This helix is projected onto the sky plane ($x_{ob}$,$y_{ob}$) where
the helical axis coordinate, $z_h$, and $x_h$ are to be projected onto the sky
plane x-axis and the helical axis makes an angle, $\alpha$, with respect to the
line of sight (which lies in the $x_h, z_h$ plane). The relation between the
observed ($x_{ob},y_{ob}$) and intrinsic ($x_h,y_h,z_h$) coordinates is the
following (Perucho et al., in preparation):
 \begin{eqnarray}\label{proj}
x_{ob} = x_h(z_h) \left[ \cos \alpha - \left( {\sin^2 \alpha \over 1 - \beta_w
\cos \alpha } \right) 
\beta_w \right]+ \nonumber \\ 
z_h {\sin \alpha \over (1 - \beta_w \cos \alpha)} \nonumber \\
  y_{ob} = y_h\,.
\end{eqnarray}
which, in the absence of motion reduces to:
 \begin{eqnarray}\label{proj2}
  x_{ob} = x_h\cos \alpha + z_h\sin \alpha \nonumber \\
  y_{ob} = y_h,
\end{eqnarray}

The observed ridge-line points could then be fitted to a helix by means of a
minimizing $\chi^2$ method, using the 
above equations. However, Equation~(\ref{proj}) embeds a key degeneracy in the
problem: the observed wavelength is a combination of the wave velocity and the
intrinsic wavelength. Thus, there are different combinations of both parameters
that may result, for a given viewing angle, in the observed wavelength. 
In order to uniquely identify the intrinsic wavelengths it is necessary to
measure the wave speed of the different
structures. A simple way to do this would be to measure motions in the
ridge-line between epochs by following visible features such as maxima or minima 
and relate those to intrinsic motions. However, we have seen
in Section~\ref{wv} how difficult this may be with the present accuracy of the VLBI 
observations. Thus, we can only find limits to the parameters implied by the structure.

An additional complication arises when more than one periodical structure is
found in a jet. In this case each periodical structure has to be independently
fitted if it propagates with different speed. The reason is that the
relativistic and projection effects, which depend on the propagation velocity,
determine the position of the displacements along the observed axis following
Equation~(\ref{proj}), which makes independent parametrization necessary. Once the velocity
and intrinsic wavelength of each wave are known, their contributions to the
ridge-line displacement, $y_{ob}$ in Equation~\ref{proj}, at a given location,
$x_{ob}$, have to be added and then compared to the observed ones. 

A more detailed description of the modeling of relativistic helices will be
published elsewhere (Perucho et al., in preparation).

\section{Jet properties derived from the observations}

\subsection{The 40~mas wave}\label{40}

In all three ridge-lines observed at 1.6~GHz, a $40$~mas wavelength between
$x\simeq 80-120$~mas (see 
Fig.~\ref{fig:ridges}), indicates a shortening of the wavelength dominating the
shape of the ridge-line and deserves some discussion. This structure can be
explained in one of the following three ways: 
a) the amplitude of a shorter wavelength increases to larger than that of a
longer one, 
b) the angle to the line of sight decreases and the long wavelength is observed
with a shorter wavelength, or 
c) the longer wavelength is decelerated. Options \textit{b} and \textit{c}
follow from Equation~\ref{wave} that 
relates the observed and intrinsic wavelengths, taking into account projection
and relativistic effects.

We discuss here the different possibilities\footnote{Considering that the changes discussed occur 
already at large de-projected distances from the core and that there are evidences of KH instability 
to dominate at distances larger than a few parsecs \citep[see, e.g.,][]{ha06,ha11,pe11}, 
we will restrict the discussion to this kind of instability.}:
\begin{itemize}

 \item \textbf{a}: The longer initial wavelength could be due to a forced change
in the jet direction that  due to its low frequency couples to a long wavelength
slowly growing KH mode. The shorter wavelength is the result of a faster
growing, possibly resonant, KH mode, whose amplitude grows sufficiently rapidly
to become larger than that of the forced oscillation. Thus, this is a viable
option.
 
 \item \textbf{b}:  We can find the required decrease in the angle to the line
of sight  for the 80~mas wavelength to change to the 40~mas wavelength using
Equation~\ref{wave}, and keeping the intrinsic wavelength and wave speed
constant.  For small viewing angles Equation~\ref{wave} becomes
$1-\beta_{w,1}\cos\alpha_1 \simeq 1-\beta_{w,2}\cos\alpha_2$ with an error of 
$< 10\%$, and:
\begin{equation}
\frac{sin\,\alpha_1}{sin\,\alpha_2}\simeq\frac{\lambda_{obs,1}}{\lambda_{obs_2}}
\simeq 2. 
\end{equation} 
This means that $\alpha_2\simeq\alpha_1/2$, and a decrease in the angle to the
line of sight 
down to one half of its original value is needed. This could be produced by a
change in the conditions of 
the ambient medium when the jet exits the ISM of the galaxy. However, this would
imply the presence of a standing shock where the jet  changes direction.  Such a
shock is not observed and there is no hint that the jet direction is appreciably
changed. Thus, this option is unlikely.

 \item \textbf{c}: A decrease in the velocity of propagation of the 80~mas wave
could also lead to the observed wavelength decrease. It is well-known that the
perturbations propagating in a flow cannot have velocities larger than that of
the underlying fluid. For example, if the flow decelerates, the wave would slow
as well.  Now using Equation~\ref{wave} and keeping the intrinsic wavelength and
the viewing angle constant we have:
 \begin{equation}
 \frac{(1-\beta_{w,2} \cos\alpha)}{(1-\beta_{w,1}
\cos\alpha)}=\frac{\lambda_{obs,1}}{\lambda_{obs,2}} \simeq 2,
 \end{equation}
which should be solved for $\beta_{w,2}$. For example, taking $\beta_{w,1}=0.9$
for the inner region, 
$\alpha=3^\circ$, $\lambda_{obs,1}=80$~mas and $\lambda_{obs,2}=40$~mas, we
obtain $\beta_{w,2}=0.8$, whereas for $\beta_{w,1}=0.6$ we get
$\beta_{w,2}=0.2$. This option allows us to place constraints on $\beta_{w,1}$
using $\lambda_{obs,1} (1-\beta_{w,1}\cos\alpha) < \lambda_{obs,2}$, to obtain
$\beta_{w,1}>0.5$. It is 
 known that a flow can be decelerated when a helical perturbation grows to
nonlinear values and disrupts the collimated flow resulting in conversion of
kinetic into thermal energy \citep[see e.g.,][]{pe+05,pe+10}. The uncollimated
structure of the 0836+710 jet at kiloparsec scales  indicates that this is a
viable option.
\end{itemize}

Figure~\ref{fig:ridgel} showed that the long wavelength is not clearly observed
in the later epochs at 
1.6~GHz and short-wavelengths seem to dominate the structure at larger distances
from the 
core at the later epochs. It seems likely that some combination of options
\emph{a} and \emph{c} are favored by the observations.  

\subsection{On the presence of a shear-layer in the jet}
\citet{pl07} presented results that pointed towards the jet in 0836+710 having a
significant velocity shear layer. For this study, the authors used as a set of jet 
parameters those resulting from a previous work by \citet{lo98}: Lorentz factor $\gamma=11$, 
Mach number $M_j=6$, and jet/ambient density ratio $\rho_j/\rho_a = 0.04$. They studied
the different characteristic wavelengths growing in a jet with a thin shear-layer ($\simeq$ 10\% of the jet
radius), and a thick one ($\simeq$ 60\% of the jet radius). The result showed that the observed wavelengths in
the jet structure could be better explained transversally stratified jet. A mistake in the equation used in 
that work \citep{pl11} led to misidentification of the modes in the original work, but the conclusions were 
shown to remain valid.

\begin{figure*}[!t]
      \centering
   \includegraphics[clip,angle=0,width=0.48\textwidth]{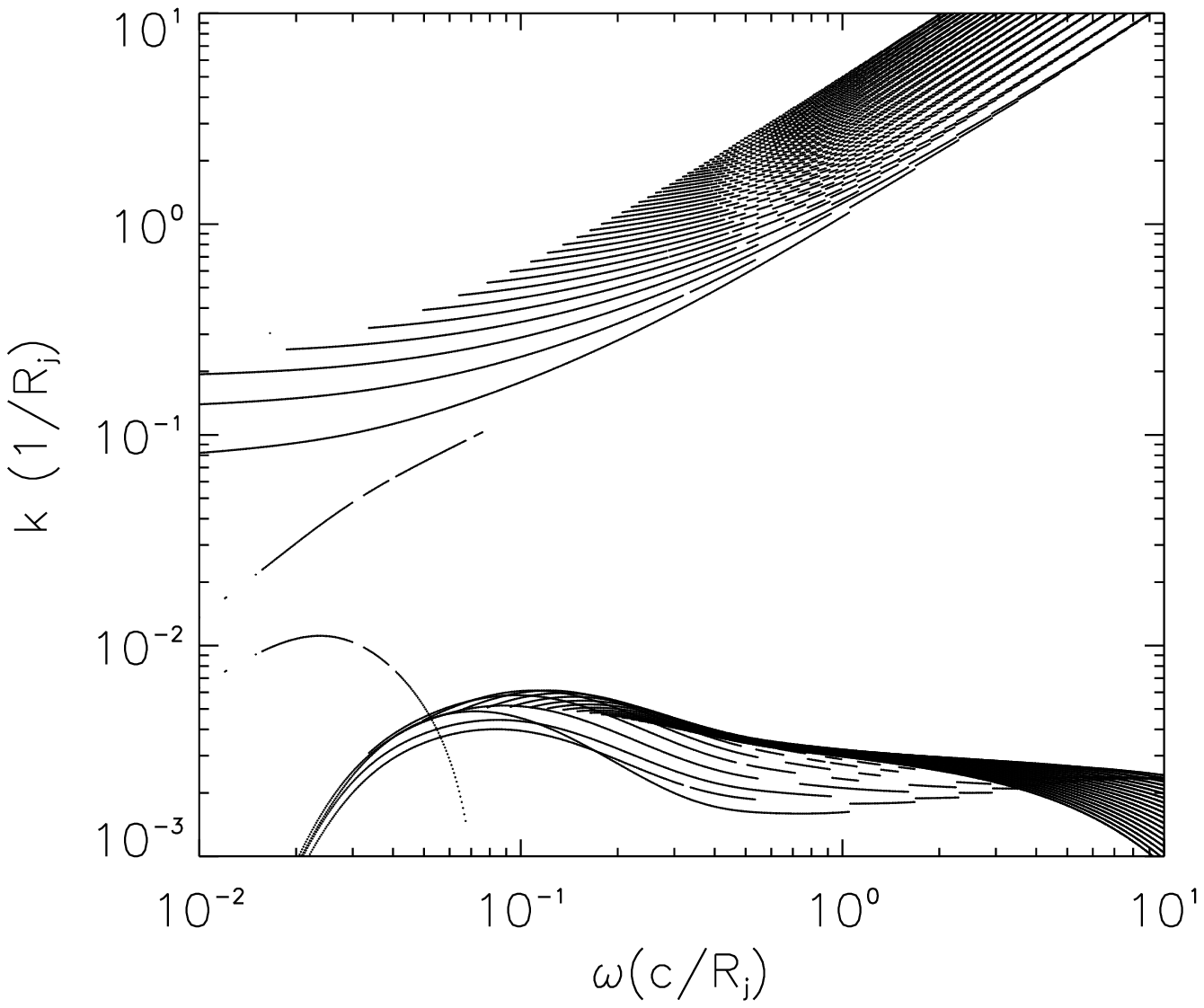}
   \includegraphics[clip,angle=0,width=0.48\textwidth]{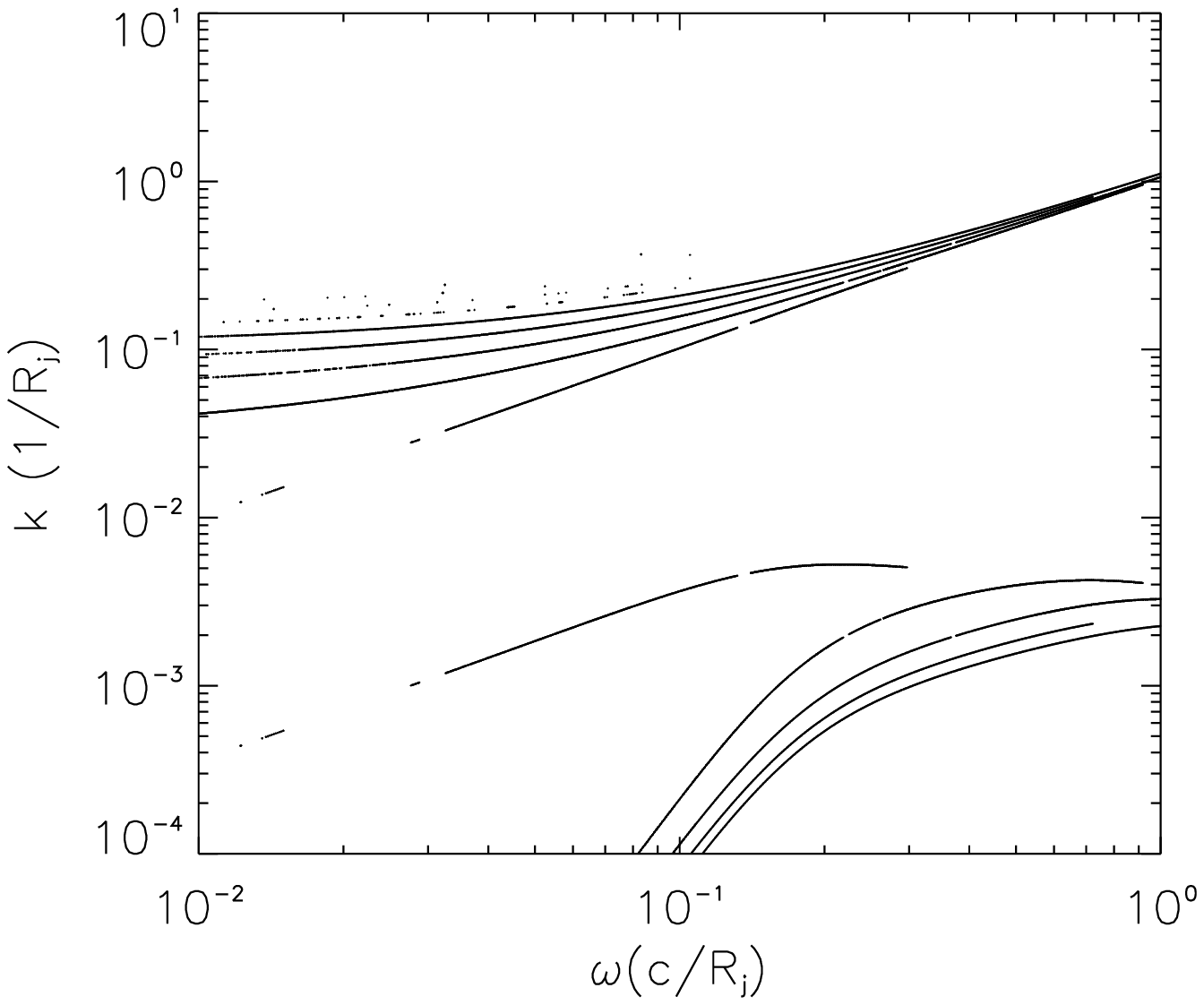}
   \caption{Solutions of the linear problem of KH instability for a
 sheared jet \citep[see][for details]{pl07}.
 The left panel shows the real (upper curves) and imaginary (lower curves) parts
 wavenumber with wave-frequency, whereas the right 
 panel shows the wave velocity (Re($\omega/k$)) with wave-frequency.}
   \label{fig:linear}
   \end{figure*}

In the present work, we have been able to accurately measure the wavelength of the
longest mode to be closer to 80~mas than to 100~mas. A correct treatment reduces the values of the wavelengths 
obtained in Table~1 of \citet{pl07} by a factor of $\simeq 3$.
This implies that their solutions both with and without a shear layer cannot
explain the observed structures, reported to be 100~mas and 7.7~mas
\citep{lo98}, with the jet parameters used to obtain the results shown in
Figure~1 of \citet{pl07}. We reproduce here the solutions for the sheared jet in 
Fig.~\ref{fig:linear}.

Taking 80~mas for the wavelength, $3^\circ$ for the viewing angle, and that the velocity of this long wave has to
be $\beta_w > 0.5$ (see Section~\ref{40}), we obtain an upper limit to
the intrinsic wavelength using Equation~\ref{wave} of
 \begin{equation}
 \lambda_{int,80\,\rm{mas}} < \left(\frac{1 - 0.5 \cos 3^\circ}{\sin
3^\circ}\right) 80\, \rm{mas} = 765\,\rm{mas}.
 \end{equation}
In terms of the solutions presented in \citet{pl07}, this is
$\lambda_{int,80\,\rm{mas}} < 45\,R_{\rm j}$, with $R_{\rm j}=17$~mas \citep{pl07}. 
Now the maximum growth rate of the surface mode occurs at $160\,R_{\rm j}$ for the sheared jet and at $105\,R_{\rm
j}$ for the vortex sheet jet, which is clearly different from the obtained intrinsic wavelength. For the sheared jet,
the maximum growth rate of the first body mode is at the frequency $\omega \simeq 0.1 c/R_j$ and 
$\lambda\simeq 42\,R_{\rm j}$. Taking into account that its velocity is $\simeq v_w\simeq 0.5\,c$ 
(computed as $\simeq Re(\omega / k)$), we obtain a value very close to $\lambda_{int,80\,\rm{mas}}$. 
We note that the maximum growth rate of the first body mode for the vortex sheet solution is at
$\lambda\simeq 21\,R_{\rm j}$, with $v_w\simeq 0.6\,c$, farther away from
the required properties to explain the observed wavelength. While the first body mode is not the fastest growing
one in the solution, it could be triggered and generate large deformations in the jet, which are not expected to be 
caused by high-order body modes \citep{pe+05,pe+07,pe+10}. We note that this interpretation is very 
dependent on the measured jet radius.

At the same triggering frequency ($\omega \simeq 0.1 c/R_j$) in the sheared case, high-order body modes could produce 
the 7 - 8~mas wavelengths \citep[identified with the 7.7~mas wavelengths studied
in][]{pl07}. This wavelength could really be $\simeq$10~mas, 
resulting an intrinsic $\lambda_{int,10\,\rm{mas}}\simeq 9\,R_{\rm j}$,
obtained using $3^\circ$ viewing angle and $v_w\simeq 0.2\,c$ ($\simeq Re(\omega / k)$). 
The fact that both growing modes would be triggered at the same frequency simplifies the nature of the 
driving mechanism or mechanisms to a single one. 
In this case, the small wavelengths can be triggered as harmonics of the longest excited one \citep{pe+05}. 
The new resulting frequency is a factor of 4 larger than that given in
PL07 for the longest wavelength. Thus, the driving period would be
now $T_{dr} \simeq 1.4 \times10^7$~yrs, and similar to the one given by \cite{ha94}
for 3C~449. 

The helical jet structure and appearance is assumed to be due to KH instability as in
\citet{har00,har03} and \citet{hr+05}. In these works, it is assumed that the velocity shear layer is very thin
compared to the jet radius and wavelength (vortex sheet approximation). We note that long wavelengths should 
be insensitive to a modest shear layer so a vortex sheet long wavelength analytical approximation
would be good enough to describe the behavior at long wavelengths. An alternative interpretation to the one 
in \citet{pl07,pl11} is that the longest wavelength of $\lambda_{h}^{ob}
\sim 80$~mas is in the long wavelength and low frequency limit of the KH
instability spectrum: $\omega <<< \omega^*$, with $\omega^*$ being the frequency
of the maximum growth rate. Also, we might assume that 
the intermediate wavelength that appears to increase from about  $\lambda_{h}^{ob} \sim
10-12$~mas ($x_{ob} \sim 0 - 5$~mas) to $\lambda_{h}^{ob} \sim 20-24$~mas
($x_{ob} \sim 5 - 15$~mas) to $\lambda_{h}^{ob} \sim 40-48$~mas ($x_{ob} \sim 15
- 35$~mas) at the different apparent distances along the jet are each operating
at about the local resonant wavelength and frequency, i.e., $\omega \sim\omega^*
\propto a_{ex}/r_j$.  Finally, the shortest wavelengths $\lambda_{h}^{ob} \sim
4.5$~mas ($x_{ob} \sim 5 - 15$~mas) and $\lambda_{h}^{ob} \sim 7-8$~mas ($x_{ob}
\sim 20 - 30$~mas) might be assumed to be a body or shear layer mode operating
at the appropriate resonant frequency and wavelength for that mode. This would require the jet 
having different parameters from those obtained in \citet{lo98}.  

As we see, different interpretations remain possible: vortex-sheet approximation for the longest 
wavelengths or changes in the jet width or in the viewing angle. 
In addition, there are still uncertainties such as the identification in 
\citet{pl11} of the longest wavelength with a first body 
helical mode, instead of the more disruptive, surface mode at the small frequency limit. Overall, 
any results of this kind are subject to sources of error associated with the measure of the jet radius, the 
thickness of the shear-layer, the viewing angle or the parameters used for the jet. 

\section{Superluminal transversal motion and nature of the core}
\label{revisit}

There exist different examples of jets showing apparent superluminal transversal
motion \citep[see, e.g.,]
[the cases of NRAO~150 and OJ~287]{ag07,ag10,ag11}. We also observe this kind of
motion for the ridge-line of 0836+710 at 43~GHz. Within the context of the
present work, there is a possible explanation for this phenomenon. We take into
account the fact that the observations at high frequencies could be showing only
the high pressure core of the jet along the ridge-line that we measure at lower
frequencies (see Sections~\ref{obs} and \ref{obshel}). This means that we may
not observe the whole of the jet, but only the region around the high pressure
maximum. As a result the radio core at 43~GHz would itself oscillate
transversely and axially over time and not be at a fixed position. If the radio
core is associated with a reconfinement shock \citep{ma08}, the presumed
oscillations could well be the triggering point of the observed growing
instabilities, except for the longest one, which could be driven by precession
at the jet formation region.   

Superluminal transversal velocities are possible for purely axial motion, a relatively large 
value for $\Delta y$ in time interval $\Delta t$,  under some conditions (Perucho et al., in preparation). 
This very special situation may only occur for very small viewing angles when $v_w$ is
close to the speed of light and we deal with high-frequency, short wavelength
perturbations. There is a second possibility that we explore in the next 
paragraphs.

\subsection{Transversal shift of the apparent core} \label{sec:tm}

One other possible reason for observed superluminal transversal motion lies in
the mistaken assumption that the radio core remains at a fixed ($x_{ob},y_{ob}$)
position. Since in our interpretation the radio core is
taken as coincident with a pressure maximum inside the jet, we should expect an
oscillation to take place in the radio core location, as the pressure maximum
oscillates transversely in the plane of the sky  and also oscillates axially
with rate and amount depending on the properties of the helical wave (see
Eq.~\ref{proj}). The oscillation should be small (much less than the jet radius)
if only a result of an instability amplitude growth but other triggering
mechanisms can be imagined that could lead to immediate helical amplitudes on
the order of the jet radius. Therefore, when aligning the ridge-line (or any
reference point) using the jet core as a fixed point at two different epochs, it
is possible that the whole ridge-line or positions of fitted components are
displaced by the same small amount as the core has moved in the transversal and
axial direction. The effect of such artificial displacement will clearly affect
any measure of velocities in the transversal direction. This oscillation has already been
reported, using phase-referencing, in at least one object \citep[M81,][]{mv11}.

For the jet in 0836+710, $1\,\rm{mas} \simeq 8.4\,\rm{pc}$ (see
Section~\ref{sec:intro}), and a measured angular motion of only
$0.012\,\rm{mas/yr}$ is equivalent to the speed of light. Thus, any small
displacement of the ridge-line anchored at the core that is not corrected for
possible core motion leads to incorrect  transversal displacement at any point
along the jet and can result in bogus measures of superluminal velocities in the
transverse direction.
   \begin{figure*}[!t]
     \centering
  \includegraphics[clip,angle=0,width=0.48\textwidth]{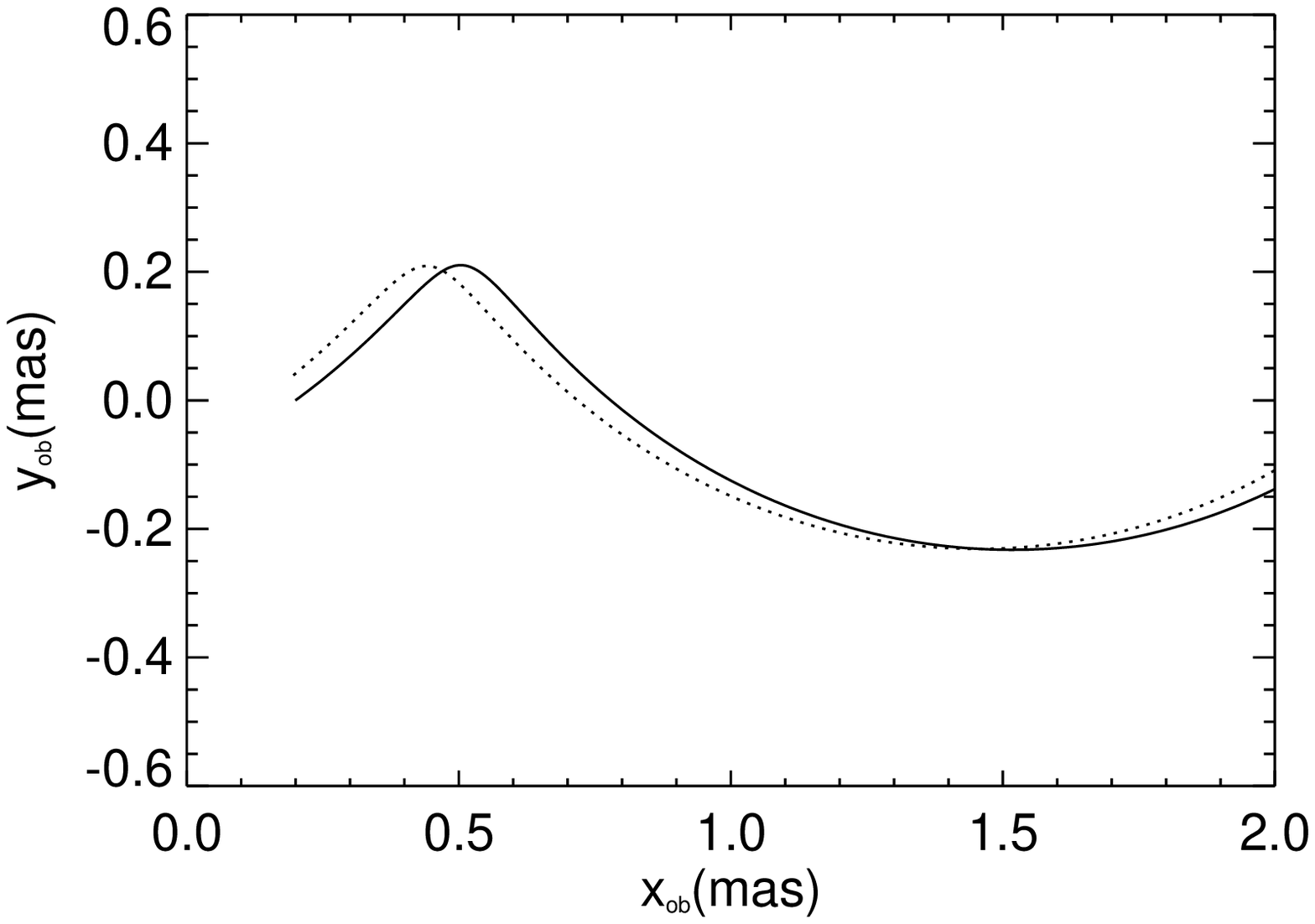}
  \includegraphics[clip,angle=0,width=0.48\textwidth]{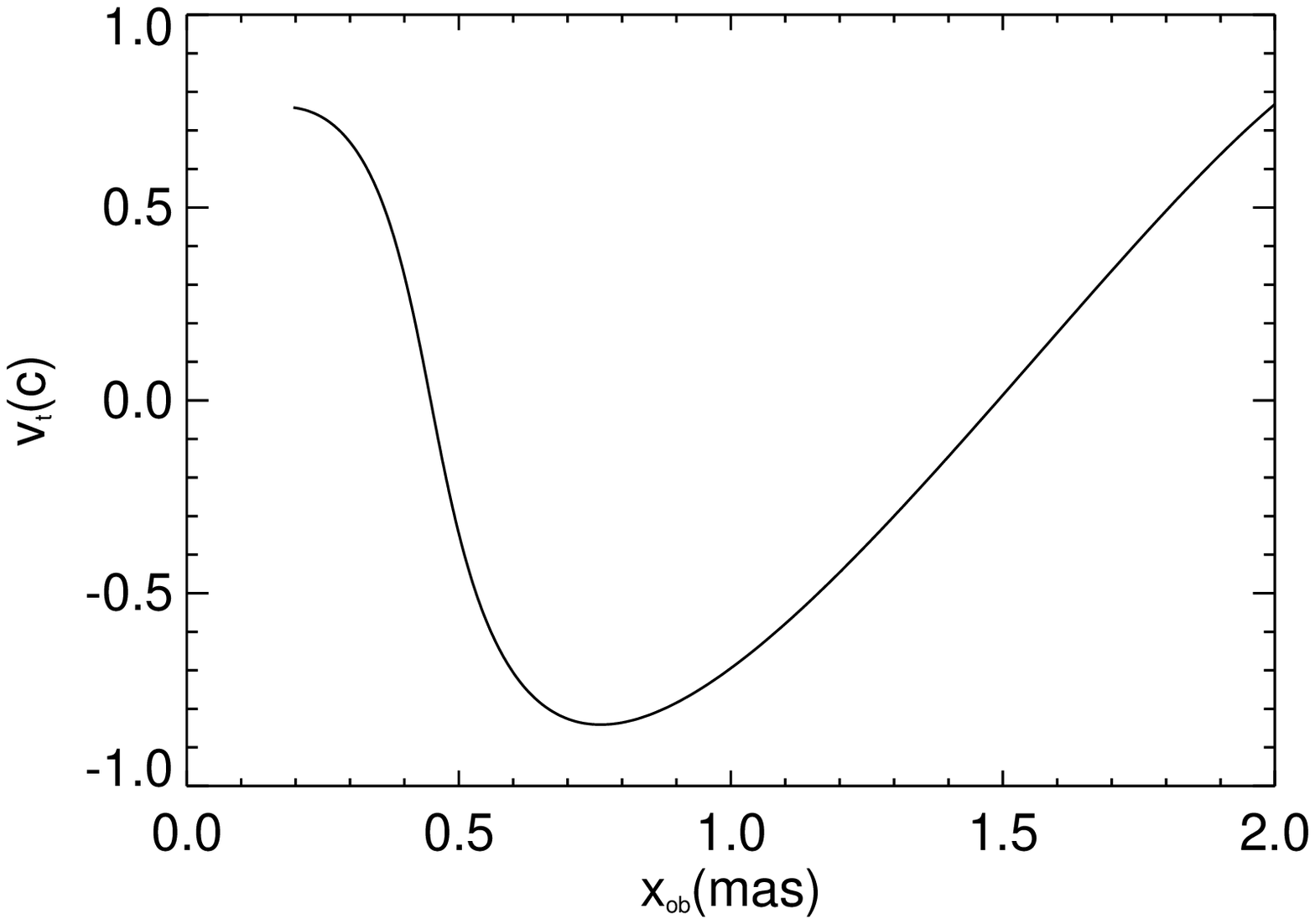}
  \caption{Absolute ridge-line position and motion at $3^\circ$ viewing angle
without core position alignment. The left panel shows the ridge-lines from the
first epoch (solid line) and second epoch (dotted line) separated by a time
interval of 2.11 years. The right panel shows the resulting transversal velocity
versus the observed (projected) axial position.}
  \label{fig1}
  \end{figure*} 
   \begin{figure*}[!t]
     \centering
  \includegraphics[clip,angle=0,width=0.48\textwidth]{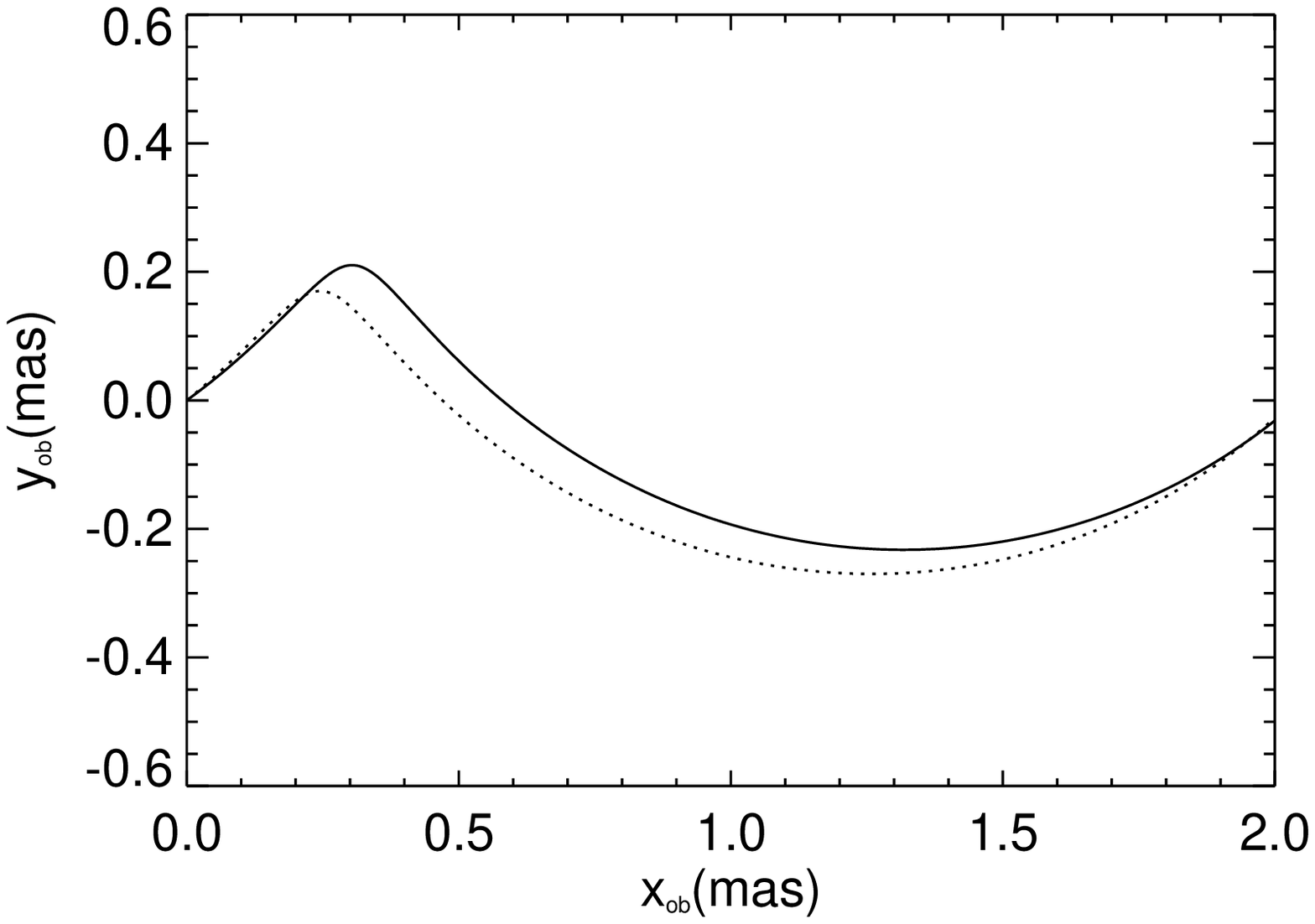}
  \includegraphics[clip,angle=0,width=0.48\textwidth]{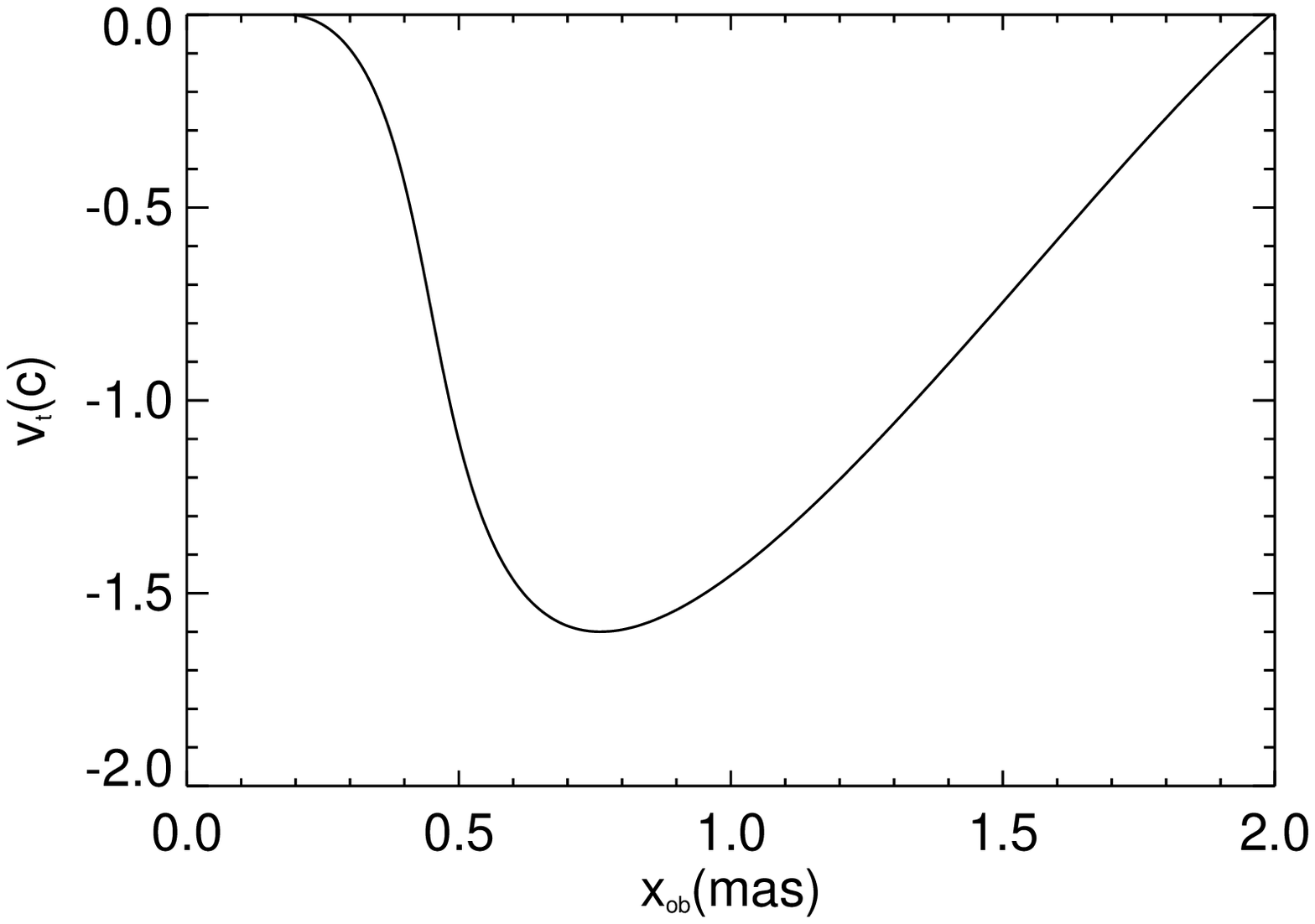}
  \caption{Ridge-line motion with at $3^\circ$ viewing angle alignment. the
ridge-line from the
first epoch is the solid line and second epoch is the dotted line. The time
interval between them is 2.11 years. 
On the right, the transversal velocity versus the observed (projected) axial
position, 
obtained from the transversal displacement measured at a fixed distance to the
core.}
  \label{fig2}
  \end{figure*} 

We have generated an artificial mode pattern of a helical wave with an intrinsic
wavelength of $\lambda = 2\,R_j$, an initial amplitude of $A_0=0.2\,R_j$, and a
wave speed of $0.96\,c$ in order to investigate the role of core shift in the
measurement of transversal velocities.  Figure~\ref{fig1} shows the resulting
ridge-lines at two 
epochs separated by two years (left panel), and the resulting transversal
velocities in units of the speed of light (right panel). Transversal velocities
have been obtained by subtracting the vertical coordinate of the helix at the
different times and dividing the result by the time interval, as we did for the 
observational data (see Fig.~\ref{fig:velu}). With these
parameters, the core moves transversally at an observed $0.006\,\rm{mas/yr}$. In
this case, the intrinsic displacement of the core is respected, and we observe
that all the velocities are subluminal with equal positive and negative maximum
values. The displacement of the ridge-line along the axis is a consequence of
the projection of the axial coordinate, $z_h$, and that perpendicular to the
plane of the sky, $x_h$, (see Eq.~\ref{proj}).

In Figure~\ref{fig2} we re-plot the ridge-lines shown in Figure~\ref{fig1} but
now with the cores at the two epochs forced to the same position at the origin.
This artificial displacement of the core results in apparent superluminal
motion.  We have investigated the dependence of this effect with viewing angle,
and axial wave velocity and amplitude. The effect does not depend on the viewing
angle, as we measure transversal motions, but the transversal velocity does grow
with wave amplitude and wave velocity. Note that positive and negative maximum
transversal velocities are not the same and Figure~\ref{fig2} shows superluminal
velocities very similar to those found between the 2008 and 2009 epochs shown in
Figure~\ref{fig:velu}. The similarities between the transversal velocities
shown in Figure~\ref{fig2} and between the 2008 and 2009 epochs in
Figure~\ref{fig:velu} provide indirect evidence for radio core displacement
accompanying a moving helical wave pattern in 0836+710.
A direct observational test for core motion would consist of phase-referencing
observations of nearby sources containing observed helical patterns \citep[see, e.g.,][]{mv11}. 
In these cases small lateral angular displacements of the core, up to $0.012$~mas/yr
if the wave-velocity is close to 
the speed of light, might be directly detectable. 
  
Core motion manages to explain the transversal velocities found in this work in
an easy way and within the framework of our modeling of helical patterns. We
note that the observed effect only requires very small displacements of the ridge-line 
in the core, well within the errors of our measurements. 
The caveats are that the wave velocity needs to be close to the speed of light
and that the initial amplitude of the wave has to be a relatively large fraction of the wavelength,
10\% in our example. However, from stability analysis we know that, at least in the case of
KH modes, short wavelengths tend to move fast and can produce
high pressures near to the jet surface. Thus, it should not be surprising to
find such high pattern velocities when observing the spine of jets at high
frequencies and close to the central engine. The proposed observational test can
be used to verify this hypothesis. 

\section{Summary}

In this work we have presented evidence for the observed helical structures in
the jet of \object{0836+710} being real, and not artifacts produced by the
observational systematic errors or to insufficient uv-coverage. The evidence
comes from the same wavelengths being found at all epochs, frequencies and
observing arrays.
 We have shown that shorter wavelengths develop on top of the longer ones and,
if their positions along the jet are considered to follow a wave structure, the
resulting location has to be the result of the superposition of the different
waves and will also depend on the dominant wave at a given position and time.

The fact that shorter wavelengths develop on top of longer wavelengths provides
a natural way to explain the misalignment between parsec and kiloparsec scale
helical jets seen at small angles to the line of sight and we conclude that at
the highest frequencies we may be seeing a small region of the jet flow
concentrated around the pressure maxima that we associated with the intensity
maximum ridge-line. This identification is bolstered by 15~GHz VLBA
observations, in which we are able to resolve the jet, that show the position of
the ridge-line. This identification is bolstered by 15~GHz VLBA
observations, in which we are able to resolve the jet, that show the position of
the ridge-line as not coincident with the center of jet. The opening
angle of the jet is found to be the same at all frequencies, with no apparent
frequency stratification.

In the light of our results, we reinterpret the results of \cite{pl07},
correcting a mistake in the calculations and 
in the possible identification of the triggered instability modes. We confirm the general 
conclusions obtained in that work to the extent that our present study allows but
we show that alternative explanations are still possible. Future work will include a detailed
discussion on this topic. 

We have pointed out that measuring the velocities of the waves is crucial to
connecting observed wavelengths to the intrinsic wavelengths responsible for the
observed structures. We have tried to measure the wave velocities but the
errors associated with the resolution at the different frequencies makes this
impractical.  
  
We have also presented a method for modeling helical patterns propagating at
relativistic speeds in extragalactic jets. Future work should include the 
application of this method to the observed structures. 
Finally, we have proposed a possible explanation for
moderate transversal superluminal motions with important implications regarding
the nature of the jet-core. In particular, we have shown that the radio core 
at any frequency could be undergoing transversal motion, implying its possible association
with wave motion. Moreover, the wave-pattern obtained for the
transversal velocity in the 0836+710 jet indicates that moving waves are present
in the jet.
 
Large angular resolution achieved by future space-VLBI projects, combined with
high precision VLBI astrometry, may help to improve the resolution and make it
possible to accurately determine the velocities of the structures in helical
jets in nearby AGN. 

\acknowledgments

The authors thank the referee for his/her
constructive criticism that has helped to improve the manuscript. 
This research has made use of data from the MOJAVE database that is maintained
by the MOJAVE team \citet{li09}. MP acknowledges
financial support by the Spanish ``Minister de Ciencia e Innovaci\'on''
(MICINN) grants AYA2010-21322-C03-01, AYA2010-21097-C03-01 and
CONSOLIDER2007-00050, and by the ``Generalitat Valenciana'' grant
``PROMETEO-2009-103''. MP acknowledges support from the postdoctoral fellowship
``Beca Postdoctoral d'Excel$\cdot$l\`encia'' of ``Generalitat Valenciana'', a
postdoctoral fellowship in  Max-Planck-Insitut for Radioastronomy in Bonn and
MICINN through a ``Juan de la Cierva'' contract. This paper was begun when YYK
was a Research Fellow of the Alexander von Humboldt Foundation. 
YYK is partly supported by the Russian Foundation for Basic Research (grant
11-02-00368). YYK thanks hospitality from the staff in the University of
Valencia and grant CONSOLIDER2007-00050. PEH acknowledges
support from  NSF award AST-0908010 and NASA award NNX08AG83G to the University
of Alabama. IA acknowledges funding support from MICINN grant AYA-2010-14844, CEIC grant P09-FQM-4784, 
NASA award NNX08AV65G, and NSF award AST-0907893. 

\facility{\emph{Facilities}: EVN, VLBA, VSOP}

\end{document}